\documentclass[longauth]{aa}
\usepackage{graphicx}
\usepackage{txfonts}
\usepackage{gensymb} 
\usepackage{xspace}
\usepackage{natbib,twoopt}
\usepackage[breaklinks=true]{hyperref} 
\bibpunct{(}{)}{;}{a}{}{,} 
\makeatletter
\newcommandtwoopt{\citeads}[3][][]{\href{http://adsabs.harvard.edu/abs/#3}%
{\def\hyper@linkstart##1##2{}%
\let\hyper@linkend\@empty\citealp[#1][#2]{#3}}}
\newcommandtwoopt{\citepads}[3][][]{\href{http://adsabs.harvard.edu/abs/#3}%
{\def\hyper@linkstart##1##2{}%
\let\hyper@linkend\@empty\citep[#1][#2]{#3}}}
\newcommandtwoopt{\citetads}[3][][]{\href{http://adsabs.harvard.edu/abs/#3}%
{\def\hyper@linkstart##1##2{}%
\let\hyper@linkend\@empty\citet[#1][#2]{#3}}}
\newcommandtwoopt{\citeyearads}[3][][]%
{\href{http://adsabs.harvard.edu/abs/#3}
{\def\hyper@linkstart##1##2{}%
\let\hyper@linkend\@empty\citeyear[#1][#2]{#3}}}
\makeatother
\def\ms{\hbox{m\,s$^{-1}$}}         
\def\m2s2{\hbox{\,m$^{2}$\,s$^{-2}$}} 
\def\kms{\hbox{\,km\,s$^{-1}$}}       
\def\Msun{\hbox{$M_{\odot}$}}             

\def\Mjup{\hbox{$\mathrm{M}_{\rm J}$}}
\def\Rjup{\hbox{$\mathrm{R}_{\rm J}$}}

\def\ten[#1]{$\;\times 10^{#1}$}

\def\logg{$\log g$}

\newcommand{\Rnom}{\hbox{$\mathcal{R}^{\rm N}_{\odot}$}} 

\newcommand{\GMnom}{\hbox{$\mathcal{(GM)}^{\rm N}_{\odot}$}}

\newcommand{\RJnom}{\hbox{$\mathcal{R}^{\rm N}_{e \rm J}$}}

\newcommand{\GMJnom}{\hbox{$\mathcal{(GM)}^{\rm N}_{\rm J}$}}

\newcommand{\reb}{{\sc \tt REBOUND}\xspace}
\newcommand{\whf}{{\sc \tt WHFast}\xspace}
\newcommand{\emcee}{{\sc \tt emcee}\xspace}
\newcommand{\juliet}{{\sc \tt juliet}\xspace}
\newcommand{\radvel}{{\sc \tt radvel}\xspace}
\newcommand{\batman}{{\sc \tt batman}\xspace}
\newcommand{\celerite}{{\sc \tt celerite}\xspace}
\newcommand{\dynesty}{{\sc \tt dynesty}\xspace}
\newcommand{\TESS}{{\sc \tt TESS}\xspace}

\newcommand{\eleanor}{{\sc \tt eleanor}\xspace}
\newcommand{\PARSEC}{{\sc \tt PARSEC}\xspace}

\begin{document} 

     \title{Photodynamical analysis of the nearly resonant planetary system WASP-148:}

    \subtitle{Accurate transit-timing variations and mutual orbital inclination}

   \author{J.M.~Almenara\inst{\ref{geneva},\ref{grenoble}}
        \and G.~H\'ebrard\inst{\ref{iap},\ref{ohp}}  
        \and R.F.~D\'{i}az\inst{\ref{ba1}}
        \and J.~Laskar\inst{\ref{imcce}}
        \and A.~C.~M.~Correia\inst{\ref{coimbra},\ref{imcce}}
        \and D.~R.~Anderson\inst{\ref{coventry1},\ref{coventry2},\ref{keele}}
        \and I.~Boisse\inst{\ref{lam}}
        \and X.~Bonfils\inst{\ref{grenoble}}
        \and D.~J.~A.~Brown\inst{\ref{coventry1},\ref{coventry2}}
        \and V.~Casanova\inst{\ref{iaa}}
        \and A.~Collier~Cameron\inst{\ref{standrews}}
        \and M.~Fern\'andez\inst{\ref{iaa}}
        \and J.M.~Jenkins\inst{\ref{ames}}
        \and F.~Kiefer\inst{\ref{iap},\ref{meudon}}
        \and A.~Lecavelier~des~\'Etangs\inst{\ref{iap}}
        \and J.J~Lissauer\inst{\ref{ames}}
        \and G.~Maciejewski\inst{\ref{copernicus}}
        \and J.~McCormac\inst{\ref{coventry1},\ref{coventry2}}
        \and H.~Osborn\inst{\ref{coventry1},\ref{coventry2},\ref{lam},\ref{bern}}
        \and D.~Pollacco\inst{\ref{coventry1},\ref{coventry2}}
        \and G.~Ricker\inst{\ref{mit}}
        \and J.~S\'anchez\inst{\ref{iaa}}
        \and S.~Seager\inst{\ref{mit},\ref{mit2},\ref{mit3}}
        \and S.~Udry\inst{\ref{geneva}}
        \and D.~Verilhac\inst{\ref{mars}}
        \and J.~Winn\inst{\ref{princeton}}
    }

      \institute{
        Observatoire de Gen\`eve, Département d’Astronomie, Universit\'e de Gen\`eve, Chemin des Maillettes 51, 1290 Versoix, Switzerland\label{geneva}
        \and Univ. Grenoble Alpes, CNRS, IPAG, F-38000 Grenoble, France\label{grenoble}
        \and Institut d'astrophysique de Paris, UMR7095 CNRS, Universit\'e Pierre \& Marie Curie, 98bis boulevard Arago, 75014 Paris, France\label{iap}
        \and Observatoire de Haute-Provence, 04670 Saint Michel l'Observatoire, France\label{ohp}
        \and International Center for Advanced Studies (ICAS) and ICIFI (CONICET), ECyT-UNSAM, Campus Miguelete, 25 de Mayo y Francia, (1650) Buenos Aires, Argentina\label{ba1}
        \and IMCCE, UMR8028 CNRS, Observatoire de Paris, PSL University, Sorbonne Univ., 77 av. Denfert-Rochereau, 75014 Paris, France\label{imcce}
        \and CFisUC, Departamento de F\'isica, Universidade de Coimbra, 3004-516 Coimbra, Portugal\label{coimbra}
        \and Centre for Exoplanets and Habitability, University of Warwick, Gibbet Hill Road, Coventry CV4 7AL, United Kingdom\label{coventry1}
        \and Department of Physics, University of Warwick, Gibbet Hill Road, Coventry CV4 7AL, United Kingdom\label{coventry2}
        \and Astrophysics Group, Keele University, Staffordshire, ST5 5BG, United Kingdom\label{keele}
        \and Laboratoire d’Astrophysique de Marseille, Univ. de Provence, UMR6110 CNRS, 38 r. F. Joliot Curie, 13388 Marseille cedex 13, France\label{lam}
        \and Instituto de Astrof\'isica de Andaluc\'ia (IAA-CSIC), Glorieta de la Astronom\'ia 3, 18008 Granada, Spain\label{iaa}
        \and School of Physics and Astronomy, Physical Science Building, North Haugh, St Andrews, United Kingdom\label{standrews}
        \and NASA Ames Research Center, Moffett Field, CA 94035, USA\label{ames}
        \and LESIA, Observatoire de Paris, Universit\'e PSL, CNRS, Sorbonne Universit\'e, Universit\'e de Paris, 92195 Meudon, France\label{meudon}
        \and Institute of Astronomy, Faculty of Physics, Astronomy and Informatics, Nicolaus Copernicus University, Grudziadzka 5, 87-100 Toru\'n, Poland\label{copernicus}
        \and Center for Space and Habitability, University of Bern, Gesellschaftsstrasse 6, 3012 Bern, Switzerland\label{bern}
        \and Department of Physics and Kavli Institute for Astrophysics and Space Research, Massachusetts Institute of Technology, Cambridge, MA 02139, USA\label{mit}
        \and Department of Earth, Atmospheric and Planetary Sciences, Massachusetts Institute of Technology, Cambridge, MA 02139, USA\label{mit2}
        \and Department of Aeronautics and Astronautics, MIT, 77 Massachusetts Avenue, Cambridge, MA 02139, USA\label{mit3}
        \and Observatoire Hubert-Reeves, 07320 Mars, France\label{mars}
        \and Department of Astrophysical Sciences, Princeton University, NJ 08544, USA\label{princeton}
        }
      \date{}

      \date{}

 
  \abstract
      {
WASP-148 is a recently announced extra-solar system harbouring at least two giant planets. The inner planet transits its host star. The planets travel on eccentric orbits and are near the 4:1 mean-motion resonance, which implies significant mutual gravitational interactions. In particular, this causes transit-timing variations of a few minutes, which were detected based on ground-based photometry. This made WASP-148 one of the few cases where such a phenomenon was detected without space-based photometry.
Here, we present a self-consistent model of WASP-148 that takes into account the gravitational interactions between all known bodies in the system. Our analysis simultaneously fits the available radial velocities and transit light curves. In particular, we used the photometry secured by the Transiting Exoplanet Survey Satellite (\TESS) and made public after the WASP-148 discovery announcement. The \TESS data confirm the transit-timing variations, but only in combination with previously measured transit times.
The system parameters we derived agree with those previously reported and have a significantly improved precision, including the mass of the non-transiting planet. We found a significant mutual inclination between the orbital planes of the two planets: $I = 41.0^{+6.2}_{-7.6}$\degree\ based on the modelling of the observations, although we found $I = 20.8 \pm 4.6$\degree\ when we imposed a constraint on the model enforcing long-term dynamical stability. When a third planet was added to the model -- based on a candidate signal in the radial velocity -- the mutual inclination between planets b and c changed significantly allowing solutions closer to coplanar. We conclude that more data are needed to establish the true architecture of the system.     
If the significant mutual inclination is confirmed, WASP-148 would become one of the only few candidate non-coplanar planetary systems. We discuss possible origins for this misalignment. 

}
      \keywords{stars: individual: \object{WASP-148} --
        stars: planetary systems --
        techniques: photometric --
        techniques: radial velocities}
   \authorrunning{J.M. Almenara et al.}
   \titlerunning{WASP-148}

   \maketitle
%

\section{Introduction}\label{section.introduction}

While the orbit of a single planet around its host star is well reproduced by a Keplerian model with a constant orbital period, multi-planetary systems have more complex orbits. Indeed, the mutual gravitational interactions between planets imply small deviations from the Keplerian orbits and in particular slight variations of the orbital periods. Such effects are negligible for most systems, but they are amplified when orbital periods are exactly or nearly commensurable. When such a resonant or nearly resonant system includes transiting planets, their orbital periods can be accurately measured, allowing their variations to be detected. This was predicted in particular by \citet{holman2005} and \citet{agol2005}, who showed how interactions in multi-planetary systems might cause transit-timing variations (TTVs). 

When they are measured, TTVs are a powerful tool to characterise planetary systems. In particular, they allow constraints to be put on masses as well as on the presence of additional planets. This was done on the first system in which TTVs were detected \citep{holman2010} as well as on dozens of other detections reported thereafter \citep[e.g.][]{lissauer2011,nesvorny2013,wang2014,hadden2014,gillon2017,freudenthal2019,Jontof-Hutter2021}. Among different methods employed to analyse TTV observations, the photodynamical modelling\footnote{A photodynamical model is a light curve model for more than a two-body system that includes the gravitational interactions between the assumed bodies in the system, as the planet-planet interactions (and not just the planet-star interaction) in a multi-planetary system.} \citep{carter2011} of light curves can be used together with radial velocities to constrain planetary system parameters without using external inputs on the masses or radii of the host stars often derived from stellar evolution models \citep{agol2005}. One of the strengths of this technique is that it fits the full transit light curves (including transit durations) instead of only using the transit timings. Photodynamical analyses have been used to characterise several TTV systems \citep[e.g.][]{almenara2018b,almenara2018}. Here, we apply it to the system WASP-148 recently reported by \citet{hebrard2020}.

Most of the confirmed TTV detections to date have been discovered using light curves obtained from space telescopes (namely \textit{Kepler} or \textit{Spitzer}). On the other hand, the TTVs in the WASP-148 system were detected using ground-based telescopes only. The WASP-148 system includes (at least) two giant planets. The inner one, WASP-148~b, is a hot Saturn of $0.72 \pm 0.06~\Rjup$ and $0.29 \pm 0.03~\Mjup$ that transits its host with an orbital period of 8.80~days. The outer planet, WASP-148~c, has an orbital period of 34.5~days and a minimum mass of $0.40 \pm 0.05~\Mjup$ and a true mass $< 0.60~\Mjup$. It was discovered and characterised using radial velocities obtained with the SOPHIE spectrograph. No transits of this planet were detected. The orbits of both planets have significant eccentricities ($e_b = 0.22 \pm 0.06$ and $e_c = 0.36 \pm 0.09$) and their orbital periods fall near the 1:4 mean-motion resonance. This particular configuration induces amplified dynamical effects, and TTVs were detected with an amplitude of about $\pm 15$~minutes and a period of roughly 460~days \citep{hebrard2020}.

Several analyses of the available data sets have been presented by \citet{hebrard2020}. The first one simultaneously fitted transit light curves and radial velocities, but it did not include any mutual interactions. The model allowed for small, artificial, ad hoc shifts in the transit times in order to reproduce the TTVs. The second model was fitted to the radial velocities taking mutual interactions into account, constrained by the average period and phase derived from the transit light curves. Both analyses gave similar results and the derived system properties imply large TTVs for both planets, whereas only those of WASP-148~b could actually be observed as no transits of WASP-148~c were detected. However, \citet{hebrard2020} did not present a complete model fitted to both radial velocity and transit light curve data sets that takes the gravitational interactions between the planets into account. Finally, \citet{hebrard2020} indicate that there is a hint of a possible third planet in the system with a period near 150~days and a minimum mass around $0.25~\Mjup$, without confirming that however.

Following the detection and characterisation of the WASP-148 system by \citet{hebrard2020}, observations of that star secured with the Transiting Exoplanet Survey Satellite (\TESS) were released. \TESS provides continuous, high-quality photometry over 28~days \citep{ricker2015}. In the case of WASP-148, it observed seven new transits of WASP-148~b, and provided a unique opportunity to search for possible transits of WASP-148~c or potential additional planets in the system. \citet{Maciejewski2020} published an analysis of the TTVs and radial velocities of WASP-148 presented by \citet{hebrard2020} in addition to \TESS data, as well as two new ground-based transit observations. They took into account the gravitational interactions between the planets, but they used only the transit timings instead of the whole transit light curves. Recently, \citet{wang2022} have shown the orbit of WASP-148~b is aligned and prograde.

Here, we present new analyses of the WASP-148 system, applying the full photodynamical approach on the available transit light curves and radial velocities. The article is organised as follows: In Section~\ref{section.observations} we describe the data used; in Section~\ref{section.stellar_parameters} we determine the stellar parameters; in Sections~\ref{section.sophie} and \ref{section.juliet} we analyse the radial velocity and transits without accounting for the gravitational interactions between the planets, respectively; in Section~\ref{section.photodynamical} we detail the photodynamical modelling; and in Section~\ref{section.results} we present the results. Finally, we discuss the results of our work in Section~\ref{section.discussion}.

\section{Observations}\label{section.observations}

We used the ground-based photometry and SOPHIE radial velocities presented in \citet{hebrard2020}. We converted the time of the photometry from BJD$_{\rm UTC}$ to BJD$_{\rm TDB}$\footnote{BJD$_{\rm TDB}$ = BJD$_{\rm UTC}$ + $\delta$, with $\delta$ being between 65.184 and 69.184~seconds depending on the date of the transit.} \citep{eastman2010}, except for the Telescopio Carlos S\'{a}nchez transit, and NITES observation on June 13, 2016, which were already in BJD$_{\rm TDB}$.
In addition to the data in \citet{hebrard2020}, we added the photometry from \TESS and four transits observed with the 1.5~m Ritchey-Chrétien Telescope at the Sierra Nevada Observatory (OSN150), the first two of which have been presented in \citet{Maciejewski2020}. 

\subsection{TESS}

\TESS observed WASP-148 in sectors 24, 25, and 26, with a total time span of 79.3~days (TIC~115524421, TOI-2064). Two transits were lost during the interruption of the observations that occur every \TESS orbit (around 14~days) at the middle of sectors 25 and 26. Each \TESS sector lasts two orbits of the satellite. Around the perigee of the \TESS orbit, data collection is paused. In total, seven new transits of WASP-148~b were observed. The photometry is available in the full-frame images (FFIs) at 30-minute cadence. The FFIs were calibrated by the Science Processing Operations Center (SPOC) at NASA Ames Research Center \citep{jenkins2016}. We used \eleanor \citep{feinstein2019} to extract the light curve of WASP-148 from the \TESS FFIs. We chose the point spread function photometry that has the lower dispersion for this object. \eleanor corrects the times for the object coordinates, which otherwise are set to the centre of the CCD in FFIs.
There is a star 4.5~mag fainter in the Gaia G-band at 26\arcsec\ of WASP-148 that partially contaminates the \TESS photometry\footnote{Assuming the flux of the contaminant star is completely inside the \TESS aperture and that the difference in magnitude in the Gaia band is the same as in the \TESS band, the dilution is about 1.6\%, which can change the planet to star radius ratio by 0.8\%, and the final planet radius by $\sim$0.4-$\sigma$.}. This is taken into account in our modelling (Sections~\ref{section.juliet} and \ref{section.photodynamical}).
The \TESS data show no evidence for any transit of WASP-148~c, which is in agreement with \citet{Maciejewski2020}.

\subsection{OSN150}

\begin{table*}[h]
\caption{Details on the observing runs.} 
\label{tab.Obs}      
\centering                  
\begin{tabular}{l c c c c c c}      
\hline\hline                
Date UT  & UT start--end  &  $X$                                & $N_{\rm{obs}}$ & $t_{\rm{exp}}$ (s) & $\Gamma$ & pnr (ppth)\\
\hline
2021 March 22  & 00:20--05:34 & $1.70 \rightarrow 1.01$ & 515 & 30 & 1.69 & 1.10  \\
2021 June 26   & 21:19--03:00 & $1.09 \rightarrow 1.01 \rightarrow 1.01 \rightarrow 1.47$ & 575 & 30 & 1.69 & 0.76  \\
\hline
\end{tabular}
\tablefoot{Date UT is given for the beginning of an observing run. $X$ shows the changes of the target's airmass during a run. $N_{\rm{obs}}$ is the number of useful scientific exposures. $t_{\rm{exp}}$ is the exposure time. $\Gamma$ is the median number of exposures per minute. $pnr$ is the photometric noise rate \citep{2011AJ....142...84F} in parts per thousand (ppth) of the normalised flux per minute of the observation.}
\end{table*}

Two new precise photometric time series for transits of WASP-148~b were acquired in March and June 2021 using the 1.5 m Ritchey-Chr\'etien telescope (OSN150) at the Sierra Nevada Observatory (OSN, Spain). The instrument was equipped with a Roper Scientific VersArray 2048B CCD camera with a $2048 \times 2048 \times 13.5\,\mu \rm{m}$ back-illuminated matrix. The field of view was $7 \farcm 92 \times 7 \farcm 92$ with the pixel scale of $0 \farcs 232$ per pixel. The instrument was mildly defocussed to allow for longer exposure times and a lower fraction of time lost for CCD readout. The observations were gathered without any filter to increase the signal-to-noise ratio for transit-timing purposes. The observing runs were scheduled following an ephemeris from \citet{Maciejewski2020}. The telescope was auto-guided to keep the star at the same position in the CCD matrix. The details on the individual runs are given in Table~\ref{tab.Obs}. The observations started about 90 minutes before a transit ingress and lasted about 90 minutes after an egress.   This out-of-transit monitoring was secured for de-trending purposes. On March 22,  2021 the observations were stopped about 12 minutes after the transit due to dawn.

Photometric data reduction was performed with the AstroImageJ software \citep{2017AJ....153...77C} following a standard calibration procedure. The science frames were de-biased and flat-field calibrated using sky flat frames. The light curves were generated with the differential aperture photometry method. The aperture size and a collection of comparison stars were set after a series of test runs to minimise the data point scatter. The fluxes were de-trended against airmass and time using out-of-transit data only and then they were normalised to unity outside the transits. Timestamps were converted into barycentric Julian dates in barycentric dynamical time $\rm{BJD_{TDB}}$.
For the homogeneity of our analysis, we also reprocessed the two transit light curves from \citet{Maciejewski2020} following the same de-trending procedure.

\section{Stellar parameters}\label{section.stellar_parameters}

To determine the stellar parameters of WASP-148, we used the precise parallax determination by Gaia \citep{gaia2016, gaiaEDR3}, which was not exploited by \cite{hebrard2020}. Stellar atmosphere models and stellar evolution models are also required to model the observed spectral energy distribution (SED). We constructed the SED of WASP-148 using the magnitudes from Gaia Early Data Release 3 \citep[Gaia EDR3,][]{riello2021}, the 2-Micron All-Sky Survey \citep[2MASS,][]{2mass,cutri2003}, and the Wide-field Infrared Survey Explorer \citep[WISE,][]{wise,cutri2013}. The measurements are listed in Table~\ref{table.sedmag}. We modelled these magnitude measurements using the procedure described by \citet{diaz2014}, with informative priors for the effective temperature ($T_{\mathrm{eff}}$), surface gravity (\logg), and metallicity ($[\rm{Fe/H}]$) from \citet{hebrard2020}, and for the distance from Gaia EDR3 \citep{lindegren2021}. We used non-informative priors for the rest of parameters. The priors are listed in Table~\ref{table.sed}.

\begin{figure}
  \centering
  \includegraphics[width=0.5\textwidth]{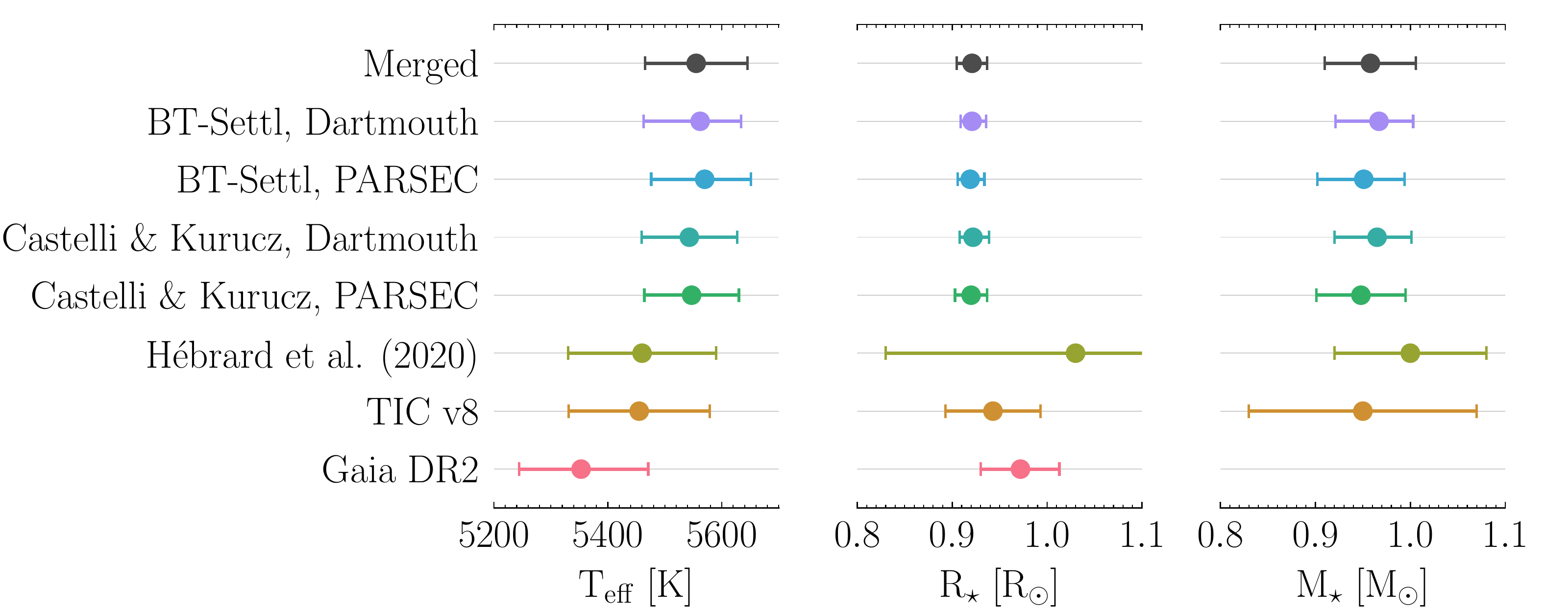}
  \caption{Stellar parameters derived in this work compared with \citet{hebrard2020}, the \TESS Input Catalog (TIC) version~8 \citep{stassun2019}, and Gaia DR2 \citep{andrae2018}. We finally adopt the values labelled as 'merged'.} \label{fig.StellarParameters}
\end{figure}

We decided to use an additive jitter for each set of photometric bands (Gaia, 2MASS, and WISE), which had the effect of slightly broadening the posteriors of the stellar parameters. We used the two stellar atmosphere models, PHOENIX/BT-Settl \citep{allard2012} and ATLAS/Castelli \& Kurucz \citep{castelli2003}, and two stellar evolution models, Dartmouth \citep{dotter2008} and \PARSEC \citep{chen2014}. 
We obtained posterior samples for the four combinations of stellar atmosphere models and stellar evolution models using the Markov Chain Monte Carlo (MCMC) algorithm from \citet{diaz2014}. The posteriors of the stellar parameters of the individual combinations agree within 1-$\sigma$ (see Figure~\ref{fig.StellarParameters}). We merged the results assuming an equal probability for each model combination (labelled {as 'merged'} in Figure~\ref{fig.StellarParameters}). The posteriors' median and 68.3\% credible intervals (CI) for model parameters and for derived physical quantities of interest are listed in Table~\ref{table.sed}. 
Those results agree with the ones reported by \citet{hebrard2020}, but 
they are more accurate, mainly due to the use of the parallax from Gaia.

\begin{figure}
  \centering
  \includegraphics[width=0.5\textwidth]{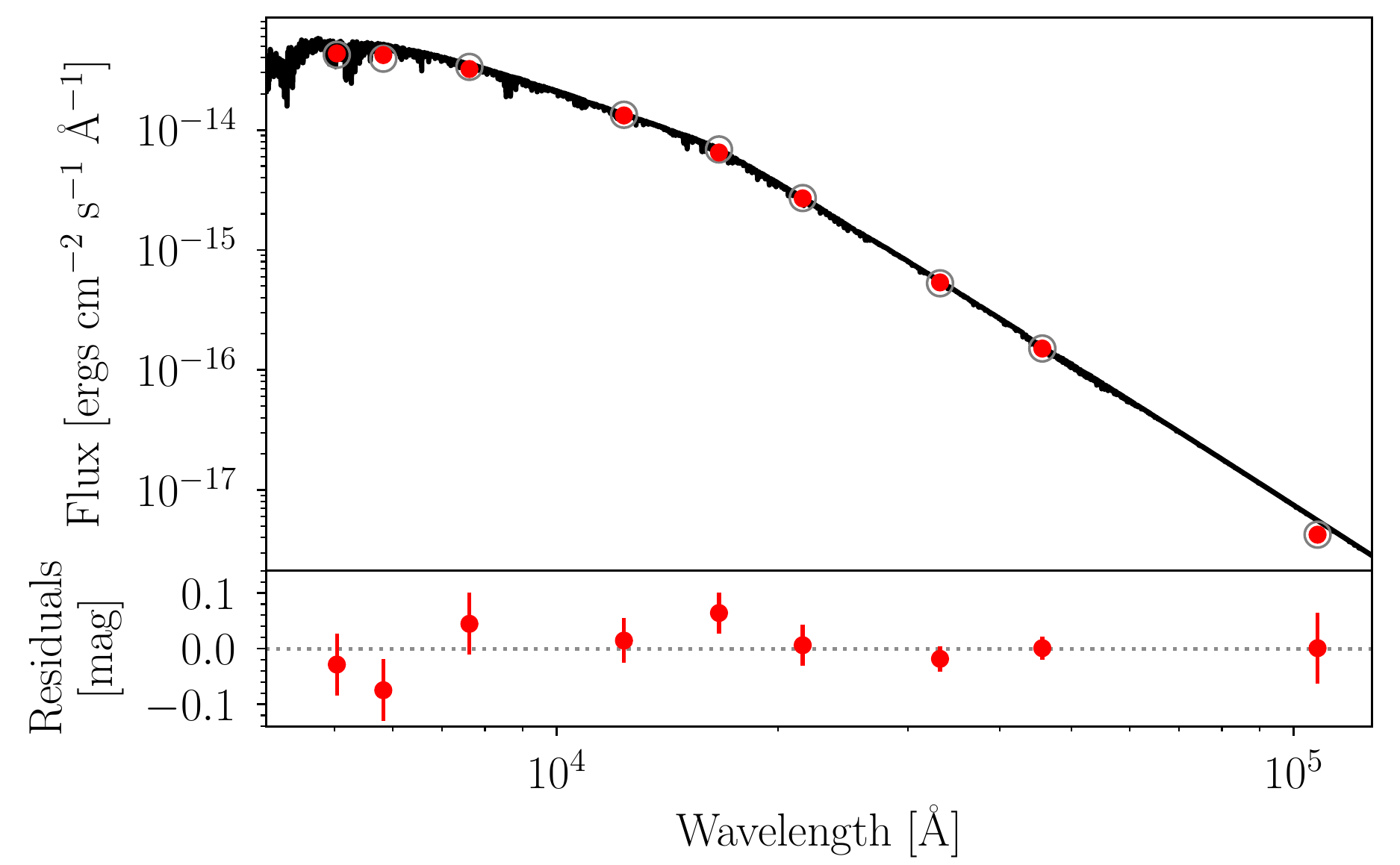}
  \caption{Spectral energy distribution of WASP-148. Top panel: The solid line is the MAP PHOENIX/BT-Settl interpolated synthetic spectrum, red circles are the absolute photometric observations, and grey open circles are the result of integrating the synthetic spectrum in the observed bandpasses. Bottom panel: Residuals of the MAP model (the jitter has been added quadratically to the data error bars).} \label{fig.sed}
\end{figure}

The data with the maximum a posteriori (MAP) stellar atmosphere model is shown in Figure~\ref{fig.sed}.
Before Gaia EDR3, we carried out the same analysis with Gaia DR2 \citep{gaia2018}, obtaining similar results. In particular, we obtained the same stellar radius error, despite a factor $\sim$3 increase in the precision of the parallax. Thus, at least for WASP-148, the precision in the parallax is not the limiting factor to improve the stellar radius determination with the SED technique.

We used the stellar rotation period, P$_{\rm rot}=26.2\pm1.3~$days, derived in \citet{hebrard2020} and the stellar mass derived in this section (Table~\ref{table.sed}) to estimate a gyrochronological age, neglecting the influence of the planets, of 4.0$^{+0.9}_{-0.7}$~Gyr \citep[using a P$_0$ between 0.12 and 3.4~days]{barnes2010,barneskim2010}, where we added a systematic 10\% error to the statistical one \citep{meibom2015}. The isochronal age (Table~\ref{table.sed}) agrees with the gyrochronological age within the uncertainties, but it is less precise.

\begin{table}
    \tiny
    \renewcommand{\arraystretch}{1.25}
    \setlength{\tabcolsep}{2pt}
\centering
\caption{Modelling of the spectral energy distribution.}\label{table.sed}
\begin{tabular}{lccc}
\hline
Parameter & & Prior & Posterior median   \\
&  & & and 68.3\% CI \\
\hline
Effective temperature, $T_{\mathrm{eff}}$ & [K]     & $N$(5460, 130)    & 5555 $\pm$ 90 \\
Surface gravity, \logg\                   & [cgs]   & $N$(4.40, 0.15)   & 4.490 $^{+0.027}_{-0.034}$ \\
Metallicity, $[\rm{Fe/H}]$                & [dex]   & $N$(0.11, 0.08)   & 0.099 $\pm$ 0.078 \\
Distance                                  & [pc]    & $N$(247.73, 0.60) & 247.73 $\pm$ 0.60 \\
$E_{\mathrm{(B-V)}}$                      & [mag]   & $U$(0, 3)         & 0.026 $^{+0.041}_{-0.019}$ \\
Jitter Gaia                               & [mag]   & $U$(0, 1)         & 0.140 $^{+0.22}_{-0.080}$ \\
Jitter 2MASS                              & [mag]   & $U$(0, 1)         & 0.056 $^{+0.072}_{-0.030}$ \\
Jitter WISE                               & [mag]   & $U$(0, 1)         & 0.033 $^{+0.076}_{-0.024}$ \\
Radius, $R_\star$                         & [R$_\odot$] &               & 0.921 $\pm$ 0.016 \\
Mass, $M_\star$                           & [M$_\odot$] &               & 0.958 $\pm$ 0.048 \\
Density, $\rho_\star$                     & [$\mathrm{g\;cm^{-3}}$] &   & 1.73 $\pm$ 0.15 \\
Isochronal age                            & [Gyr] &                     & 3.6 $^{+4.0}_{-2.7}$ \\
Luminosity                                & [L$_\odot$] &               & 0.726 $\pm$0.036 \smallskip\\
\hline
\end{tabular}
\tablefoot{$N$($\mu$,$\sigma$): Normal distribution prior with mean $\mu$, and standard deviation $\sigma$. $U$(l,u): Uniform distribution prior in the range [l, u].}
\end{table}

\section{SOPHIE-only analysis}\label{section.sophie}

When radial velocity and transit observations are modelled simultaneously \citep[as was done by][]{hebrard2020}, both data sets constrain the eccentricity. In this section, we aim for a planetary eccentricity determination using only radial velocity data. We analysed the SOPHIE radial velocities with a two-planet Keplerian model, without taking mutual gravitational interactions into account and using \juliet \citep{espinoza2019}, \radvel \citep{fulton2018}, and \dynesty \citep{speagle2020}. We used normal priors from \citet{hebrard2020} for the period and the time of inferior conjunction of planet~b, and non-informative priors for the rest of the parameters. We included a jitter parameter that was added quadratically to the velocity uncertainties. We performed a second analysis using a Gaussian process (GP) regression model with a quasi-periodic (QP) kernel to model the error terms. We employed the QP kernel included in \celerite \citep{foreman-mackey2017}, with non-informative priors for the hyperparameters. We found an odds ratio of $\ln\left(\mathcal{Z}_{\rm jitter}/\mathcal{Z}_{\rm jitter + QP}\right) = 0.5 \pm 0.4$. Neither of the models is strongly favoured over the other \citep{kass1995}, and we therefore adopted the simpler model of uncorrelated error terms with an additive jitter. The posterior median and 68.3\% credible interval are shown in Table~\ref{table.RV}. 
They agree with the results presented by \citet{hebrard2020}.

We investigated the significance of the main peak in the residuals of the two-planet model \citep[Fig. 3 in][]{hebrard2020}, thus we repeated the analysis adding a third Keplerian with a period of $\sim$150~days. We found a minimum mass of $0.24 \pm 0.05~\Mjup$, similar to the 0.25~\Mjup\ value reported by \citet{hebrard2020}. The model comparison gives an odds ratio of $\ln\left(\mathcal{Z}_{\rm 3~planets}/\mathcal{Z}_{\rm 2~planets}\right) = 1.6 \pm 0.4$, which is positive evidence in favour of the three-planet model but below the {strong} evidence cutoff to be preferred over the two-planet model, so we continued with the adopted two-planet model.

\begin{table}[h]
    \tiny
\caption{Two-Keplerian fit to the SOPHIE data.}\label{table.RV}
\begin{tabular}{cccc}
\hline
Keplerian && WASP-148~b & WASP-148~c \\
\hline
P&[d]& 8.803809 $\pm$ 0.000043 & 34.524 $\pm$ 0.029 \\
T$_{\rm c}$&[BJD]& 2\;457\;957.4876 $\pm$ 0.0060 & 2\;457\;935.6 $\pm$ 1.2\\
e&& 0.183 $\pm$ 0.070 & 0.352 $\pm$ 0.085 \\
$\omega$&[\degree]& 59 $^{+15}_{-20}$ & 9 $^{+17}_{-14}$ \\
K&[\ms]& 28.8 $\pm$ 2.0 & 27.0 $\pm$ 2.9 \\
\hline
$\gamma_{\rm SOPHIE}$&[\kms]&\multicolumn{2}{c}{-5.6174 $\pm$ 0.0014}\\
jitter &[\ms]&\multicolumn{2}{c}{10.9 $\pm$ 1.3}\\
\hline
\end{tabular}
\tablefoot{The parameters are: Orbital period, time of conjunction, eccentricity, argument of pericentre, radial velocity semi-amplitude, systemic velocity, and jitter.}
\end{table}

\section{Transit-only analysis}\label{section.juliet}

We derived the transit times of planet~b without considering planet~c or the radial velocities. This modelling is independent of the two-planet system hypothesis adopted in the photodynamical modelling (Section~\ref{section.photodynamical}), and it can be used to verify if it is an appropriate hypothesis given the data.

In \citet{hebrard2020}, the error term of the transit photometry data were modelled using a simple additive jitter term. However, if systematics are not correctly accounted for, the posteriors of the transit modelling can be biased \citep{barros2013}. This should be particularly severe for incomplete transits. Here, we reanalyse the transits presented in \citet{hebrard2020} and the \TESS observations using a more sophisticated error model.

\begin{figure*}
  \centering
  \includegraphics[width=1.0\textwidth]{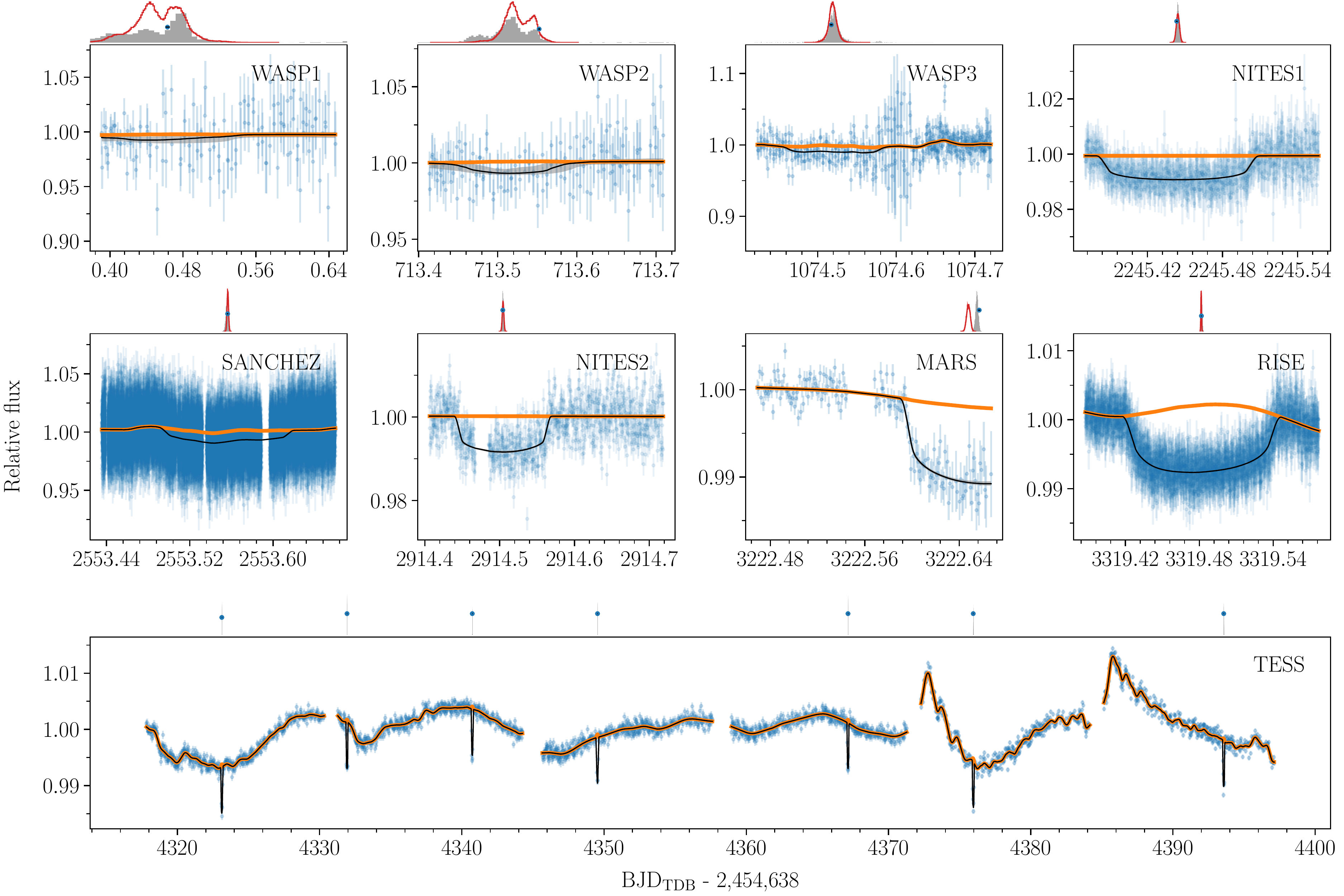}
  \caption{Modelling of the planet~b transits presented in \citet{hebrard2020} and observed by \TESS (blue error bars). Each panel is labelled with the observatory or instrument name and sequential night of observation if there were more than one. Black lines and intervals in grey show the model median and 68.3\% credible interval computed with 1000 random samples of the posterior. The mean of the predictive distribution of the kernel model is shown in orange. On top of each panel, the transit-timing posterior is shown for \citet{hebrard2020} (red line histogram), the modelling with \juliet (Section~\ref{section.juliet}, grey histogram), and the photodynamical modelling (blue points with black error bars, although the errors are barely visible).} \label{fig.juliet}
\end{figure*}

To model the transits of planet~b simultaneously with the systematics, we used \juliet \citep{espinoza2019}, using \batman \citep{kreidberg2015} for the transit model, and we chose the approximate Matern kernel  GP included in \celerite \citep{foreman-mackey2017}. We used different GPs for each ground-based transit, and for each \TESS sector. The timing of each individual transit is a free parameter. 
We used a prior for the stellar density from Section~\ref{section.stellar_parameters}, and for $\sqrt{e}\cos{\omega}$ and $\sqrt{e}\sin{\omega}$ from \citet{hebrard2020}. Otherwise, we adopted non-informative or large priors for the rest of the parameters. We oversampled \citep{kipping2010} the model of the MARS transit by a factor 3, which has an 88~second cadence, and the \TESS data by a factor 30, which have about a 30~minute cadence. The remaining transit observations have cadences between 0.9 and 37~seconds, and we deemed them unnecessary to oversample their model. We set a dilution factor for the wide-field observations: WASP (one dilution for the three WASP transits), and \TESS (one dilution per sector). To reduce the number of free parameters, we made the choice to use one set of limb darkening parameters for all transit observations without a filter ('clear'), but they certainly do not correspond to exactly the same instrument bandpass. In total, there are 76 free parameters. To sample from the posterior, we used the nested sampling code \dynesty \citep{speagle2020}.
The complete list of parameters, priors, and posteriors are shown in Table~\ref{table.juliet}.
Figure~\ref{fig.juliet} shows the data and the model posterior.
We compare the timing of the ground-based observations with the results in \citet{hebrard2020}, and found differences within 1-$\sigma$, except for the partial transit observed with MARS where it is $12.0\pm2.5$~min later (see Figure~\ref{fig.ttcomparison}). The transit normalisation can affect the transit-timing determination \citep{barros2013}. The MARS transit can be more affected because it is a partial transit and the baseline is not well-defined.

\begin{figure*}
  \centering
  \includegraphics[width=1.0\textwidth]{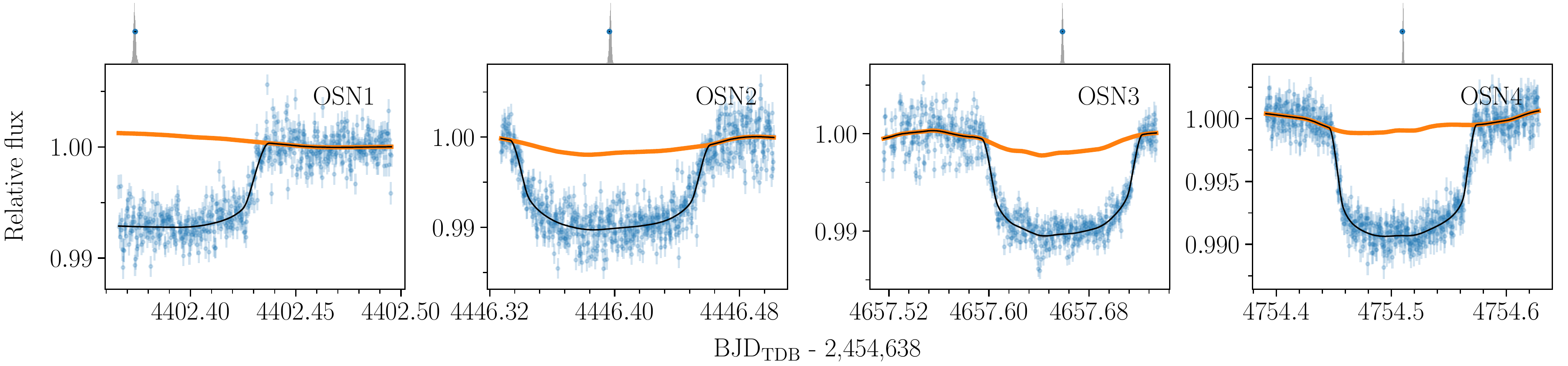}
  \caption{Idem. as Figure~\ref{fig.juliet}, but for the OSN150 transits.} \label{fig.OSN}
\end{figure*}

\begin{figure}
  \centering
  \includegraphics[width=0.5\textwidth]{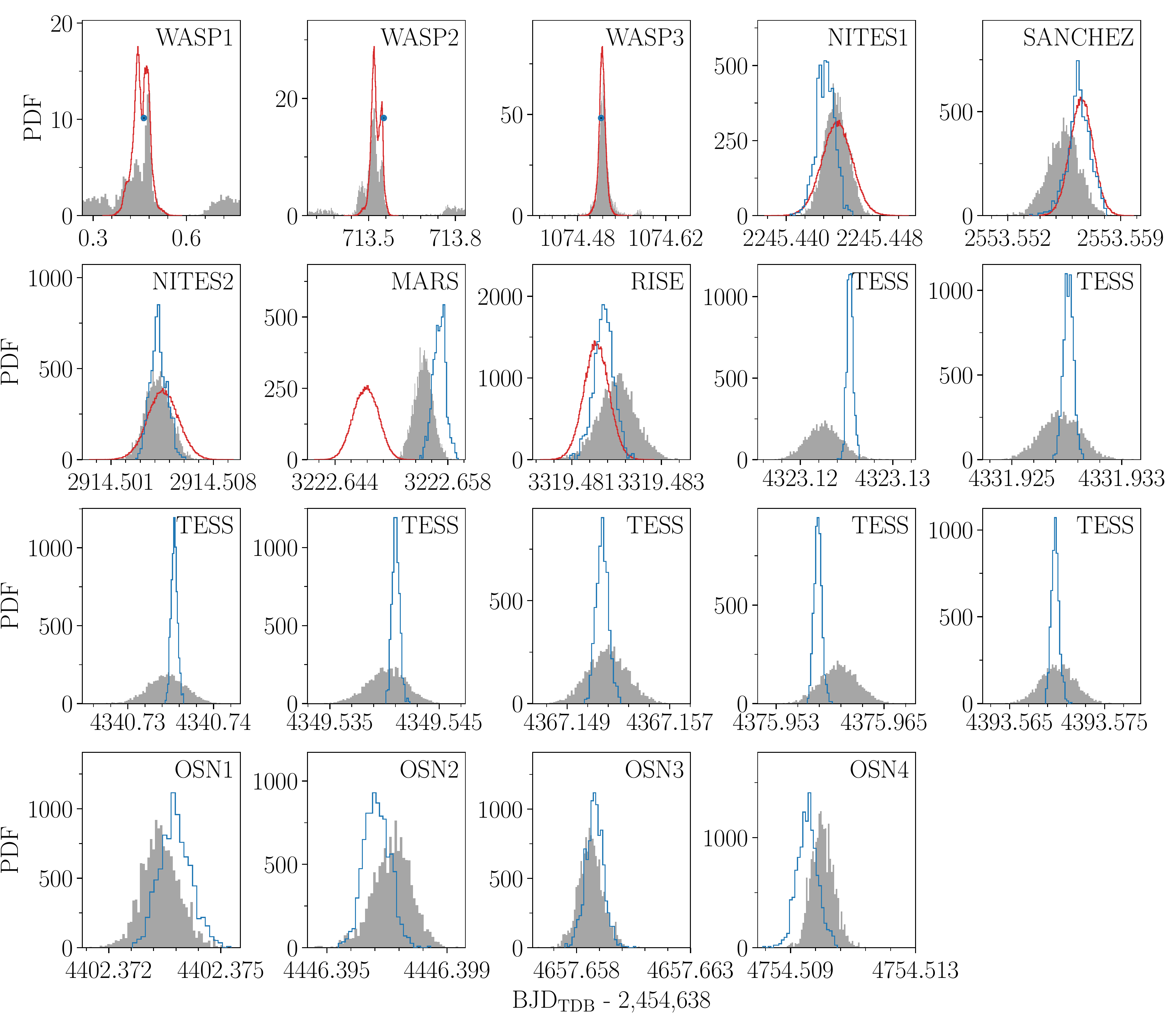}
  \caption{Comparison between transit-time posteriors of \citet{hebrard2020} (red line histograms), the modelling with \juliet (Section~\ref{section.juliet}, grey histograms), and the photodynamical modelling (Section~\ref{section.photodynamical}, blue points with black error bars and blue histograms).} \label{fig.ttcomparison}
\end{figure}

We repeat this analysis for the four OSN150 transits, which were included at a later stage in this work. The results are presented in Table~\ref{table.OSN} and Figure~\ref{fig.OSN}. 

This analysis could be used to de-trend the transits for the photodynamical modelling. However, in this process, one set of parameters of the transit model needs to be fixed, including a transit time. Therefore, if the de-trended transits are then modelled, the results can be biased. Instead, we decided to include the GP in the photodynamical modelling.

\section{Photodynamical modelling}\label{section.photodynamical}

While Sections~\ref{section.sophie} and \ref{section.juliet} do not take mutual interactions into account, here we report our fits of the observed photometry and radial velocity measurements accounting for the gravitational interactions between the three bodies known in the system using a photodynamical model. Its positions and velocities in time were obtained through an n-body integration. The sky-projected positions were used to compute the light curve \citep{mandelagol2002} using a quadratic limb-darkening law \citep{manduca1977}, which we parametrised following \citet{kipping2013}. To account for the integration time, the model was oversampled by a factor of 30 and 3 for the \TESS and MARS data, respectively, and then binned back to match the cadence of the data points \citep{kipping2010}. The line-of-sight projected velocity of the star issued from the n-body integration was used to model the radial velocity measurements. We used the n-body code \reb \citep{rein2012} with the \whf integrator \citep{rein2015} and an integration step of 0.01~days, which results in a maximum error of $\sim$20~ppm for the photometric model. The light-time effect \citep{irwin1952} is included, although with an amplitude of $\sim$0.05~s in the transit timing; this is a negligible effect for this system. The model was parametrised using the stellar mass and radius, planet-to-star mass ratios, planet~b-to-star radius ratio, and Jacobi orbital elements (Table~\ref{table.results}) at the reference time, $t_{\mathrm{ref}} = 2\;457\;957.48167$~BJD$_{\mathrm{TDB}}$, close to the RISE transit centre. Due to the symmetry of the problem, we fixed the longitude of the ascending node of the interior planet $\Omega_{\mathrm{b}}$ at $t_{\mathrm{ref}}$ to 180\degree, and we limited the inclination of the outer one $i_c>90$\degree. 
We used a GP regression model, with an approximate Matern kernel \citep[\celerite,][]{foreman-mackey2017} for the model of the error terms of the transit light curves. We used different kernel hyperparameters for each transit, except for the \TESS data for which we used different kernel hyperparameters for each sector. We added one dilution factor\footnote{The definition of the dilution factor and transit normalisation factor is different from the one in \juliet. For the photodynamical model, we used $M = f_0 (f_t T + f_c) / (f_t + f_c)$, with $M$ being the model to be compared with the observations, $f_0$ being the transit normalisation factor (the light curve flux level out of transit), $f_t$ being the flux of the target star in the aperture, $T$ being the transit model, and $f_c$ being the flux of the contaminant star in the aperture. The dilution factor is defined as $d = f_c / (f_t + f_c)$.} for each \TESS sector, and another one for the WASP transits. For each photometric data set, we added a transit normalisation factor and an additive jitter parameter. For the radial velocity, we added a systemic radial velocity and an additive jitter parameter. In total, the model has 90 free parameters. We used normal priors for the stellar mass and radius from Section~\ref{section.stellar_parameters}, a non-informative sinusoidal prior for the orbital inclinations, and non-informative uniform prior distributions for the rest of the parameters. The joint posterior distribution was sampled using the \emcee\ algorithm \citep{goodmanweare2010, emcee} with 1\;000~walkers with starting points based on the results of \citet{hebrard2020} and Section~\ref{section.juliet}. 

\begin{table*}
    \small
\renewcommand{\arraystretch}{1.2}
\centering
\caption{Inferred system parameters.}\label{table.results}
\begin{tabular}{lccccc}
\hline
Parameter & Units & Prior & Median and 68.3\% CI & Stable MAP & Stable median and 68.3\% CI \\
\hline
\emph{\bf Star} \\
Stellar mass, $M_\star$                  & [\Msun]     & $N(0.958, 0.048)$   &  0.967 $\pm$ 0.049 & 0.954 & 0.954 $^{+0.030}_{-0.052}$ \\
Stellar radius, $R_\star$                & [\Rnom]     & $N(0.921, 0.016)$   &  0.920 $\pm$ 0.016 & 0.913 & 0.912 $\pm$ 0.014 \\
Stellar mean density, $\rho_{\star}$ & [$\mathrm{g\;cm^{-3}}$] &             &  1.76 $\pm$ 0.12   & 1.77 & 1.76 $\pm$ 0.12 \\
Surface gravity, \logg\                  & [cgs]           &                 &  4.497 $\pm$ 0.025 & 4.497 & 4.496 $^{+0.022}_{-0.028}$ \smallskip\\

\citet{kipping2013} $q_1, q_2$ CLEAR     &         & $U(0, 1)$ &  0.42 $^{+0.13}_{-0.11}$, 0.240 $^{+0.11}_{-0.094}$ & 0.304, 0.421 & 0.388 $^{+0.15}_{-0.083}$, 0.279 $^{+0.093}_{-0.16}$ \\
\citet{kipping2013} $q_1, q_2$ Johnson-R &         & $U(0, 1)$ &  0.230 $^{+0.13}_{-0.096}$, 0.48 $^{+0.32}_{-0.26}$ & 0.106, 0.98 & 0.162 $^{+0.16}_{-0.056}$, 0.59 $^{+0.28}_{-0.32}$ \\
\citet{kipping2013} $q_1, q_2$ RISE      &         & $U(0, 1)$ &  0.80 $\pm$ 0.11, 0.269 $\pm$ 0.062                 & 0.802, 0.272 & 0.802 $^{+0.091}_{-0.13}$, 0.278 $^{+0.056}_{-0.063}$ \\
\citet{kipping2013} $q_1, q_2$ TESS      &         & $U(0, 1)$ &  0.38 $^{+0.21}_{-0.15}$, 0.40 $^{+0.33}_{-0.24}$   & 0.26, 0.72 & 0.38 $^{+0.17}_{-0.12}$, 0.40 $^{+0.31}_{-0.20}$ \smallskip\\

\emph{\bf Planet~b} \\
Semi-major axis, $a$                      & [au]                   &             &  0.0825 $\pm$ 0.0014        & 0.08214 & 0.08215 $^{+0.00086}_{-0.0015}$ \\
Eccentricity, $e$                         &                        &             &  0.214 $^{+0.021}_{-0.018}$ & 0.208 & 0.208 $^{+0.020}_{-0.025}$ \\
Argument of pericentre, $\omega$          & [\degree]              &             &  63.7 $\pm$ 8.3             & 62.6 & 60.6 $^{+5.0}_{-6.0}$ \\
Inclination, $i$                          & [\degree]              & $S(0, 180)$ &  89.61 $^{+0.21}_{-0.27}$   & 89.32 & 89.30 $\pm$ 0.24 \\
Longitude of the ascending node, $\Omega$ & [\degree]              &             &  180 (fixed at $t_{\mathrm{ref}}$) & 180 & 180  \\
Mean anomaly, $M_0$                       & [\degree]              &             &  16.7 $^{+5.8}_{-5.1}$      & 17.8 & 19.2 $^{+4.0}_{-3.5}$ \smallskip\\

$\sqrt{e}\cos{\omega}$                    &                        & $U(-1, 1)$ &  0.204 $\pm$ 0.056           & 0.210 & 0.219 $\pm$ 0.041 \\
$\sqrt{e}\sin{\omega}$                    &                        & $U(-1, 1)$ &  0.413 $\pm$ 0.039           & 0.405 & 0.394 $^{+0.028}_{-0.031}$ \\
Mass ratio, $M_{\mathrm{p}}/M_\star$      &                        & $U(0, 1)$  &  0.000284 $\pm$ 0.000019     & 0.000309 & 0.000292 $^{+0.000012}_{-0.000014}$ \\
Radius ratio, $R_{\mathrm{p}}/R_\star$    &                        & $U(0, 1)$  &  0.08436 $\pm$ 0.00058       & 0.08532 & 0.08498 $^{+0.00064}_{-0.00080}$ \\
Scaled semi-major axis, $a/R_{\star}$     &                        &            &  19.31 $^{+0.41}_{-0.46}$    & 19.34 & 19.33 $^{+0.39}_{-0.44}$ \\
Impact parameter, $b$                     &                        &            &  0.105 $^{+0.073}_{-0.056}$  & 0.185 & 0.199 $^{+0.050}_{-0.070}$ \\
$T_0'$\;-\;2\;450\;000                    & [BJD$_{\mathrm{TDB}}$] & $U(6957, 8957)$ &  7957.48172 $\pm$ 0.00022 & 7957.48190 & 7957.48166 $^{+0.00025}_{-0.00015}$ \\
$P'$                                      & [d]                    & $U(0, 1000)$ &  8.80369 $^{+0.00019}_{-0.00017}$ & 8.80366 & 8.80354 $^{+0.00011}_{-0.00016}$ \\
$K'$                                      & [\ms]                  &             &  29.7 $\pm$ 2.0                & 32.0 & 30.1 $^{+1.6}_{-1.5}$ \smallskip\\

Planet mass, $M_{\mathrm{p}}$             &[\Mjup]                 &             &  0.288 $\pm$ 0.021             & 0.308 & 0.287 $^{+0.022}_{-0.016}$ \\
Planet radius, $R_{\mathrm{p}}$           &[\RJnom]                &             &  0.756 $\pm$ 0.014             & 0.758 & 0.756 $^{+0.013}_{-0.017}$ \\
Planet mean density, $\rho_{\mathrm{p}}$  &[$\mathrm{g\;cm^{-3}}$] &             &  0.829 $\pm$ 0.077             & 0.878 & 0.831 $^{+0.087}_{-0.069}$ \\
Planet surface gravity, $\log$\,$g_{\mathrm{p}}$ &[cgs]            &             &  3.096 $^{+0.032}_{-0.036}$    & 3.124 & 3.098 $^{+0.037}_{-0.033}$ \smallskip\\

\emph{\bf Planet~c} \\
Semi-major axis, $a$                      & [au]                   &   &  0.2053 $\pm$ 0.0034        & 0.2044 & 0.2044 $^{+0.0021}_{-0.0038}$ \\
Eccentricity, $e$                         &                        &   &  0.228 $^{+0.014}_{-0.019}$ & 0.1791 & 0.1809 $^{+0.018}_{-0.0072}$ \\
Argument of pericentre, $\omega$          & [\degree]              &   &  22.9 $^{+6.8}_{-5.1}$       & 25.7 & 26.1 $\pm$ 9.4\\
Inclination, $i$                          & [\degree]              & $S(90, 180)$  &  120.6 $\pm$ 7.3 & 106.7 & 104.9 $^{+4.6}_{-7.3}$ \\
Longitude of the ascending node, $\Omega$ & [\degree]              & $U(90, 270)$  &  207.4 $^{+4.4}_{-6.4}$ & 191.6 & 192.2 $^{+6.0}_{-2.7}$ \\
Mean anomaly, $M_0$                       & [\degree]              &   &  260.5 $\pm$ 3.5        & 260.4 & 258.0 $^{+5.0}_{-4.0}$ \smallskip\\

$\sqrt{e}\cos{\omega}$                    &                        & $U(-1, 1)$ &  0.437 $^{+0.025}_{-0.032}$   & 0.381 & 0.382 $\pm$ 0.030 \\
$\sqrt{e}\sin{\omega}$                    &                        & $U(-1, 1)$ &  0.185 $^{+0.051}_{-0.037}$   & 0.183 & 0.187 $\pm$ 0.062 \\
Mass ratio, $M_{\mathrm{p}}/M_\star$      &                        & $U(0, 1)$  &  0.000417 $\pm$ 0.000040       & 0.000406 & 0.000394 $\pm$ 0.000027 \\
Scaled semi-major axis, $a/R_{\star}$     &                        &   &  48.0 $\pm$ 1.1    & 48.12 & 48.10 $^{+0.98}_{-1.1}$ \\
Impact parameter, $b$                     &                        &   &  21.2 $\pm$ 4.5       & 12.4 & 11.0 $^{+3.4}_{-5.3}$ \\
$T_0'$\;-\;2\;450\;000                    & [BJD$_{\mathrm{TDB}}$] & $U(6957, 8957)$ &  7971.27 $^{+0.33}_{-0.28}$           & 7971.54 & 7971.62 $^{+0.31}_{-0.26}$ \\
$P'$                                      & [d]                    & $U(0, 1000)$ &  34.5412 $\pm$ 0.0028          & 34.54626 & 34.54619 $^{+0.00094}_{-0.0016}$ \\
$K'$                                      & [\ms]                  &   &  23.7 $\pm$ 2.2                     & 25.47 & 24.85 $^{+1.5}_{-0.97}$ \smallskip\\

Planet mass, $M_{\mathrm{p}}$             &[\Mjup]               &   &  0.424 $\pm$ 0.046       & 0.406 & 0.392 $^{+0.023}_{-0.027}$ \\
$M_{\mathrm{p}}\sin{i}$                   &[\Mjup]               &   &  0.361 $\pm$ 0.036       & 0.389 & 0.378 $^{+0.021}_{-0.024}$ \medskip\\

Mutual inclination, $I$                   & [\degree]            &   &  41.0 $^{+6.2}_{-7.6}$ & 20.7 & 20.8 $\pm$ 4.6 \smallskip\\

\hline

\end{tabular}
\tablefoot{The table lists: Prior, posterior median, and 68.3\% credible interval (CI) for the photodynamical analysis (Section~\ref{section.photodynamical}), maximum a posteriori (MAP) and median, and 68.3\% CI for the stable samples (Section~\ref{sec.stability}). The Jacobi orbital elements are given for the reference time $t_{\mathrm{ref}}~=~2\;457\;957.48167$~BJD$_{\mathrm{TDB}}$. Additional parameters are in Table~\ref{table.results2}. The 'stable MAP' solution is the nominal solution used in Sections~\ref{sec.stability} and \ref{sec.secular}. We note that $P'$ and $T_0'$ should not be confused with the period or the time of conjunction, respectively, and they were only used to reduce the correlations between jump parameters, replacing the semi-major axis and the mean anomaly at $t_{\mathrm{ref}}$. $T'_0 \equiv t_{\mathrm{ref}} - \frac{P'}{2\pi}\left(M_0-E+e\sin{E}\right)$ with $E=2\arctan{\left\{\sqrt{\frac{1-e}{1+e}}\tan{\left[\frac{1}{2}\left(\frac{\pi}{2}-\omega\right)\right]}\right\}}$, $P' \equiv \sqrt{\frac{4\pi^2a^{3}}{\mathcal G M_{\star}}}$, $K' \equiv \frac{M_p \sin{i}}{M_\star^{2/3}\sqrt{1-e^2}}\left(\frac{2 \pi \mathcal G}{P'}\right)^{1/3}$. CODATA 2018: $\mathcal G$ = 6.674$\;$30\ten[-11]~$\rm{m^3\;kg^{-1}\;s^{-2}}$. IAU 2012: \rm{au} = 149$\;$597$\;$870$\;$700~\rm{m}$\;$. IAU 2015: \Rnom = 6.957\ten[8]~\rm{m}, \GMnom = 1.327$\;$124$\;$4\ten[20]~$\rm{m^3\;s^{-2}}$, $\RJnom$ = 7.149$\;$2\ten[7]~\rm{m}, \GMJnom = 1.266$\;$865$\;$3\ten[17]~$\rm{m^3\;s^{-2}}$. $\Msun$ = \GMnom/$\mathcal G$, \Mjup = \GMJnom/$\mathcal G$, $k^2$ = \GMnom$\;(86\;400~\rm{s})^2$/$\rm{au}^3$. $N(\mu, \sigma)$: Normal distribution with mean $\mu$ and standard deviation $\sigma$. $U(a, b)$: A uniform distribution defined between a lower $a$ and upper $b$ limit. $S(a, b)$: A sinusoidal distribution defined between a lower $a$ and upper $b$ limit.}
\end{table*}

\section{Results}\label{section.results}

\begin{figure}
  \includegraphics[width=0.235\textwidth]{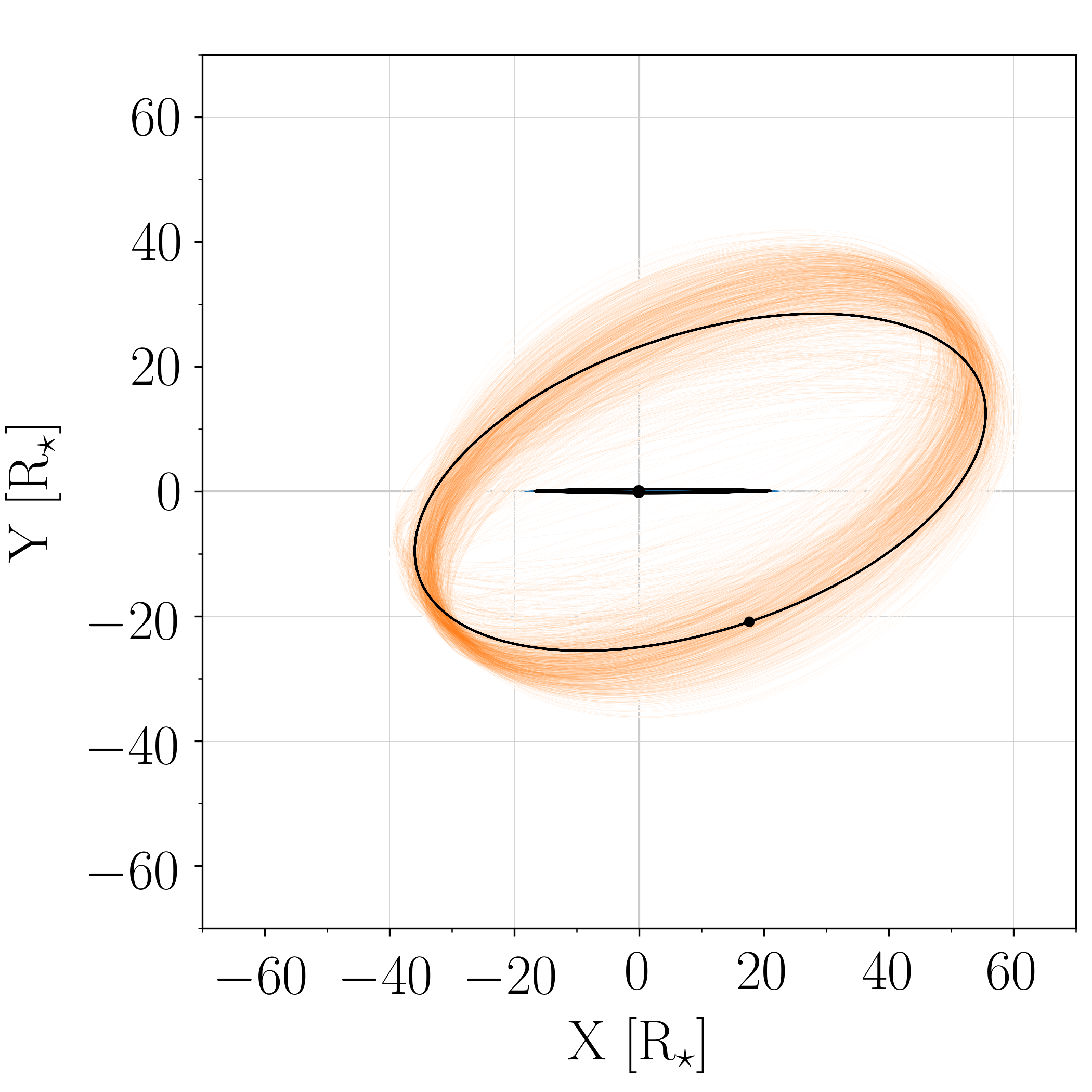}
  \hspace{0.2cm}\includegraphics[width=0.235\textwidth]{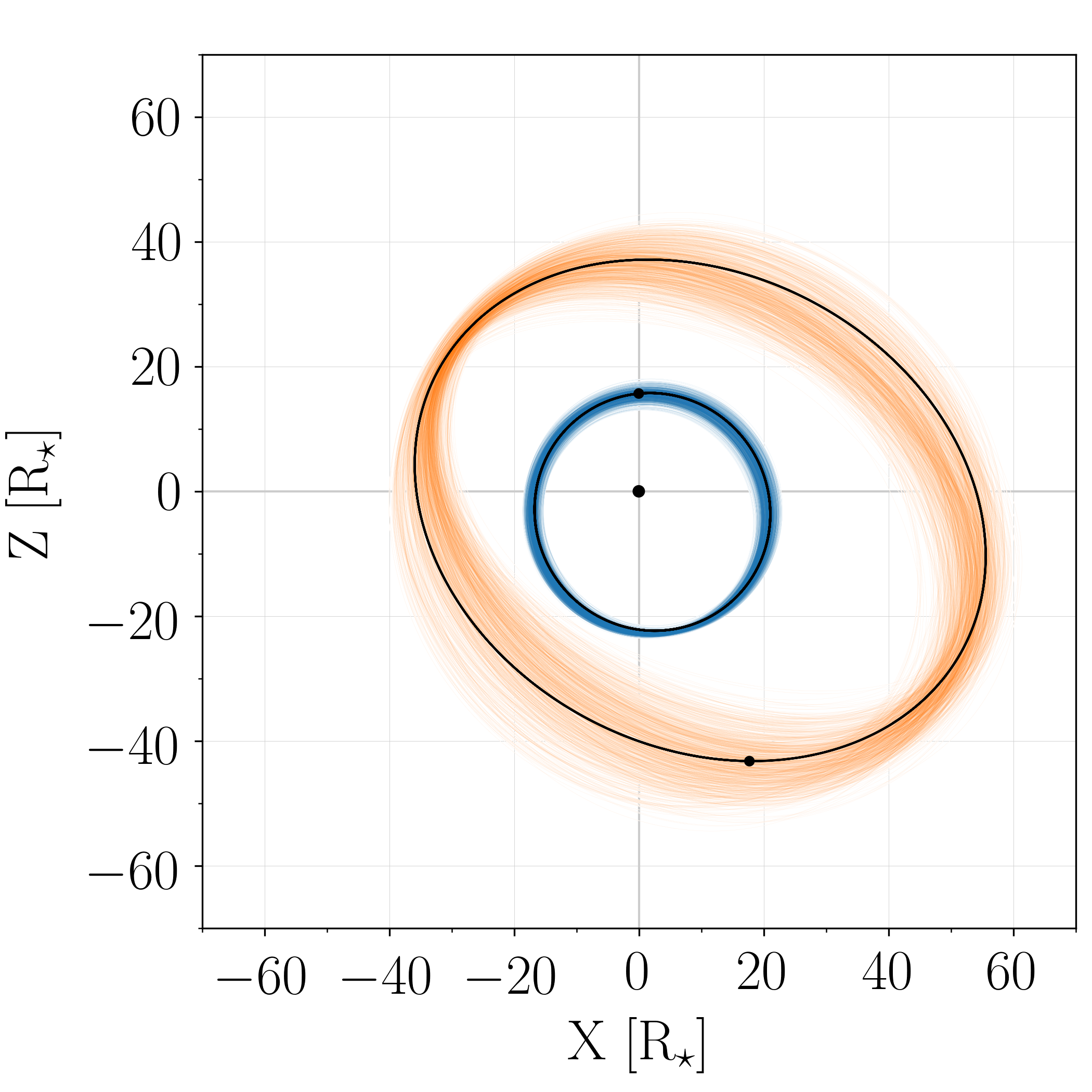}\\
  \includegraphics[width=0.235\textwidth]{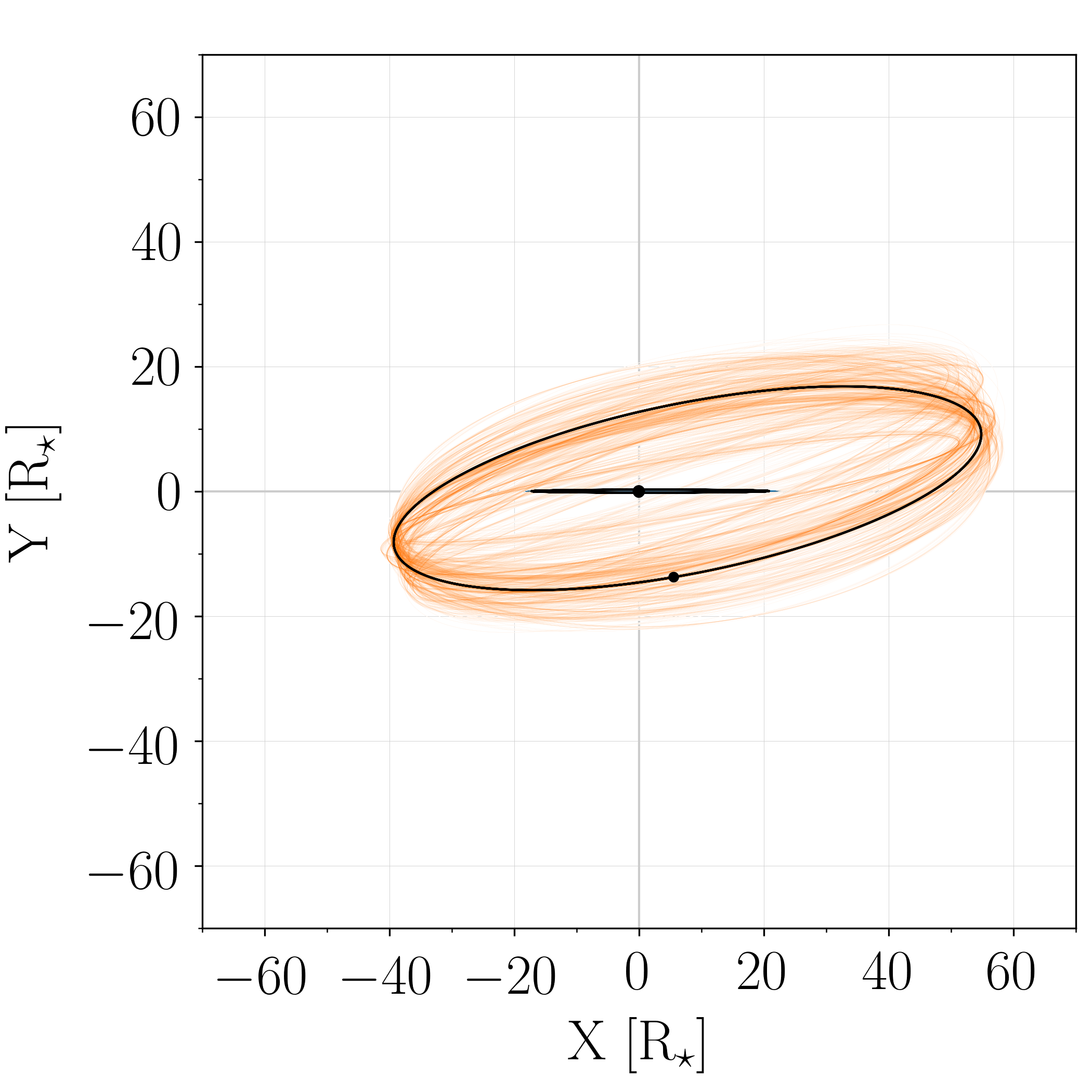}
  \hspace{0.2cm}\includegraphics[width=0.235\textwidth]{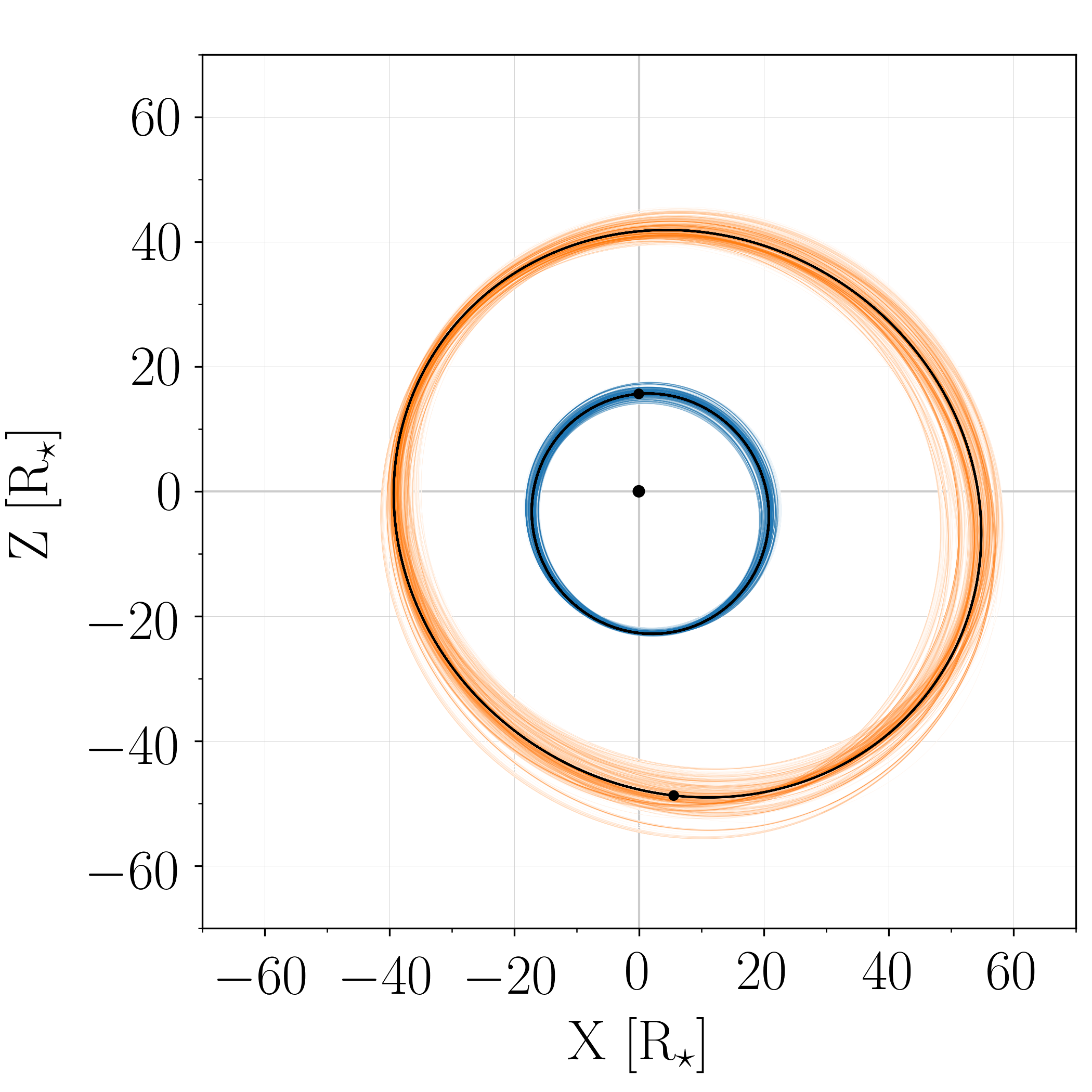}
  \caption{Orbital projections. Top: Orbital projections around the time of the RISE transit for planet~b (blue) and planet~c (orange). The origin is the system barycentre, and the orbits are projected in the sky plane seen by the observer (X-Y, left) and X-Z plane (right, system top view, the movement is clockwise, the positive Z-axis points towards the observer). A thousand random orbits were drawn from the posterior samples, and the MAP is shown as a black orbit. The black points mark the position of the star (size to scale) and the planets at the central time of the RISE transit for the MAP (the size of planet~b is enlarged by a factor of 10, and the size of planet~c is not known). Bottom: Idem. as for 1000 stable solutions (Section~\ref{sec.stability}).} \label{fig.orbits}
\end{figure}

In Table~\ref{table.results} we list the prior, the median, and the 68\% credible interval of the inferred system parameters' marginal distributions. The one- and two-dimensional projections of the posterior sample are shown in Figure~\ref{fig.pyramid}. The MAP model is plotted in Figures~\ref{fig.PH} and \ref{fig.RV}. Figure~\ref{fig.orbits} shows the posterior of the planet orbits. With an inferred impact parameter of $21.2 \pm 4.5$, the transits of planet~c are highly disfavoured, which is in agreement with the null result of the transit search in \citet{Maciejewski2020}. The derived parameters agree with the ones reported by \citet{hebrard2020}, but they have a significantly improved precision. A difference, however, is the slightly larger radius ratio $R_{\mathrm{p}}/R_\star$. Instead of $\rm{M_p\sin{i}}$, the photodynamical modelling allows the true mass of planet~c to be measured at $M_{\mathrm{c}} = 0.424 \pm 0.046$~\Mjup, which is in agreement with the dynamical upper limit of 0.60~\Mjup\ reported by \citet{hebrard2020}.
Most of the posterior samples have the apses of the orbits that librate around alignment.  
From tests we determined that the improved precision in model posteriors over the earlier study of \citet{hebrard2020} is due to both the additional data and the photodynamical modelling that account for the interactions between the planets.

\subsection{Transit-timing variations}
\begin{figure}
  \centering
  \includegraphics[width=0.49\textwidth]{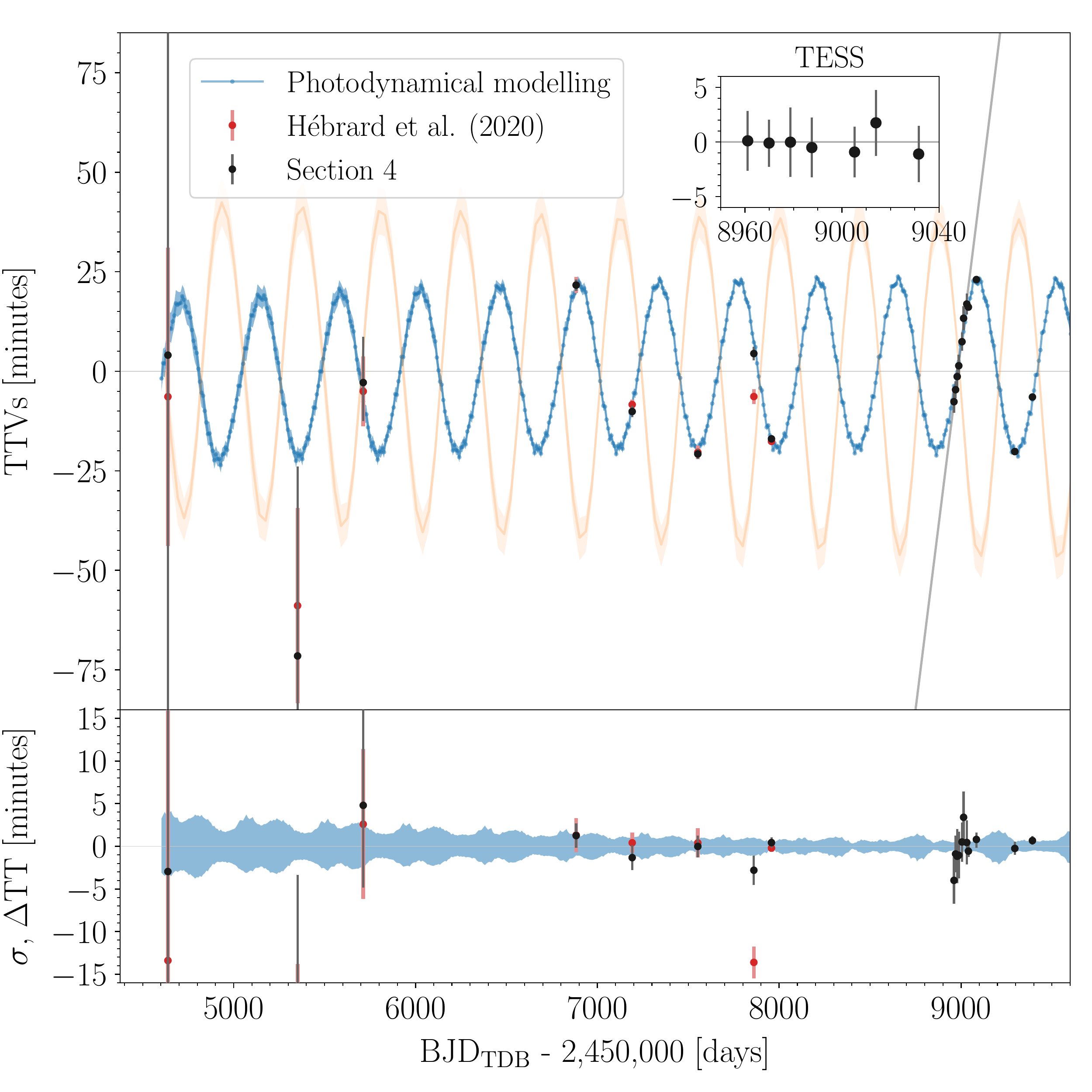}
  \caption{Posterior TTV predictions of planet~b computed relative to a linear ephemeris\protect\footnotemark\ are shown in blue. A thousand random draws from the posterior distribution were used to estimate the TTV median value and its uncertainty (68.3\% credible interval). In the upper panel, the median TTV values are shown and compared with individual transit-time determinations (\citealt{hebrard2020} in red, and Section~\ref{section.juliet} in black). The thick grey line represents a linear ephemeris computed using only the transits observed by \TESS (whose residuals are shown in the small panel in the upper right). In the lower panel, the posterior median transit-timing value was subtracted to visualise the uncertainty of the distribution. The posterior median transit time was also subtracted from each observed epoch for the individual transit-time determinations to allow for better comparison with the photodynamical modelling. The orange curve in the upper panel represents the variation in the times of inferior conjunction for planet~c.} \label{fig.TTVs}
\end{figure}

Figures~\ref{fig.TTVs} and \ref{fig.TTVs2100} show the posterior TTVs of planet~b, obtained from the time of conjunction. The posterior transit times, which rely on the three-body system hypothesis, agree with the individually derived transit times (Section~\ref{section.juliet}). In addition, the periodicity of the TTVs agrees with the $\approx$450~days super-period \citep{lithwick2012} for two planets near the 4:1 mean-motion resonance\footnote{$P_{\rm sup} = \frac{P_c}{4 |\Delta|}$, $\Delta = \frac{P_c}{4 P_b} - 1$, with $P_{b,c}$ being the planet orbital period of the corresponding subscript.}. The periodicity of the TTVs also agrees with the one reported by \citet{hebrard2020}.

The posterior timing from the photodynamical modelling is quite wide at some epochs (lower panel of Figure~\ref{fig.TTVs} and \ref{fig.TTVs2100}). The uncertainty in the posterior transit times is related with the knowledge of the system parameters. Future transit observations should favour the epochs where the posterior transit uncertainty is large to further improve the characterisation of the system. There is room for improvement with $\sim$1~minute transit-timing precision observations. Predictions of transit times up to 2026 are listed in Table~\ref{table.transits}.

If only the \TESS observations are considered, TTVs would not have been detected (Figure~\ref{fig.TTVs}). \TESS nearly continuous photometry observations of WASP-148, with a time span of approximately nine planet~b periods, are insufficient to detect the TTVs of planet~b. This is due to the particular configuration of the \TESS observations, which accidentally only cover a part of increasing TTVs (see upper panel Figure~\ref{fig.TTVs}). Thus, it is possible that a similar situation is occurring in other systems on which TTVs are also missed.
The period derived using only the \TESS data lasts $8.80604\pm0.00014$~days, and from all the observed transit times it is $8.8038083\pm0.0000026$~days (Section~\ref{sec.ephemeris}). This is a difference of $193\pm12$~seconds. Using the \TESS ephemeris to predict future transit timing induces an offset of $\sim$2.2~hours/year.

The posterior TTVs' evolution in the near future is shown in Figure~\ref{fig.TTVs2100}. The posterior transit duration variations are shown in Figure.~\ref{fig.TDVs}.

\footnotetext{Computed using the posterior median of the mid-transit times between the first and last transit observation, 2457957.493798 + 8.803808\;$\times$\;Epoch [BJD$_{\rm TDB}$]}

\subsection{Model-independent linear ephemeris}\label{sec.ephemeris}

The predictions issued from the photodynamical model and presented in Table~\ref{table.transits} have the disadvantage of having been obtained under the assumption that the two-planet model is correct. The results from the transit-only analysis presented in Section~\ref{section.juliet} can be used to provide a model independent ephemeris which should provide valid predictions, albeit less precise ones, even if the system is discovered to contain additional planets in the future.

A value of a mean period and time of transit can be straightforwardly produced by fitting a slope to the transit times in Table~\ref{table.juliet}. However, in the presence of transit-timing variations, a slope is not a flexible enough model to describe the transit times. Therefore, the parameters inferred from such a model are likely biased \citep[e.g.][]{bishop2007}. \citet{hebrard2020} dealt with this by inflating the error bars of the individual transit times to reach $\chi^2 \sim 1$. Here we decided to use a non-parametric model to describe the variation of the transit times over the linear ephemeris model. 

More precisely, we chose a GP regression model whose mean function was specified to be a linear function with parameters --slope and intercept-- to be inferred from the data \citep[see][Section~2.7]{rasmussenwilliams2005}. If normal priors are chosen for these parameters, the computation of the marginal likelihood can be performed analytically. We chose diffuse priors with widths of 100~days, centred at 8.804~days for the slope (period) and at the observed time of the second \TESS transit observed in sector 25 for the intercept.

We tried several kernel functions to define the covariance function of the GP. We used a modified version of the implementation in \texttt{scikit-learn} \citep{scikit-learn} and optimised the hyperparameters using the L-BFGS-B \citep{byrd1995, zhu1997} and the Sequential Least SQuares Programming (SLSQP) algorithms \citep{kraft1988}, implemented in the \texttt{scipy} package \citep{scipy2020}. All tested kernels produced almost identical results for the parameters of the linear ephemeris. We discuss below the results issued from the kernel choice that provided the largest value of optimised marginal likelihood.

\begin{figure}
  \centering
\includegraphics{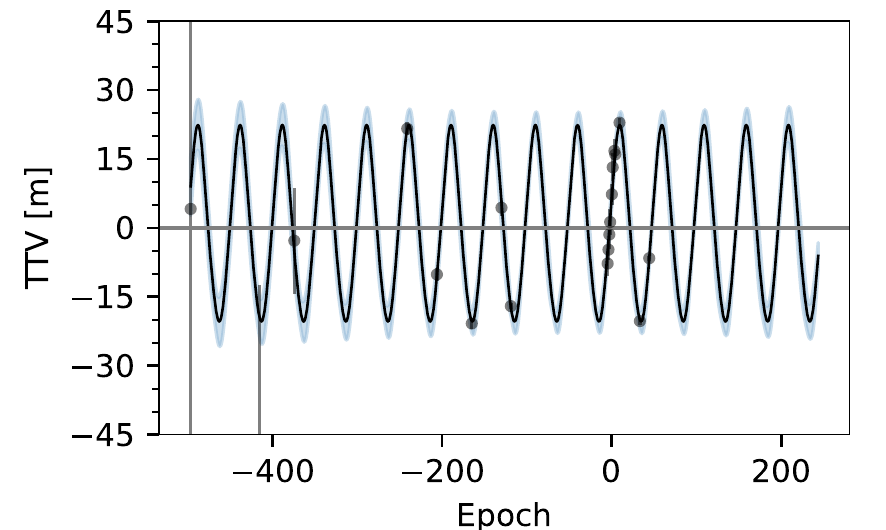}
\caption{Model-independent analysis of the timings of the observed transits (black circles). The black line is the mean prediction of the posterior GP model, and the blue band is the 68.3\% credible interval.} \label{fig.linearGPR}
\end{figure}

The kernel function, $k(x, x^\prime)$, producing the largest marginal likelihood value was a exp-sine-squared (ESS) kernel, without exponential decay:

\begin{equation}
\arraycolsep=0pt
\def\arraystretch{2.2}
\begin{array}{rl}
k(x, x^\prime) = &A_\mathrm{ESS} \exp{\left\{-\frac{2 \sin^2{\left[\pi (x - x^\prime)^2 / P\right]}}{\epsilon^2}\right\}},
\end{array}
\def\arraystretch{1}
\end{equation}
where $P$ and $\epsilon$ are the kernel hyperparameters that correspond to the period and length-scale, respectively, and $A_\mathrm{ESS}$ is the amplitude of the covariance function. The posterior mean and 68.3\% intervals for the TTVs are shown in Fig.~\ref{fig.linearGPR}. A kernel function with an additional decay term produced similar results, with a decay timescale far exceeding the time span of the observations. 

    Under the assumption of normal priors, the linear model parameter posterior is also normal. We find the posterior ephemeris:
    $$
    \begin{array}{lcl}
        P &=& 8.8038083 \pm 0.0000026 \;\text{d}\\
        T_c &=& 2459005.148 \pm 0.016\;\left[\mathrm{BJD}_\mathrm{TDB}\right]\;\;,
    \end{array}
    $$
    with covariance term, $\mathrm{cov}(P, T_c) = 6.2\times10^{-10}$. The large uncertainties in $T_c$ are misleading. The predicted transit times issued from this model have a covariance which includes additional terms coming from the interaction between the mean function and the non-parametric part of the model. In fact, the predictions of the transit times remain precise to better than 10~minutes for over 2640 transits of planet~b, that is to say over 60~years. 
Comparison with the transit times reported in Table~\ref{table.transits} shows that this method predicts transit times that are in agreement with the fully photodynamical one to better than 5 minutes up to the end of 2026.

In comparison, an ordinary least squares (OLS) fit to the transit times, with uncertainties scaled to have a reduced $\chi^2$ of one as presented in \citet{hebrard2020}, produces a period of $P = 8.803824 \pm 0.000029$~days, which is in agreement with the result from above and with the values presented in \citet{hebrard2020}. The OLS intercept estimator has a standard deviation six times smaller than the one in our model, which is probably unrealistic. However, the predictions from the OLS model remain precise to better than 10~minutes for less than four years.

The optimised covariance amplitude is $A_\mathrm{ESS}=25.3$~minutes, and the period $P$ is 49.7~orbits, corresponding to 437~days. The length-scale $\epsilon$ was fixed to 2.0. The fact that the model with the largest marginal likelihood does not include a long-term evolution of the transit times means that with the current data, such a trend is not detected.

\subsection{Mutual inclination}\label{section.mutualInclination}

\begin{figure}
  \centering
  \includegraphics[width=0.5\textwidth]{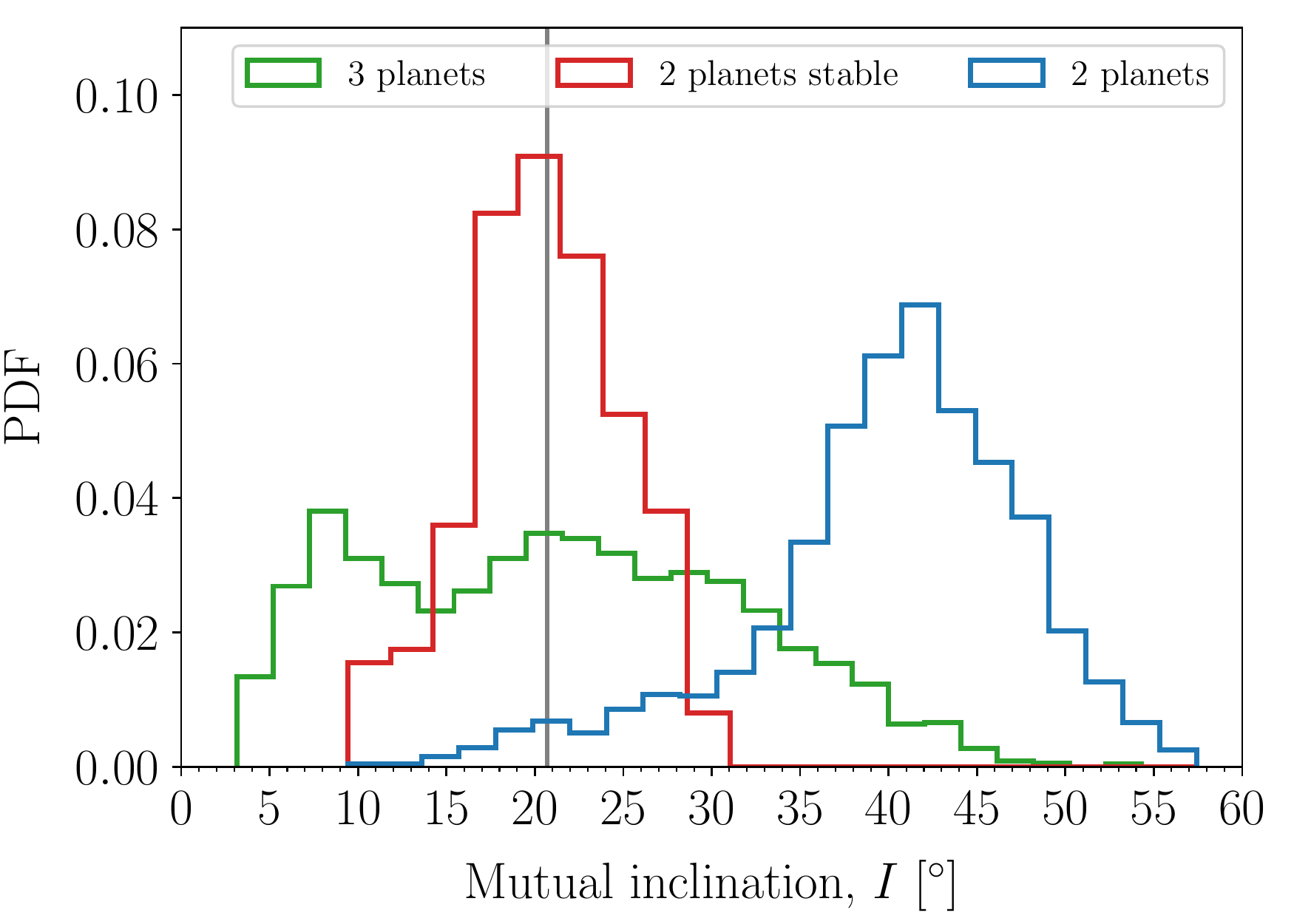}
  \caption{Posterior of the mutual inclination at $t_{\mathrm{ref}}$ from the photodynamical modelling (Section~\ref{section.photodynamical}), the stable samples (Section~\ref{sec.stability}, the stable MAP value is marked by a vertical grey line), and the three-planet model (Section~\ref{sec.3planets}).} \label{fig.irel}
\end{figure}

\begin{figure}
  \centering
  \includegraphics[width=0.5\textwidth]{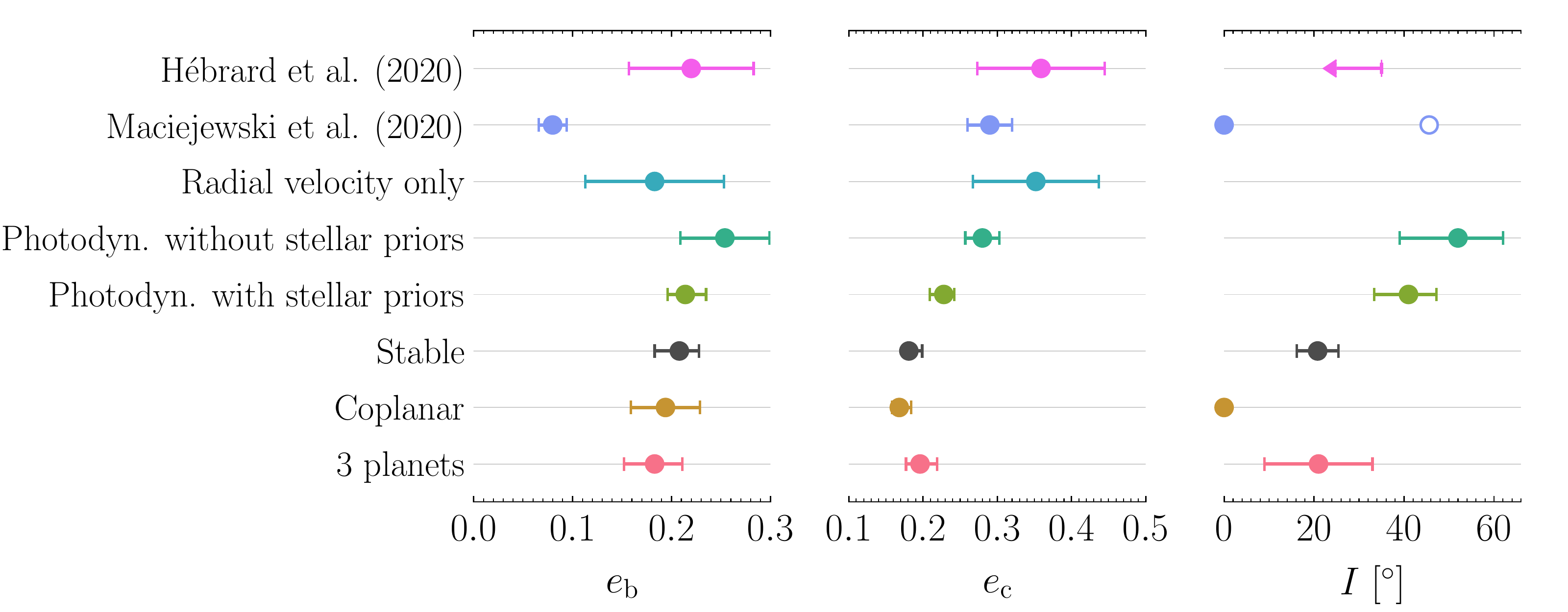}
  \caption{Planet~b and c eccentricity, as well as the mutual inclination for the different analysis presented in this work, and in previous literature. \citet{Maciejewski2020} used a $t_{\mathrm{ref}}~=~2\;459\;048.747$~BJD$_{\mathrm{TDB}}$, which is different from the one used in this work, but the difference in the inferred orbital parameters cannot be explained by their evolution between the two epochs (Figure~\ref{fig.LongTermEvolution}). We adopted the values labelled as 'stable'.}\label{fig.comparison}
\end{figure}

We inferred a mutual inclination between the planets\footnote{$\cos I = \cos i_b\cos i_c + \sin i_b\sin i_c \cos\left({\Omega_b-\Omega_c}\right)$}, $I$, at $t_{\mathrm{ref}}$ of 41.0$^{+6.2}_{-7.6}$~\degree\ (median and 68.3\% credible interval), and a 95\% highest density interval (HDI) of [22.8\degree, 54.8\degree] (Figure~\ref{fig.irel}). With these values, a coplanar system, as assumed in \citet{Maciejewski2020}, was discarded. \citet{Maciejewski2020} tried a non-coplanar model and found a best fit solution (with $\Omega_c - \Omega_b \approx -17\degree$ and $i_c = 47\degree$, or $\Omega_c - \Omega_b \approx +17\degree$ and $i_c = 133\degree$) which corresponds to a mutual inclination of $\sim$46\degree, although they also found that the Bayesian information criterion (BIC) disfavours the non-coplanar solution. However, for high-dimensional models such as these ones, the BIC is known to provide unreliable results \citep{diaz2016, nelson2020}. The coplanar solution in \citet{Maciejewski2020} has a lower eccentricity for planet~b (Figure~\ref{fig.comparison}).
From a stability analysis of the orbital solution they derived, \citet{hebrard2020} found an upper limit of 35\degree\ for the mutual inclination. We note that the reduced-$\chi^2$ level curves of the Newtonian fit plotted in Fig.~9 of \citet{hebrard2020} favoured mutual inclinations around 30\degree.

We tried to investigate which observable favours the significant mutual inclination. For this, we repeated the photo-dynamical analysis (Section 6) assuming coplanar orbits; we fixed the longitude of the ascending node of both planets to the same value, and we matched the orbital inclination of planet~c to the one of planet~b, with the latter still being a free parameter. The results are presented in Table~\ref{table.coplanar}, and Figures~\ref{fig.comparison}, \ref{fig.TTVs2100}, and \ref{fig.TDVs}. The TTVs' posteriors of coplanar and inclined orbits are roughly the same for the observed transits (Figure~\ref{fig.TTVs2100}). On the other hand, the transit duration posteriors are different, the transit duration of the observed transits remains almost constant for the coplanar model, whereas it decreased for the model with inclined orbits (Figure~\ref{fig.TDVs}). However, we found that the precision of individually determined transit durations\footnote{We analysed each transit observation individually to determine the transit duration (with the techniques described in Section~\ref{section.juliet}), excluding partial transits, as well as transits observed by SuperWASP (poor precision) and \TESS (30-minute cadence). The transit duration was computed numerically from 1,000 models from the posterior predictive distribution sampled to one-second cadence. This values differs from the transit duration computed in \citet{hebrard2020} using the approximations in \citet{tingley2005}, whose limitations were discussed by \citet{kipping2010b}.} is not enough to confirm  the results of the modelling with inclined orbits independently. The photodynamical modelling is in principle more sensible than the analysis on individual transits \citep{almenara2015}, but the heterogeneity of the transit observations analysed in this work call for caution. Small variations in transit shape or timing intrinsic to the different observations could be wrongly interpreted by photodynamical modelling as an evolution in the orbital parameters\footnote{For example, transit depth variations, which have not been discussed in this work, can be due to changes in the impact parameter, but also due to a different type of contamination in the photometric aperture.}.
If not correctly taken into account, the limb darkening dependence with the observation bandpass could be misinterpreted as changes in the impact parameter, and therefore high mutual inclinations between the planets. However, no correlation is seen between the mutual inclination and the limb-darkening parameters. In addition, the posterior distributions of the limb-darkening parameters agree with those expected from theoretical computation \citep{claret2011,claret2017}, although they are much wider.

In addition, models starting with a co-planar configuration were run. We found that the co-planar region of parameter space is left quickly by the MCMC walkers. The detection of a non-negative mutual inclination does not seem to be produced by an inadequate exploration of a space parameter, or by bad mixing or lack of convergence of the MCMCs.
Future observations focussed on distinguishing between the increasingly and differently predicted transit duration could conclude about the mutual inclination.

\subsection{Photodynamical modelling without stellar priors}\label{section.noprior}

The photodynamical modelling of photometry and radial velocity data allows one to measure the absolute radius and mass in multi-planetary systems \citep{agol2005,almenara2015}. To test the precision of this determination in this system, we ran the same analysis as in Section~\ref{section.photodynamical}, but with non-informative priors for the stellar mass and radius. Masses, radii, densities, and orbital parameters are listed in Table~\ref{table.absolute}. The absolute parameters' precision is poor. The radii were determined with a precision of 32\% relative uncertainty for the star and planet~b. The bulk densities of the star and planet~b were determined with a precision of 20\% and 39\%, respectively. The masses of the star, planet~b, and planet~c were determined with a precision of 89\%, 58\%, and 82\%, respectively. The precision on planet masses outperform the one on the star \citep{almenara2018}.
The posterior eccentricities are also more precise for the photodynamical modelling with the stellar priors (Figure~\ref{fig.comparison}). This means that the photoeccentric effect \citep{dawson2012} puts constraints on the eccentricities in addition to the ones coming from the TTVs. The same is true for the mutual inclination.

\subsection{Stability analysis}\label{sec.stability}

\begin{figure}
  \centering
  \includegraphics[width=0.5\textwidth]{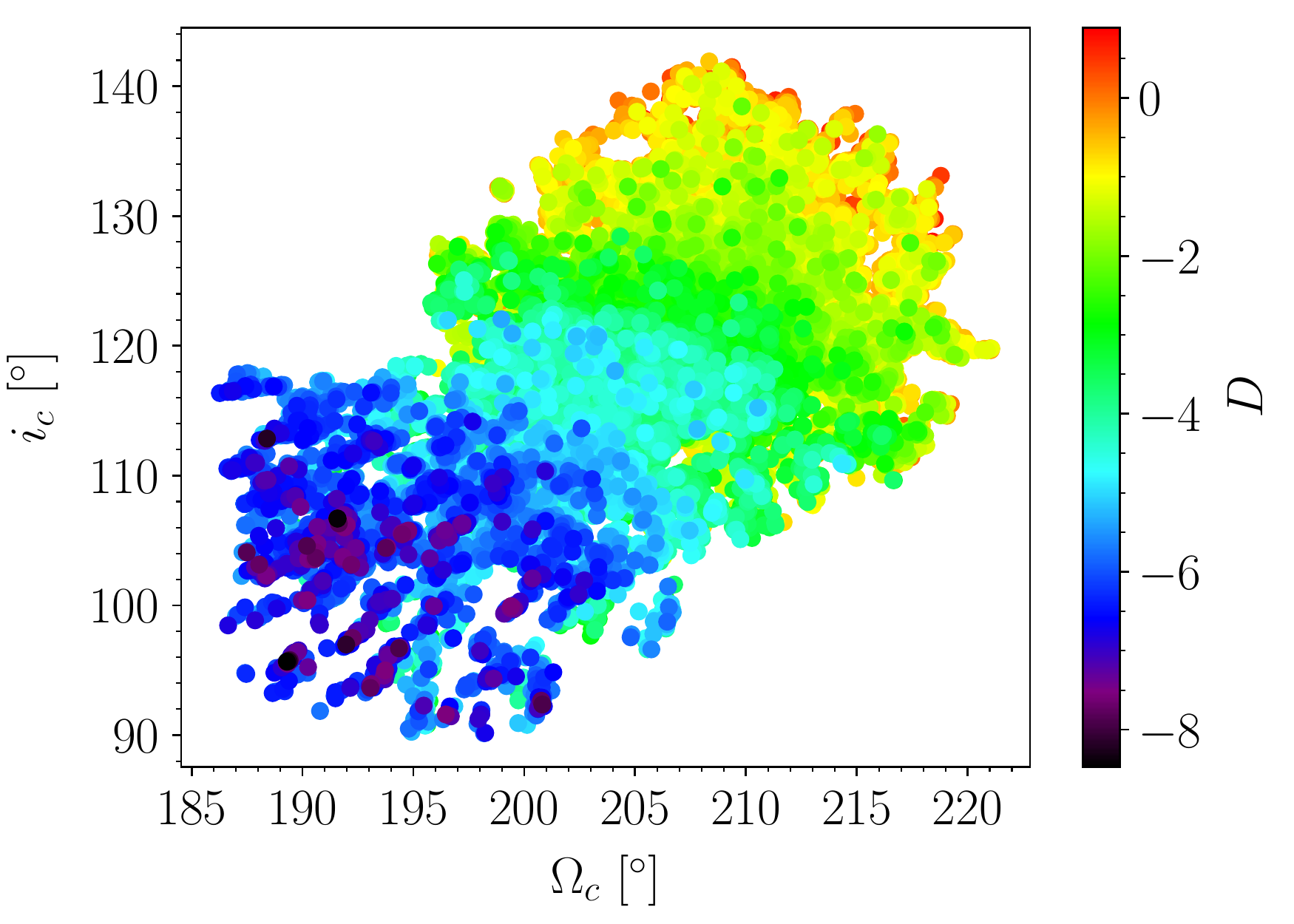}
  \caption{Stability analysis of 50,000 samples of the posterior distribution obtained from the photodynamical modelling (Section~\ref{section.photodynamical}), projected in a $(\Omega_c, i_c)$ diagram. The colour scale corresponds to values between --8.5 (black) and 0.9 (red) for the decimal logarithm of the stability index $D$ used in \citet{correia2010}. The red zones correspond to highly unstable orbits, while the dark blue region can be assumed to be stable on a billion-year timescale.} \label{scatterD}
\end{figure}

The dynamical analysis of the WASP-148 orbital solution reported in \citet{hebrard2020} has shown that the system is stable, despite significant mutual gravitational interactions between the planets. For this study, we repeated a similar stability study, but instead we performed a global frequency analysis \citep{laskar1990,laskar1993} on 50,000 samples of the posterior distribution obtained from our new photodynamical model. We used the symplectic integrator SABA1064 of \citet{Farres_etal_2013}, with a step size of $5\times 10^{-3}$~yr and general relativity corrections. Each initial condition was integrated over 50~kyr, and a stability indicator $D$ was derived with the frequency analysis of the mean longitude, that is the variation in the measured mean motion over the two consecutive 25~kyr intervals of time \citep[for more details, see][]{couetdic2010}. 
For regular motion there is no significant variation in the mean motion along the trajectory, while it can significantly vary for chaotic trajectories.
In Figure~\ref{scatterD} we show the distribution of the 50,000 samples projected in a $(\Omega_c, i_c)$ diagram, which corresponds to the two less constrained parameters.
The colour index gives the value of $D$ for each solution.
The values of $\log_{10} D < -6$ for both WASP-148~b and c correspond to stable systems on scales of billions of years \citep{correia2010}. 
Only a small region from this diagram is stable for $\Omega_c < 205^\circ$ and $i_c < 120^\circ$, corresponding to 1239 solutions (2.5\% of the total). In Figure~\ref{fig.irel} we show the  probability density function (PDF) of this stable subset of solutions.
The solutions cluster around a mutual inclination of $20.8 \pm 4.6$~\degree.
We hence conclude from that three-body analysis that WASP-148 can only be stable for mutual inclinations below about $30^\circ$.
In the last two columns in Table~\ref{table.results}, we provide the system parameters corresponding to the MAP, the median, and the 68\% credible interval of the stable solutions.

\begin{figure}
  \centering
  \includegraphics[width=0.5\textwidth]{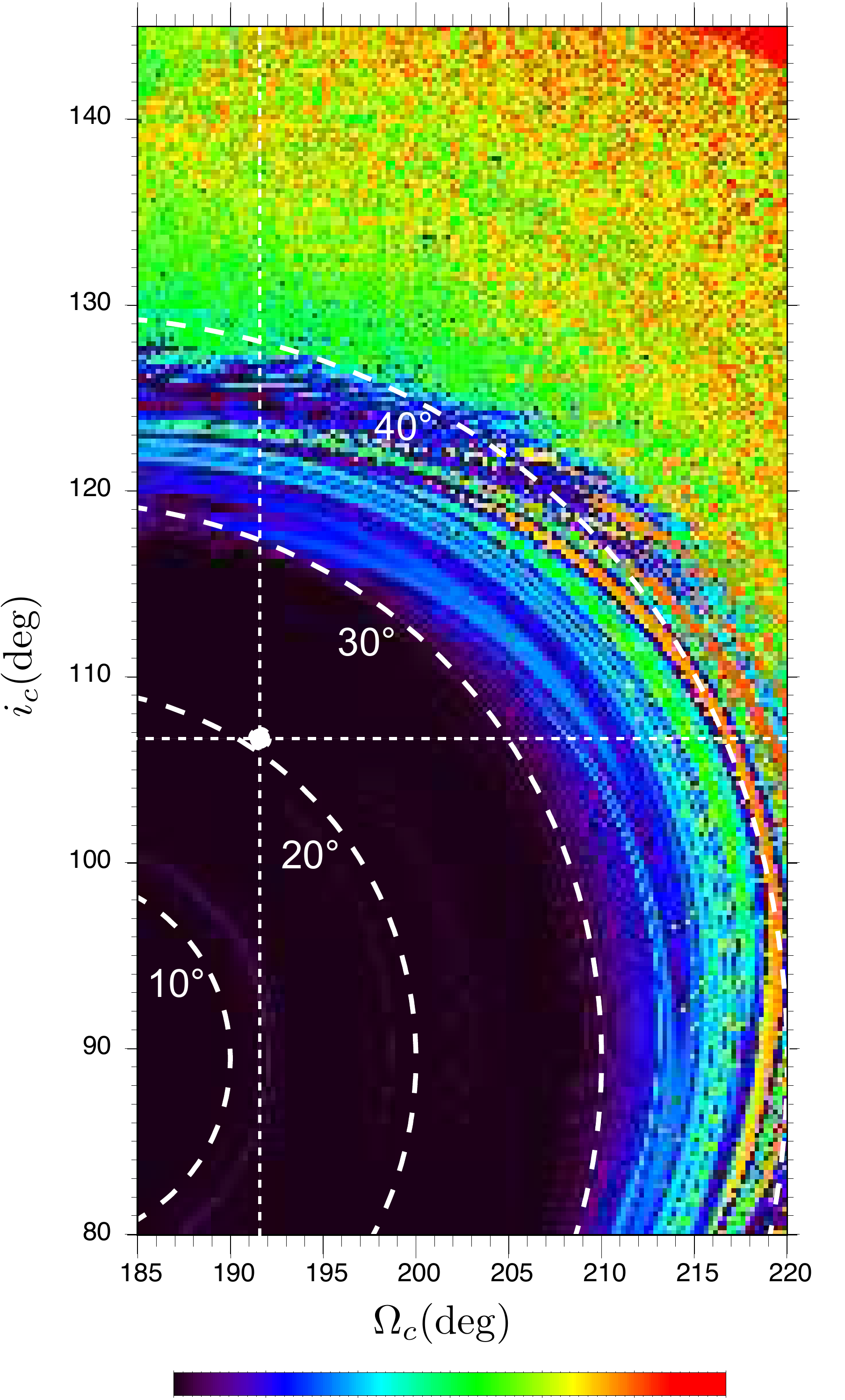}
  \caption{Global stability analysis of the WASP-148 planetary system. We fixed all orbital parameters of the stable MAP solution (Table~\ref{table.results}), and we varied the inclination, $i_c$, and the longitude of the ascending node, $\Omega_c$, of planet-$c$. The step size is $0.25\degree$ in both axes. For each initial condition, the system was integrated over 50~kyr and a stability criterion was derived with the frequency analysis of the mean longitude. White dashed curves give the isolines of constant mutual inclination $I = 10\degree$, $20\degree$, $30\degree$, and $40\degree$. The white dot marks the position of the stable MAP solution from Table~\ref{table.results}. The colour bar corresponds to the one in Figure~\ref{scatterD}.}\label{fig.stability}
\end{figure}

As in \citet{hebrard2020}, we also explored the stability around the stable MAP solution (Table~\ref{table.results}), which we refer to as the nominal solution henceforward. As expected, in the $(a_c,e_c)$ domain, we confirm that it lies in a stable area, with the orbits close to the 4:1 mean-motion resonance. 
In the $(i_c, \Omega_c)$ domain (Figure~\ref{fig.stability}), we confirm that stable orbits must have a mutual inclination of $I \lesssim 30\degree$.  
In comparison to the stability map shown in Fig.~9 in \citet{hebrard2020} with $i_c < 90\degree$, here we only zoom into the most stable regions, and imposed $i_c > 90\degree$: this is equivalent from a dynamical point of view, but now the stable solutions are centred around $\Omega_c = 180\degree$ rather than $\Omega_c=0\degree$. The main difference in the new analysis is the fact that the nodes and the inclination are now constrained by the observations.

The two stability analyses presented above use different approaches: Fig.~\ref{scatterD} presents a stability analysis carried out on different solutions allowing all parameters to vary, whereas Fig.~\ref{fig.stability} studies the stability with only two free parameters ($i_c$, $\Omega_c$) and the other parameters are fixed at their MAP values. We show here that they provide similar results in terms of the derived parameters for stability, which supports the reliability of both approaches and their results.

\subsection{Orbital evolution}\label{sec.orbitalEvolution}

To explore the dynamics of the system, we analyse 1000 stable samples from the joint parameter posterior distribution of the photodynamical model and performed numerical integrations\footnote{With the same n-body integrator and time step used in Section~\ref{section.photodynamical}.} for 1~kyr after $t_{\mathrm{ref}}$. The results for the selected parameters are plotted in Figure~\ref{fig.LongTermEvolution} and \ref{fig.LongTermEvolutionCorrelations}. 
The posterior median and 68.3\% credible interval of the mutual inclination over 1~kyr is $20.3 \pm 4.5\degree$, and for the eccentricities it is $e_b = 0.126 \pm 0.076$ and $e_c = 0.217 ^{+0.018}_{-0.027}$.
The mutual inclination remains above $\sim$10\degree\ at the 95.4\% credible interval.
Over the 1~kyr integration, the orbital inclination of planet~c is too low for transits to occur for most of the samples. Interestingly, planet~b only transits for a small fraction of that time (Figure~\ref{fig.LongTermEvolution}); in particular, it will not transit anymore after about 200 years, then it will transit again in about 600 years.

\subsection{Tidal evolution}\label{sec.secular}

\begin{figure}
  \centering
  \includegraphics[width=0.5\textwidth]{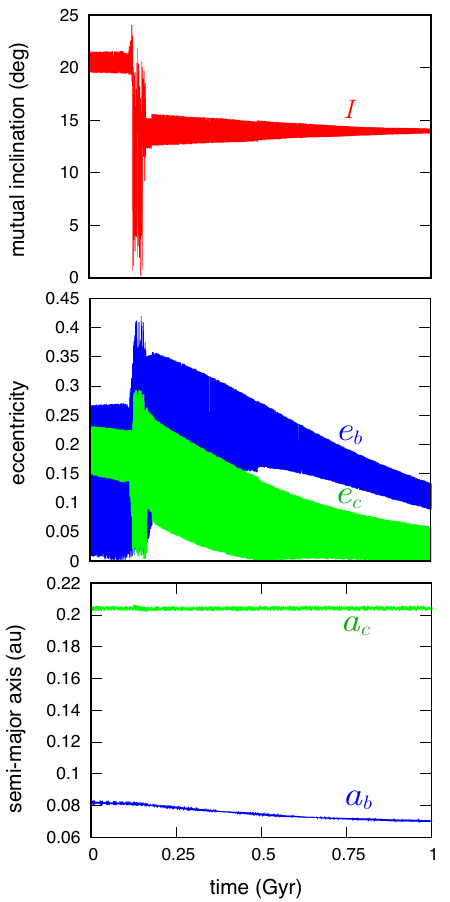}
  \caption{Tidal evolution of the WASP-148 system. Starting with the nominal solution from Table~\ref{table.results}, we show the evolution of the mutual inclination (top), the eccentricities (middle), and semi-major axes (bottom) over 1~Gyr. We used a direct three-body model with linear tides \citep{Correia_2018} with general relativity corrections, and adopted $k_2=0.5$ and $\Delta t = 1$~s for both planets.}\label{tidal_evolution}
\end{figure}

The semi-major axis of the innermost planet is only 0.082~au, which means that the planet is close enough to the star to undergo some tidal evolution.
As a result, the eccentricity can be damped and the inner orbit circularised \citep[e.g.][]{Hut_1981}.
Adopting a value identical to Jupiter's value \citep{Lainey_etal_2009} for the tidal quality factor $Q=10^5$, we get a characteristic timescale $\sim$10~Gyr \citep[e.g.][]{Correia_Laskar_2010B} for the circularisation, which is comparable to the lifetime of the system.
Therefore, it is not surprising that at present the innermost planet still shows a significant eccentricity.
Moreover, the orbits of the two planets strongly interact, and secular or resonant effects can also excite the eccentricity of the innermost planet \citep[e.g.][]{Correia_etal_2012, Correia_etal_2013}.
To check this scenario, using a direct three-body model with linear tides \citep{Correia_2018}, and general relativity corrections, we ran a simulation over 1~Gyr, starting with the initial conditions of the nominal solution from Table~\ref{table.results}. We adopted a Love number of $k_2 = 0.5$ and a time lag of $\Delta t= 1$~s (equivalent to $Q\sim10^5$).
We observed that the eccentricity of the innermost planet undergoes large oscillations owing to secular interactions. Moreover, after 130~Myr, the system crosses a resonance which pumps the inner planet eccentricity and damps the mutual inclination (Figure~\ref{tidal_evolution}). 
As a result, at the end of the simulation, the average inner planet eccentricity is higher than the initial one. 
We conclude that the presently observed non-zero value is compatible with the tidal evolution of the WASP-148 system.
As expected, the semi-major axis of the innermost planet also slightly decreases, and we have $a_b \approx 0.073$~au after 1~Gyr. 
Since the age of the star at present is already 4~Gyr, we can assume that the initial semi-major axis was slightly larger than the present value, and so the two planets could even be trapped in the 4:1 mean motion resonance.
This is an interesting formation scenario that deserves more attention in future work on the system.

\subsection{Three-planet model}\label{sec.3planets}
All these results rely on the three-body system hypothesis. If the dynamics of planet~b is affected by additional planets other than planet~c, increasing the mutual inclination could provide the additional variability in the model required to fit the data. The photodynamical model is flexible and could overfit the data. The argument of an inaccurately specified model is discussed by \citet{petit2020} in the context of the determination of the eccentricity of the K2-19 planets.

To test the influence in the system parameters of an unaccounted for additional planet in the system, we repeated the photodynamical analysis (Section~\ref{section.photodynamical}) with a third planet, which we started at the period of 150~days, which is the peak in the residuals of a 2~Keplerian fit of the radial velocities \citep[Sect.~4.1 and Fig.~3 in][]{hebrard2020}. The results are presented in Table~\ref{table.3planets} and Figures~\ref{fig.irel}, \ref{fig.comparison}, \ref{fig.TTVs2100}, and \ref{fig.TDVs}. The third planet converges to a period of $151.2 ^{+2.1}_{-1.7}$~days and a mass of $0.262 ^{+0.10}_{-0.076}$~\Mjup, which is in agreement with the estimations reported in \citet{hebrard2020}. Its mutual inclination relative to the planets b and c is not well-constrained. The mutual inclination between planet~b and c is reduced from $41.0 ^{+6.2}_{-7.6}$~\degree\ in the two-planet model (without imposing long-term stability) to $21 \pm 12$~\degree\ in the three-planet model. The masses of planets b and c as well as the radius of planet~b are compatible within $1~\sigma$ to the two-planet model.
The transit timing and duration variations do not allow one to distinguish between the two- and three-planet models with the current data (Figures~\ref{fig.TTVs2100} and \ref{fig.TDVs}). More data are needed to assert the presence of additional planets in the system.

\section{Discussion}\label{section.discussion}

The WASP-148 system is composed of a G5V star orbited by a hot Saturn ($0.287 ^{+0.022}_{-0.016}$~\Mjup, $0.756 ^{+0.013}_{-0.017}$~\Rjup, and $0.08215 ^{+0.00086}_{-0.0015}$~au) and a warm Jupiter ($0.392^{+0.023}_{-0.027}$~\Mjup and $0.2044 ^{+0.0021}_{-0.0038}$~au) near a 4:1 mean-motion resonance\footnote{The period ratio at $t_{\mathrm{ref}}$ is $3.92411 ^{+0.00014}_{-0.00017}$ and computed as $(a_c/a_b)^{3/2}$, with $a_c/a_b = 2.487868 ^{+0.000061}_{-0.000073}$. For the stable MAP solution, the period ratio (computed over a 1000~orbits of the planet~c) is 3.922.}. The planets have eccentricities $e_b = 0.208 ^{+0.020}_{-0.025}$ and $e_c = 0.1809 ^{+0.018}_{-0.0072}$, and a mass ratio of $M_c/M_b = 1.351^{+0.11}_{-0.076}$. Also, assuming only two planets in the system, their orbits have a mutual inclination of $20.8 \pm 4.6$~\degree.

\begin{figure}
  \centering
  \includegraphics[width=0.49\textwidth]{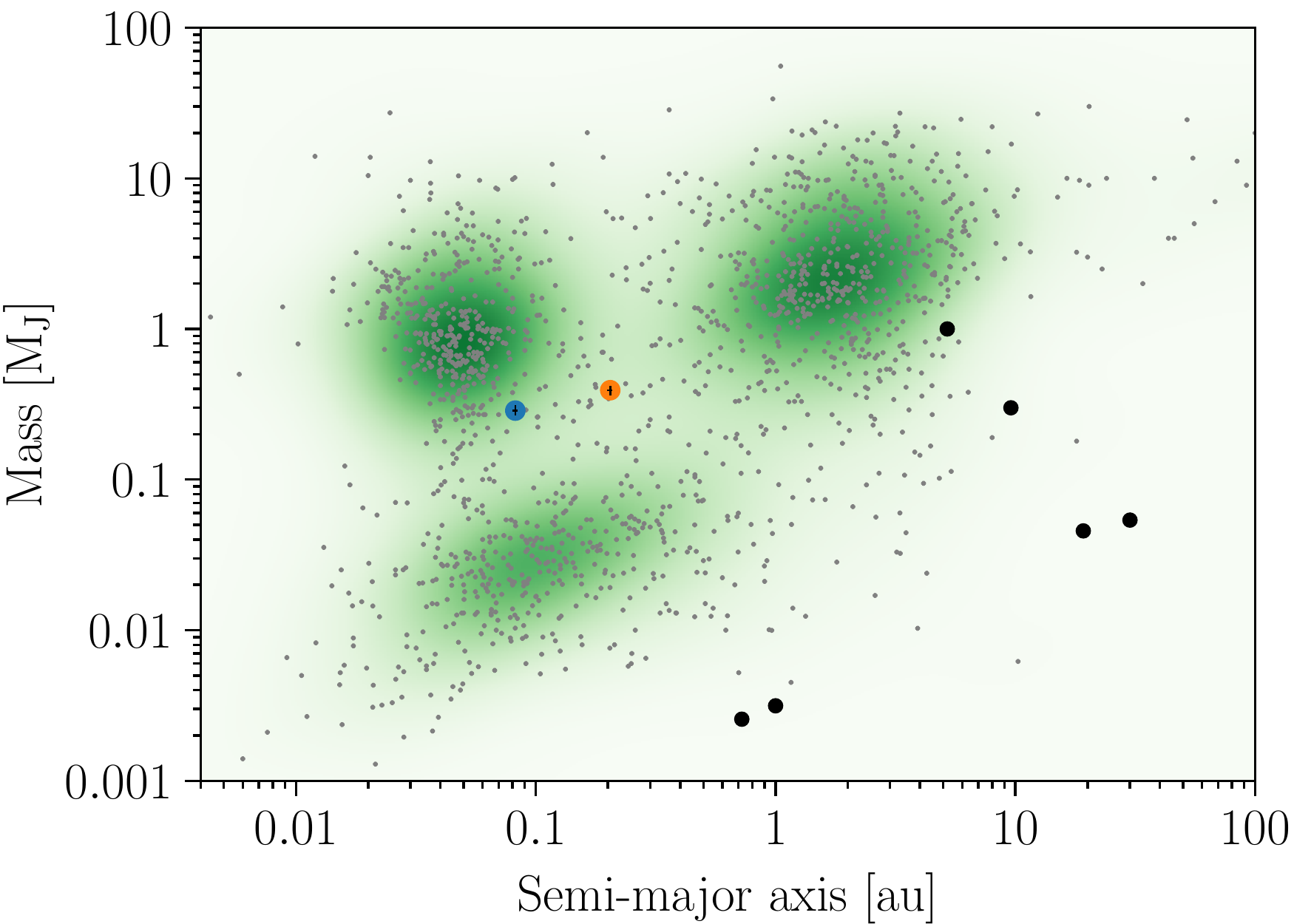}
  \caption{Planet mass versus orbital semi-major axis in logarithmic scale. Error bars mark the position of planet~b (blue) and planet~c (orange). Grey dots are planets listed in the NASA Exoplanet Archive accessed through DACE API (\protect\url{dace.unige.ch}). A Gaussian kernel density estimate is shown in different intensities of green. Solar System planets (black points, from left to right: Venus, Earth, Jupiter, Saturn, Uranus, and Neptune) from NASA.} \label{fig.mass_separation}
\end{figure}

Both planets are located in a scarcely populated part of the mass-separation diagram (Figure~\ref{fig.mass_separation}), which cannot be fully explained by a detection bias\footnote{While the detection of systems similar to WASP-148 is accessible for high-precision stable radial velocities' instruments, they are difficult to detect with transit searches. Due to the mutually inclined orbits and the secular oscillation of the orbital inclination, the probability that both planets transit for a given observer is low (Section~\ref{sec.orbitalEvolution}).}, thus indicating a low probability outcome for the planetary formation process. The scarcity of planets such as WASP-148~b can be partially explained by the low efficiency of the Lidov-Kozai mechanism \citep{lidov1962, kozai1962} to form hot Saturns because most migrating planets are tidally disrupted \citep{anderson2016}. WASP-148~c was placed in the 'period valley' \citep{udry2003} between the hot and warm Jupiters clumps.

The WASP-148 giant planets are expected to form in a proto-planetary disk beyond the snow line, which is located at a few au \citep{lecar2006}, and then migrate inwards \citep{goldreich1980}. Convergent migration leads to planets being captured in mean-motion resonances, but usually of first order. High order resonances such as the 4:1 one require high initial orbital eccentricities \citep{rein2010}. However, disk migration does not favour significant orbital eccentricity excitation because planet-disk interactions tend to damp eccentricities \citep{bitsch2013, dunhill2013}. After the disk disappears, planet-planet scattering can cause high eccentricity migration for the inner planet \citep{rasio1996}, or the eccentricities can be excited by chaotic secular interactions \citep{wu2011}. Both mechanisms are compatible with the non-coplanarity of the observed orbits. 

Another possibility that does not require planet-planet scattering or secular chaos is explored in \citet{lee2002}: the planets are captured in the mean-motion resonance and are massive enough to open gaps in the disk. With the mechanism for damping the eccentricity being reduced, the orbits can become elliptical during the inward migration within the resonance \citep{artymowicz1992}. Capture in the 4:1 mean-motion resonance is then able to produce some inclination excitation given that the inner planet is not too massive \citep{Thommes2003, libert2009}.

Finally, at a later stage, tidal interactions with the star should shrink the inner planet orbit  afterwards and pull the system out of resonance. Tidal interactions usually also damp the eccentricity, but the presence of the outer planet delays this process due to secular and resonant interactions.

There are only a few other planetary systems for which a large mutual inclination has been reported: 102.3$^{+7.4}_{-8.0}$\degree\ in HD\,3167 \citep{dalal2019,bourrier2021}, 30$\pm$1\degree\ in $\upsilon$ Andromedae \citep{mcarthur2010}, 24$^{+11}_{-8}$\degree\ in Kepler-108 \citep{mills2017}, [34.5, 140.6]\degree\ (95\% credible interval) in $\pi$ Mensae \citep{derosa2020}, and some systems with ultra short period planets with mutual inclinations larger than 10\degree\ reported in \citet{dai2018}. HD\,3167 is a system with three small-sized planets, including two transiting that allowed Rossiter-McLaughlin measurements. $\pi$ Mensae is a hierarchical system with a period ratio of $\sim$330, and a large difference in planet masses. $\upsilon$ Andromedae is composed of three planets, and the mutual inclination is measured for the outer pair of super Jovian planets with a period ratio of 5.3. Kepler-108, composed of two Saturn-mass planets with a period ratio of 3.88, is the one that resembles WASP-148 the most, but with longer period planets in lower eccentricity orbits. Interestingly, the period ratio is also close to the 4:1 mean-motion resonance. 

As shown in Section~\ref{sec.3planets}, additional planets not accounted for in the modelling can affect the determination of the system parameters, in particular the mutual inclination. Thus more data are needed to conclude on the true architecture of the system.

\begin{acknowledgements}
  
  This paper includes data collected by the \TESS mission. Funding for the \TESS mission is provided by the NASA Explorer Program.
  
  Resources supporting this work were provided by the NASA High-End Computing (HEC) Program through the NASA Advanced Supercomputing (NAS) Division at Ames Research Center for the production of the SPOC data products.
  
  This work has made use of data from the European Space Agency (ESA) mission
{\it Gaia} (\url{https://www.cosmos.esa.int/gaia}), processed by the {\it Gaia}
Data Processing and Analysis Consortium (DPAC,
\url{https://www.cosmos.esa.int/web/gaia/dpac/consortium}). Funding for the DPAC
has been provided by national institutions, in particular the institutions
participating in the {\it Gaia} Multilateral Agreement.
  
  Simulations in this paper made use of the \reb\ code which can be downloaded freely at \texttt{http://github.com/hannorein/rebound}. 
  
  Part of these simulations have been run on the {\it Lesta} cluster kindly provided by the Observatoire de Gen\`eve. 
  
  This research has made use of the NASA Exoplanet Archive, which is operated by the California Institute of Technology, under contract with the National Aeronautics and Space Administration under the Exoplanet Exploration Program.
  
  AC acknowledges support by CFisUC projects (UIDB/04564/2020 and UIDP/04564/2020), GRAVITY (PTDC/FIS-AST/7002/2020), ENGAGE SKA (POCI-01-0145-FEDER-022217), and PHOBOS (POCI-01-0145-FEDER-029932), funded by COMPETE 2020 and FCT, Portugal. 
  
  GM acknowledges the financial support from the National Science Centre, Poland through grant no. 2016/23/B/ST9/00579. MF acknowledges financial support from grant PID2019-109522GB-C5X/AEI/10.13039/501100011033 of the Spanish Ministry of Science and Innovation (MICINN). MF, VC and JS acknowledge financial support from the State Agency for Research of the Spanish MCIU through the \textit{Center of Excellence Severo Ochoa} award to the Instituto de Astrof\'isica de Andaluc\'ia (SEV-2017-0709).

  JMA and XB acknowledges funding from the European Research Council under the ERC Grant Agreement n. 337591-ExTrA.
  
  We thank L. Kreidberg for her Mandel \& Agol code.  
\end{acknowledgements}

\bibliographystyle{aa}
\bibliography{WASP-148}

\begin{thebibliography}{107}
\expandafter\ifx\csname natexlab\endcsname\relax\def\natexlab#1{#1}\fi

\bibitem[{{Agol} {et~al.}(2005){Agol}, {Steffen}, {Sari}, \&
  {Clarkson}}]{agol2005}
{Agol}, E., {Steffen}, J., {Sari}, R., \& {Clarkson}, W. 2005, \mnras, 359, 567

\bibitem[{{Allard} {et~al.}(2012){Allard}, {Homeier}, \&
  {Freytag}}]{allard2012}
{Allard}, F., {Homeier}, D., \& {Freytag}, B. 2012, Philosophical Transactions
  of the Royal Society of London Series A, 370, 2765

\bibitem[{{Almenara} {et~al.}(2018{\natexlab{a}}){Almenara}, {D{\'\i}az},
  {Dorn}, {Bonfils}, \& {Udry}}]{almenara2018b}
{Almenara}, J.~M., {D{\'\i}az}, R.~F., {Dorn}, C., {Bonfils}, X., \& {Udry}, S.
  2018{\natexlab{a}}, \mnras, 478, 460

\bibitem[{{Almenara} {et~al.}(2018{\natexlab{b}}){Almenara}, {D{\'\i}az},
  {H{\'e}brard}, {Mardling}, {Damiani}, {Santerne}, {Bouchy}, {Barros},
  {Boisse}, {Bonfils}, {Bonomo}, {Courcol}, {Demangeon}, {Deleuil}, {Rey},
  {Udry}, \& {Wilson}}]{almenara2018}
{Almenara}, J.~M., {D{\'\i}az}, R.~F., {H{\'e}brard}, G., {et~al.}
  2018{\natexlab{b}}, \aap, 615, A90

\bibitem[{{Almenara} {et~al.}(2015){Almenara}, {D{\'{\i}}az}, {Mardling},
  {Barros}, {Damiani}, {Bruno}, {Bonfils}, \& {Deleuil}}]{almenara2015}
{Almenara}, J.~M., {D{\'{\i}}az}, R.~F., {Mardling}, R., {et~al.} 2015, \mnras,
  453, 2644

\bibitem[{{Anderson} {et~al.}(2016){Anderson}, {Storch}, \&
  {Lai}}]{anderson2016}
{Anderson}, K.~R., {Storch}, N.~I., \& {Lai}, D. 2016, \mnras, 456, 3671

\bibitem[{{Andrae} {et~al.}(2018){Andrae}, {Fouesneau}, {Creevey}, {Ordenovic},
  {Mary}, {Burlacu}, {Chaoul}, {Jean-Antoine-Piccolo}, {Kordopatis}, {Korn},
  {Lebreton}, {Panem}, {Pichon}, {Th{\'e}venin}, {Walmsley}, \&
  {Bailer-Jones}}]{andrae2018}
{Andrae}, R., {Fouesneau}, M., {Creevey}, O., {et~al.} 2018, \aap, 616, A8

\bibitem[{{Artymowicz}(1992)}]{artymowicz1992}
{Artymowicz}, P. 1992, \pasp, 104, 769

\bibitem[{{Barnes}(2010)}]{barnes2010}
{Barnes}, S.~A. 2010, \apj, 722, 222

\bibitem[{{Barnes} \& {Kim}(2010)}]{barneskim2010}
{Barnes}, S.~A. \& {Kim}, Y.-C. 2010, \apj, 721, 675

\bibitem[{{Barros} {et~al.}(2013){Barros}, {Bou{\'e}}, {Gibson}, {Pollacco},
  {Santerne}, {Keenan}, {Skillen}, \& {Street}}]{barros2013}
{Barros}, S.~C.~C., {Bou{\'e}}, G., {Gibson}, N.~P., {et~al.} 2013, \mnras,
  430, 3032

\bibitem[{Bishop(2007)}]{bishop2007}
Bishop, C.~M. 2007, Pattern Recognition and Machine Learning (Information
  Science and Statistics), 1st edn. (Springer)

\bibitem[{{Bitsch} {et~al.}(2013){Bitsch}, {Crida}, {Libert}, \&
  {Lega}}]{bitsch2013}
{Bitsch}, B., {Crida}, A., {Libert}, A.~S., \& {Lega}, E. 2013, \aap, 555, A124

\bibitem[{{Bourrier} {et~al.}(2021){Bourrier}, {Lovis}, {Cretignier}, {Allart},
  {Dumusque}, {Delisle}, {Deline}, {Sousa}, {Adibekyan}, {Alibert}, {Barros},
  {Borsa}, {Cristiani}, {Demangeon}, {Ehrenreich}, {Figueira}, {Gonz{\'a}lez
  Hern{\'a}ndez}, {Lendl}, {Lillo-Box}, {Lo Curto}, {Di Marcantonio},
  {Martins}, {M{\'e}gevand}, {Mehner}, {Micela}, {Molaro}, {Oshagh}, {Palle},
  {Pepe}, {Poretti}, {Rebolo}, {Santos}, {Scandariato}, {Seidel}, {Sozzetti},
  {Su{\'a}rez Mascare{\~n}o}, \& {Zapatero Osorio}}]{bourrier2021}
{Bourrier}, V., {Lovis}, C., {Cretignier}, M., {et~al.} 2021, \aap, 654, A152

\bibitem[{Byrd {et~al.}(1995)Byrd, Lu, Nocedal, \& Zhu}]{byrd1995}
Byrd, R.~H., Lu, P., Nocedal, J., \& Zhu, C. 1995, SIAM Journal on scientific
  computing, 16, 1190

\bibitem[{{Carter} {et~al.}(2011){Carter}, {Fabrycky}, {Ragozzine}, {Holman},
  {Quinn}, {Latham}, {Buchhave}, {Van Cleve}, {Cochran}, {Cote}, {Endl},
  {Ford}, {Haas}, {Jenkins}, {Koch}, {Li}, {Lissauer}, {MacQueen}, {Middour},
  {Orosz}, {Rowe}, {Steffen}, \& {Welsh}}]{carter2011}
{Carter}, J.~A., {Fabrycky}, D.~C., {Ragozzine}, D., {et~al.} 2011, Science,
  331, 562

\bibitem[{{Castelli} \& {Kurucz}(2003)}]{castelli2003}
{Castelli}, F. \& {Kurucz}, R.~L. 2003, in IAU Symposium, Vol. 210, Modelling
  of Stellar Atmospheres, ed. N.~{Piskunov}, W.~W. {Weiss}, \& D.~F. {Gray},
  A20

\bibitem[{{Chen} {et~al.}(2014){Chen}, {Girardi}, {Bressan}, {Marigo},
  {Barbieri}, \& {Kong}}]{chen2014}
{Chen}, Y., {Girardi}, L., {Bressan}, A., {et~al.} 2014, \mnras, 444, 2525

\bibitem[{{Claret}(2017)}]{claret2017}
{Claret}, A. 2017, \aap, 600, A30

\bibitem[{{Claret} \& {Bloemen}(2011)}]{claret2011}
{Claret}, A. \& {Bloemen}, S. 2011, \aap, 529, A75

\bibitem[{{Collins} {et~al.}(2017){Collins}, {Kielkopf}, {Stassun}, \&
  {Hessman}}]{2017AJ....153...77C}
{Collins}, K.~A., {Kielkopf}, J.~F., {Stassun}, K.~G., \& {Hessman}, F.~V.
  2017, \aj, 153, 77

\bibitem[{{Correia}(2018)}]{Correia_2018}
{Correia}, A.~C.~M. 2018, \icarus, 305, 250

\bibitem[{{Correia} {et~al.}(2012){Correia}, {Bou{\'e}}, \&
  {Laskar}}]{Correia_etal_2012}
{Correia}, A.~C.~M., {Bou{\'e}}, G., \& {Laskar}, J. 2012, \apjl, 744, L23

\bibitem[{{Correia} {et~al.}(2013){Correia}, {Bou{\'e}}, {Laskar}, \&
  {Morais}}]{Correia_etal_2013}
{Correia}, A.~C.~M., {Bou{\'e}}, G., {Laskar}, J., \& {Morais}, M.~H.~M. 2013,
  \aap, 553, A39

\bibitem[{{Correia} {et~al.}(2010){Correia}, {Couetdic}, {Laskar}, {Bonfils},
  {Mayor}, {Bertaux}, {Bouchy}, {Delfosse}, {Forveille}, {Lovis}, {Pepe},
  {Perrier}, {Queloz}, \& {Udry}}]{correia2010}
{Correia}, A.~C.~M., {Couetdic}, J., {Laskar}, J., {et~al.} 2010, \aap, 511,
  A21

\bibitem[{{Correia} \& {Laskar}(2010)}]{Correia_Laskar_2010B}
{Correia}, A.~C.~M. \& {Laskar}, J. 2010, {Tidal Evolution of Exoplanets}, ed.
  S.~Seager (University of Arizona Press, Tucson), 239--266

\bibitem[{{Couetdic} {et~al.}(2010){Couetdic}, {Laskar}, {Correia}, {Mayor}, \&
  {Udry}}]{couetdic2010}
{Couetdic}, J., {Laskar}, J., {Correia}, A.~C.~M., {Mayor}, M., \& {Udry}, S.
  2010, \aap, 519, A10

\bibitem[{{Cutri} \& {et al.}(2013)}]{cutri2013}
{Cutri}, R.~M. \& {et al.} 2013, VizieR Online Data Catalog, II/328

\bibitem[{{Cutri} {et~al.}(2003){Cutri}, {Skrutskie}, {van Dyk}, {Beichman},
  {Carpenter}, {Chester}, {Cambresy}, {Evans}, {Fowler}, {Gizis}, {Howard},
  {Huchra}, {Jarrett}, {Kopan}, {Kirkpatrick}, {Light}, {Marsh}, {McCallon},
  {Schneider}, {Stiening}, {Sykes}, {Weinberg}, {Wheaton}, {Wheelock}, \&
  {Zacarias}}]{cutri2003}
{Cutri}, R.~M., {Skrutskie}, M.~F., {van Dyk}, S., {et~al.} 2003, VizieR Online
  Data Catalog, II/246

\bibitem[{{Dai} {et~al.}(2018){Dai}, {Masuda}, \& {Winn}}]{dai2018}
{Dai}, F., {Masuda}, K., \& {Winn}, J.~N. 2018, \apjl, 864, L38

\bibitem[{{Dalal} {et~al.}(2019){Dalal}, {H{\'e}brard}, {Lecavelier des
  {\'E}tangs}, {Petit}, {Bourrier}, {Laskar}, {K{\"o}nig}, \&
  {Correia}}]{dalal2019}
{Dalal}, S., {H{\'e}brard}, G., {Lecavelier des {\'E}tangs}, A., {et~al.} 2019,
  \aap, 631, A28

\bibitem[{{Dawson} \& {Johnson}(2012)}]{dawson2012}
{Dawson}, R.~I. \& {Johnson}, J.~A. 2012, \apj, 756, 122

\bibitem[{{De Rosa} {et~al.}(2020){De Rosa}, {Dawson}, \&
  {Nielsen}}]{derosa2020}
{De Rosa}, R.~J., {Dawson}, R., \& {Nielsen}, E.~L. 2020, \aap, 640, A73

\bibitem[{{D{\'\i}az} {et~al.}(2014){D{\'\i}az}, {Almenara}, {Santerne},
  {Moutou}, {Lethuillier}, \& {Deleuil}}]{diaz2014}
{D{\'\i}az}, R.~F., {Almenara}, J.~M., {Santerne}, A., {et~al.} 2014, \mnras,
  441, 983

\bibitem[{{D{\'\i}az} {et~al.}(2016){D{\'\i}az}, {S{\'e}gransan}, {Udry},
  {Lovis}, {Pepe}, {Dumusque}, {Marmier}, {Alonso}, {Benz}, {Bouchy},
  {Coffinet}, {Collier Cameron}, {Deleuil}, {Figueira}, {Gillon}, {Lo Curto},
  {Mayor}, {Mordasini}, {Motalebi}, {Moutou}, {Pollacco}, {Pompei}, {Queloz},
  {Santos}, \& {Wyttenbach}}]{diaz2016}
{D{\'\i}az}, R.~F., {S{\'e}gransan}, D., {Udry}, S., {et~al.} 2016, \aap, 585,
  A134

\bibitem[{{Dotter} {et~al.}(2008){Dotter}, {Chaboyer}, {Jevremovi{\'c}},
  {Kostov}, {Baron}, \& {Ferguson}}]{dotter2008}
{Dotter}, A., {Chaboyer}, B., {Jevremovi{\'c}}, D., {et~al.} 2008, \apjs, 178,
  89

\bibitem[{{Dunhill} {et~al.}(2013){Dunhill}, {Alexander}, \&
  {Armitage}}]{dunhill2013}
{Dunhill}, A.~C., {Alexander}, R.~D., \& {Armitage}, P.~J. 2013, \mnras, 428,
  3072

\bibitem[{{Eastman} {et~al.}(2010){Eastman}, {Siverd}, \&
  {Gaudi}}]{eastman2010}
{Eastman}, J., {Siverd}, R., \& {Gaudi}, B.~S. 2010, \pasp, 122, 935

\bibitem[{{Espinoza}(2018)}]{espinoza2018}
{Espinoza}, N. 2018, Research Notes of the American Astronomical Society, 2,
  209

\bibitem[{{Espinoza} {et~al.}(2019){Espinoza}, {Kossakowski}, \&
  {Brahm}}]{espinoza2019}
{Espinoza}, N., {Kossakowski}, D., \& {Brahm}, R. 2019, \mnras, 490, 2262

\bibitem[{{Farr{\'e}s} {et~al.}(2013){Farr{\'e}s}, {Laskar}, {Blanes}, {Casas},
  {Makazaga}, \& {Murua}}]{Farres_etal_2013}
{Farr{\'e}s}, A., {Laskar}, J., {Blanes}, S., {et~al.} 2013, Celestial
  Mechanics and Dynamical Astronomy, 116, 141

\bibitem[{{Feinstein} {et~al.}(2019){Feinstein}, {Montet}, {Foreman-Mackey},
  {Bedell}, {Saunders}, {Bean}, {Christiansen}, {Hedges}, {Luger}, {Scolnic},
  \& {Cardoso}}]{feinstein2019}
{Feinstein}, A.~D., {Montet}, B.~T., {Foreman-Mackey}, D., {et~al.} 2019,
  \pasp, 131, 094502

\bibitem[{{Foreman-Mackey} {et~al.}(2017){Foreman-Mackey}, {Agol},
  {Ambikasaran}, \& {Angus}}]{foreman-mackey2017}
{Foreman-Mackey}, D., {Agol}, E., {Ambikasaran}, S., \& {Angus}, R. 2017, \aj,
  154, 220

\bibitem[{{Foreman-Mackey} {et~al.}(2013){Foreman-Mackey}, {Hogg}, {Lang}, \&
  {Goodman}}]{emcee}
{Foreman-Mackey}, D., {Hogg}, D.~W., {Lang}, D., \& {Goodman}, J. 2013, \pasp,
  125, 306

\bibitem[{{Freudenthal} {et~al.}(2019){Freudenthal}, {von Essen}, {Ofir},
  {Dreizler}, {Agol}, {Wedemeyer}, {Morris}, {Becker}, {Deeg}, {Hoyer},
  {Mallonn}, {Poppenhaeger}, {Herrero}, {Ribas}, {Boumis}, \&
  {Liakos}}]{freudenthal2019}
{Freudenthal}, J., {von Essen}, C., {Ofir}, A., {et~al.} 2019, \aap, 628, A108

\bibitem[{{Fulton} {et~al.}(2018){Fulton}, {Petigura}, {Blunt}, \&
  {Sinukoff}}]{fulton2018}
{Fulton}, B.~J., {Petigura}, E.~A., {Blunt}, S., \& {Sinukoff}, E. 2018, \pasp,
  130, 044504

\bibitem[{{Fulton} {et~al.}(2011){Fulton}, {Shporer}, {Winn}, {Holman},
  {P{\'a}l}, \& {Gazak}}]{2011AJ....142...84F}
{Fulton}, B.~J., {Shporer}, A., {Winn}, J.~N., {et~al.} 2011, \aj, 142, 84

\bibitem[{{Gaia Collaboration} {et~al.}(2018){Gaia Collaboration}, {Brown},
  {Vallenari}, {Prusti}, {de Bruijne}, {Babusiaux}, {Bailer-Jones}, {Biermann},
  {Evans}, {Eyer}, {Jansen}, {Jordi}, {Klioner}, {Lammers}, {Lindegren},
  {Luri}, {Mignard}, {Panem}, {Pourbaix}, {Randich}, {Sartoretti}, {Siddiqui},
  {Soubiran}, {van Leeuwen}, {Walton}, {Arenou}, {Bastian}, {Cropper},
  {Drimmel}, {Katz}, {Lattanzi}, {Bakker}, {Cacciari}, {Casta{\~n}eda},
  {Chaoul}, {Cheek}, {De Angeli}, {Fabricius}, {Guerra}, {Holl}, {Masana},
  {Messineo}, {Mowlavi}, {Nienartowicz}, {Panuzzo}, {Portell}, {Riello},
  {Seabroke}, {Tanga}, {Th{\'e}venin}, {Gracia-Abril}, {Comoretto},
  {Garcia-Reinaldos}, {Teyssier}, {Altmann}, {Andrae}, {Audard},
  {Bellas-Velidis}, {Benson}, {Berthier}, {Blomme}, {Burgess}, {Busso},
  {Carry}, {Cellino}, {Clementini}, {Clotet}, {Creevey}, {Davidson}, {De
  Ridder}, {Delchambre}, {Dell'Oro}, {Ducourant},
  {Fern{\'a}ndez-Hern{\'a}ndez}, {Fouesneau}, {Fr{\'e}mat}, {Galluccio},
  {Garc{\'\i}a-Torres}, {Gonz{\'a}lez-N{\'u}{\~n}ez}, {Gonz{\'a}lez-Vidal},
  {Gosset}, {Guy}, {Halbwachs}, {Hambly}, {Harrison}, {Hern{\'a}ndez},
  {Hestroffer}, {Hodgkin}, {Hutton}, {Jasniewicz}, {Jean-Antoine-Piccolo},
  {Jordan}, {Korn}, {Krone-Martins}, {Lanzafame}, {Lebzelter}, {L{\"o}ffler},
  {Manteiga}, {Marrese}, {Mart{\'\i}n-Fleitas}, {Moitinho}, {Mora}, {Muinonen},
  {Osinde}, {Pancino}, {Pauwels}, {Petit}, {Recio-Blanco}, {Richards},
  {Rimoldini}, {Robin}, {Sarro}, {Siopis}, {Smith}, {Sozzetti}, {S{\"u}veges},
  {Torra}, {van Reeven}, {Abbas}, {Abreu Aramburu}, {Accart}, {Aerts},
  {Altavilla}, {{\'A}lvarez}, {Alvarez}, {Alves}, {Anderson}, {Andrei},
  {Anglada Varela}, {Antiche}, {Antoja}, {Arcay}, {Astraatmadja}, {Bach},
  {Baker}, {Balaguer-N{\'u}{\~n}ez}, {Balm}, {Barache}, {Barata}, {Barbato},
  {Barblan}, {Barklem}, {Barrado}, {Barros}, {Barstow}, {Bartholom{\'e}
  Mu{\~n}oz}, {Bassilana}, {Becciani}, {Bellazzini}, {Berihuete}, {Bertone},
  {Bianchi}, {Bienaym{\'e}}, {Blanco-Cuaresma}, {Boch}, {Boeche}, {Bombrun},
  {Borrachero}, {Bossini}, {Bouquillon}, {Bourda}, {Bragaglia}, {Bramante},
  {Breddels}, {Bressan}, {Brouillet}, {Br{\"u}semeister}, {Brugaletta},
  {Bucciarelli}, {Burlacu}, {Busonero}, {Butkevich}, {Buzzi}, {Caffau},
  {Cancelliere}, {Cannizzaro}, {Cantat-Gaudin}, {Carballo}, {Carlucci},
  {Carrasco}, {Casamiquela}, {Castellani}, {Castro-Ginard}, {Charlot},
  {Chemin}, {Chiavassa}, {Cocozza}, {Costigan}, {Cowell}, {Crifo}, {Crosta},
  {Crowley}, {Cuypers}, {Dafonte}, {Damerdji}, {Dapergolas}, {David}, {David},
  {de Laverny}, {De Luise}, {De March}, {de Martino}, {de Souza}, {de Torres},
  {Debosscher}, {del Pozo}, {Delbo}, {Delgado}, {Delgado}, {Di Matteo},
  {Diakite}, {Diener}, {Distefano}, {Dolding}, {Drazinos}, {Dur{\'a}n},
  {Edvardsson}, {Enke}, {Eriksson}, {Esquej}, {Eynard Bontemps}, {Fabre},
  {Fabrizio}, {Faigler}, {Falc{\~a}o}, {Farr{\`a}s Casas}, {Federici},
  {Fedorets}, {Fernique}, {Figueras}, {Filippi}, {Findeisen}, {Fonti},
  {Fraile}, {Fraser}, {Fr{\'e}zouls}, {Gai}, {Galleti}, {Garabato},
  {Garc{\'\i}a-Sedano}, {Garofalo}, {Garralda}, {Gavel}, {Gavras}, {Gerssen},
  {Geyer}, {Giacobbe}, {Gilmore}, {Girona}, {Giuffrida}, {Glass}, {Gomes},
  {Granvik}, {Gueguen}, {Guerrier}, {Guiraud}, {Guti{\'e}rrez-S{\'a}nchez},
  {Haigron}, {Hatzidimitriou}, {Hauser}, {Haywood}, {Heiter}, {Helmi}, {Heu},
  {Hilger}, {Hobbs}, {Hofmann}, {Holland}, {Huckle}, {Hypki}, {Icardi},
  {Jan{\ss}en}, {Jevardat de Fombelle}, {Jonker}, {Juh{\'a}sz}, {Julbe},
  {Karampelas}, {Kewley}, {Klar}, {Kochoska}, {Kohley}, {Kolenberg},
  {Kontizas}, {Kontizas}, {Koposov}, {Kordopatis}, {Kostrzewa-Rutkowska},
  {Koubsky}, {Lambert}, {Lanza}, {Lasne}, {Lavigne}, {Le Fustec}, {Le
  Poncin-Lafitte}, {Lebreton}, {Leccia}, {Leclerc}, {Lecoeur-Taibi},
  {Lenhardt}, {Leroux}, {Liao}, {Licata}, {Lindstr{\o}m}, {Lister}, {Livanou},
  {Lobel}, {L{\'o}pez}, {Managau}, {Mann}, {Mantelet}, {Marchal}, {Marchant},
  {Marconi}, {Marinoni}, {Marschalk{\'o}}, {Marshall}, {Martino}, {Marton},
  {Mary}, {Massari}, {Matijevi{\v{c}}}, {Mazeh}, {McMillan}, {Messina},
  {Michalik}, {Millar}, {Molina}, {Molinaro}, {Moln{\'a}r}, {Montegriffo},
  {Mor}, {Morbidelli}, {Morel}, {Morris}, {Mulone}, {Muraveva}, {Musella},
  {Nelemans}, {Nicastro}, {Noval}, {O'Mullane}, {Ord{\'e}novic},
  {Ord{\'o}{\~n}ez-Blanco}, {Osborne}, {Pagani}, {Pagano}, {Pailler},
  {Palacin}, {Palaversa}, {Panahi}, {Pawlak}, {Piersimoni}, {Pineau}, {Plachy},
  {Plum}, {Poggio}, {Poujoulet}, {Pr{\v{s}}a}, {Pulone}, {Racero}, {Ragaini},
  {Rambaux}, {Ramos-Lerate}, {Regibo}, {Reyl{\'e}}, {Riclet}, {Ripepi}, {Riva},
  {Rivard}, {Rixon}, {Roegiers}, {Roelens}, {Romero-G{\'o}mez}, {Rowell},
  {Royer}, {Ruiz-Dern}, {Sadowski}, {Sagrist{\`a} Sell{\'e}s}, {Sahlmann},
  {Salgado}, {Salguero}, {Sanna}, {Santana-Ros}, {Sarasso}, {Savietto},
  {Schultheis}, {Sciacca}, {Segol}, {Segovia}, {S{\'e}gransan}, {Shih},
  {Siltala}, {Silva}, {Smart}, {Smith}, {Solano}, {Solitro}, {Sordo}, {Soria
  Nieto}, {Souchay}, {Spagna}, {Spoto}, {Stampa}, {Steele},
  {Steidelm{\"u}ller}, {Stephenson}, {Stoev}, {Suess}, {Surdej}, {Szabados},
  {Szegedi-Elek}, {Tapiador}, {Taris}, {Tauran}, {Taylor}, {Teixeira},
  {Terrett}, {Teyssand ier}, {Thuillot}, {Titarenko}, {Torra Clotet}, {Turon},
  {Ulla}, {Utrilla}, {Uzzi}, {Vaillant}, {Valentini}, {Valette}, {van Elteren},
  {Van Hemelryck}, {van Leeuwen}, {Vaschetto}, {Vecchiato}, {Veljanoski},
  {Viala}, {Vicente}, {Vogt}, {von Essen}, {Voss}, {Votruba}, {Voutsinas},
  {Walmsley}, {Weiler}, {Wertz}, {Wevers}, {Wyrzykowski}, {Yoldas},
  {{\v{Z}}erjal}, {Ziaeepour}, {Zorec}, {Zschocke}, {Zucker}, {Zurbach}, \&
  {Zwitter}}]{gaia2018}
{Gaia Collaboration}, {Brown}, A.~G.~A., {Vallenari}, A., {et~al.} 2018, \aap,
  616, A1

\bibitem[{{Gaia Collaboration} {et~al.}(2021){Gaia Collaboration}, {Brown},
  {Vallenari}, {Prusti}, {de Bruijne}, {Babusiaux}, {Biermann}, {Creevey},
  {Evans}, {Eyer}, {Hutton}, {Jansen}, {Jordi}, {Klioner}, {Lammers},
  {Lindegren}, {Luri}, {Mignard}, {Panem}, {Pourbaix}, {Randich}, {Sartoretti},
  {Soubiran}, {Walton}, {Arenou}, {Bailer-Jones}, {Bastian}, {Cropper},
  {Drimmel}, {Katz}, {Lattanzi}, {van Leeuwen}, {Bakker}, {Cacciari},
  {Casta{\~n}eda}, {De Angeli}, {Ducourant}, {Fabricius}, {Fouesneau},
  {Fr{\'e}mat}, {Guerra}, {Guerrier}, {Guiraud}, {Jean-Antoine Piccolo},
  {Masana}, {Messineo}, {Mowlavi}, {Nicolas}, {Nienartowicz}, {Pailler},
  {Panuzzo}, {Riclet}, {Roux}, {Seabroke}, {Sordo}, {Tanga}, {Th{\'e}venin},
  {Gracia-Abril}, {Portell}, {Teyssier}, {Altmann}, {Andrae}, {Bellas-Velidis},
  {Benson}, {Berthier}, {Blomme}, {Brugaletta}, {Burgess}, {Busso}, {Carry},
  {Cellino}, {Cheek}, {Clementini}, {Damerdji}, {Davidson}, {Delchambre},
  {Dell'Oro}, {Fern{\'a}ndez-Hern{\'a}ndez}, {Galluccio}, {Garc{\'\i}a-Lario},
  {Garcia-Reinaldos}, {Gonz{\'a}lez-N{\'u}{\~n}ez}, {Gosset}, {Haigron},
  {Halbwachs}, {Hambly}, {Harrison}, {Hatzidimitriou}, {Heiter},
  {Hern{\'a}ndez}, {Hestroffer}, {Hodgkin}, {Holl}, {Jan{\ss}en}, {Jevardat de
  Fombelle}, {Jordan}, {Krone-Martins}, {Lanzafame}, {L{\"o}ffler}, {Lorca},
  {Manteiga}, {Marchal}, {Marrese}, {Moitinho}, {Mora}, {Muinonen}, {Osborne},
  {Pancino}, {Pauwels}, {Petit}, {Recio-Blanco}, {Richards}, {Riello},
  {Rimoldini}, {Robin}, {Roegiers}, {Rybizki}, {Sarro}, {Siopis}, {Smith},
  {Sozzetti}, {Ulla}, {Utrilla}, {van Leeuwen}, {van Reeven}, {Abbas}, {Abreu
  Aramburu}, {Accart}, {Aerts}, {Aguado}, {Ajaj}, {Altavilla}, {{\'A}lvarez},
  {{\'A}lvarez Cid-Fuentes}, {Alves}, {Anderson}, {Anglada Varela}, {Antoja},
  {Audard}, {Baines}, {Baker}, {Balaguer-N{\'u}{\~n}ez}, {Balbinot}, {Balog},
  {Barache}, {Barbato}, {Barros}, {Barstow}, {Bartolom{\'e}}, {Bassilana},
  {Bauchet}, {Baudesson-Stella}, {Becciani}, {Bellazzini}, {Bernet}, {Bertone},
  {Bianchi}, {Blanco-Cuaresma}, {Boch}, {Bombrun}, {Bossini}, {Bouquillon},
  {Bragaglia}, {Bramante}, {Breedt}, {Bressan}, {Brouillet}, {Bucciarelli},
  {Burlacu}, {Busonero}, {Butkevich}, {Buzzi}, {Caffau}, {Cancelliere},
  {C{\'a}novas}, {Cantat-Gaudin}, {Carballo}, {Carlucci}, {Carnerero},
  {Carrasco}, {Casamiquela}, {Castellani}, {Castro-Ginard}, {Castro Sampol},
  {Chaoul}, {Charlot}, {Chemin}, {Chiavassa}, {Cioni}, {Comoretto}, {Cooper},
  {Cornez}, {Cowell}, {Crifo}, {Crosta}, {Crowley}, {Dafonte}, {Dapergolas},
  {David}, {David}, {de Laverny}, {De Luise}, {De March}, {De Ridder}, {de
  Souza}, {de Teodoro}, {de Torres}, {del Peloso}, {del Pozo}, {Delbo},
  {Delgado}, {Delgado}, {Delisle}, {Di Matteo}, {Diakite}, {Diener},
  {Distefano}, {Dolding}, {Eappachen}, {Edvardsson}, {Enke}, {Esquej}, {Fabre},
  {Fabrizio}, {Faigler}, {Fedorets}, {Fernique}, {Fienga}, {Figueras},
  {Fouron}, {Fragkoudi}, {Fraile}, {Franke}, {Gai}, {Garabato},
  {Garcia-Gutierrez}, {Garc{\'\i}a-Torres}, {Garofalo}, {Gavras}, {Gerlach},
  {Geyer}, {Giacobbe}, {Gilmore}, {Girona}, {Giuffrida}, {Gomel}, {Gomez},
  {Gonzalez-Santamaria}, {Gonz{\'a}lez-Vidal}, {Granvik},
  {Guti{\'e}rrez-S{\'a}nchez}, {Guy}, {Hauser}, {Haywood}, {Helmi}, {Hidalgo},
  {Hilger}, {H{\l}adczuk}, {Hobbs}, {Holland}, {Huckle}, {Jasniewicz},
  {Jonker}, {Juaristi Campillo}, {Julbe}, {Karbevska}, {Kervella}, {Khanna},
  {Kochoska}, {Kontizas}, {Kordopatis}, {Korn}, {Kostrzewa-Rutkowska},
  {Kruszy{\'n}ska}, {Lambert}, {Lanza}, {Lasne}, {Le Campion}, {Le Fustec},
  {Lebreton}, {Lebzelter}, {Leccia}, {Leclerc}, {Lecoeur-Taibi}, {Liao},
  {Licata}, {Lindstr{\o}m}, {Lister}, {Livanou}, {Lobel}, {Madrero Pardo},
  {Managau}, {Mann}, {Marchant}, {Marconi}, {Marcos Santos}, {Marinoni},
  {Marocco}, {Marshall}, {Martin Polo}, {Mart{\'\i}n-Fleitas}, {Masip},
  {Massari}, {Mastrobuono-Battisti}, {Mazeh}, {McMillan}, {Messina},
  {Michalik}, {Millar}, {Mints}, {Molina}, {Molinaro}, {Moln{\'a}r},
  {Montegriffo}, {Mor}, {Morbidelli}, {Morel}, {Morris}, {Mulone}, {Munoz},
  {Muraveva}, {Murphy}, {Musella}, {Noval}, {Ord{\'e}novic}, {Orr{\`u}},
  {Osinde}, {Pagani}, {Pagano}, {Palaversa}, {Palicio}, {Panahi}, {Pawlak},
  {Pe{\~n}alosa Esteller}, {Penttil{\"a}}, {Piersimoni}, {Pineau}, {Plachy},
  {Plum}, {Poggio}, {Poretti}, {Poujoulet}, {Pr{\v{s}}a}, {Pulone}, {Racero},
  {Ragaini}, {Rainer}, {Raiteri}, {Rambaux}, {Ramos}, {Ramos-Lerate}, {Re
  Fiorentin}, {Regibo}, {Reyl{\'e}}, {Ripepi}, {Riva}, {Rixon}, {Robichon},
  {Robin}, {Roelens}, {Rohrbasser}, {Romero-G{\'o}mez}, {Rowell}, {Royer},
  {Rybicki}, {Sadowski}, {Sagrist{\`a} Sell{\'e}s}, {Sahlmann}, {Salgado},
  {Salguero}, {Samaras}, {Sanchez Gimenez}, {Sanna}, {Santove{\~n}a},
  {Sarasso}, {Schultheis}, {Sciacca}, {Segol}, {Segovia}, {S{\'e}gransan},
  {Semeux}, {Shahaf}, {Siddiqui}, {Siebert}, {Siltala}, {Slezak}, {Smart},
  {Solano}, {Solitro}, {Souami}, {Souchay}, {Spagna}, {Spoto}, {Steele},
  {Steidelm{\"u}ller}, {Stephenson}, {S{\"u}veges}, {Szabados}, {Szegedi-Elek},
  {Taris}, {Tauran}, {Taylor}, {Teixeira}, {Thuillot}, {Tonello}, {Torra},
  {Torra}, {Turon}, {Unger}, {Vaillant}, {van Dillen}, {Vanel}, {Vecchiato},
  {Viala}, {Vicente}, {Voutsinas}, {Weiler}, {Wevers}, {Wyrzykowski}, {Yoldas},
  {Yvard}, {Zhao}, {Zorec}, {Zucker}, {Zurbach}, \& {Zwitter}}]{gaiaEDR3}
{Gaia Collaboration}, {Brown}, A.~G.~A., {Vallenari}, A., {et~al.} 2021, \aap,
  649, A1

\bibitem[{{Gaia Collaboration} {et~al.}(2016){Gaia Collaboration}, {Prusti},
  {de Bruijne}, {Brown}, {Vallenari}, {Babusiaux}, {Bailer-Jones}, {Bastian},
  {Biermann}, {Evans}, {Eyer}, {Jansen}, {Jordi}, {Klioner}, {Lammers},
  {Lindegren}, {Luri}, {Mignard}, {Milligan}, {Panem}, {Poinsignon},
  {Pourbaix}, {Randich}, {Sarri}, {Sartoretti}, {Siddiqui}, {Soubiran},
  {Valette}, {van Leeuwen}, {Walton}, {Aerts}, {Arenou}, {Cropper}, {Drimmel},
  {H{\o}g}, {Katz}, {Lattanzi}, {O'Mullane}, {Grebel}, {Holland}, {Huc},
  {Passot}, {Bramante}, {Cacciari}, {Casta{\~n}eda}, {Chaoul}, {Cheek}, {De
  Angeli}, {Fabricius}, {Guerra}, {Hern{\'a}ndez}, {Jean-Antoine-Piccolo},
  {Masana}, {Messineo}, {Mowlavi}, {Nienartowicz}, {Ord{\'o}{\~n}ez-Blanco},
  {Panuzzo}, {Portell}, {Richards}, {Riello}, {Seabroke}, {Tanga},
  {Th{\'e}venin}, {Torra}, {Els}, {Gracia-Abril}, {Comoretto},
  {Garcia-Reinaldos}, {Lock}, {Mercier}, {Altmann}, {Andrae}, {Astraatmadja},
  {Bellas-Velidis}, {Benson}, {Berthier}, {Blomme}, {Busso}, {Carry},
  {Cellino}, {Clementini}, {Cowell}, {Creevey}, {Cuypers}, {Davidson}, {De
  Ridder}, {de Torres}, {Delchambre}, {Dell'Oro}, {Ducourant}, {Fr{\'e}mat},
  {Garc{\'\i}a-Torres}, {Gosset}, {Halbwachs}, {Hambly}, {Harrison}, {Hauser},
  {Hestroffer}, {Hodgkin}, {Huckle}, {Hutton}, {Jasniewicz}, {Jordan},
  {Kontizas}, {Korn}, {Lanzafame}, {Manteiga}, {Moitinho}, {Muinonen},
  {Osinde}, {Pancino}, {Pauwels}, {Petit}, {Recio-Blanco}, {Robin}, {Sarro},
  {Siopis}, {Smith}, {Smith}, {Sozzetti}, {Thuillot}, {van Reeven}, {Viala},
  {Abbas}, {Abreu Aramburu}, {Accart}, {Aguado}, {Allan}, {Allasia},
  {Altavilla}, {{\'A}lvarez}, {Alves}, {Anderson}, {Andrei}, {Anglada Varela},
  {Antiche}, {Antoja}, {Ant{\'o}n}, {Arcay}, {Atzei}, {Ayache}, {Bach},
  {Baker}, {Balaguer-N{\'u}{\~n}ez}, {Barache}, {Barata}, {Barbier}, {Barblan},
  {Baroni}, {Barrado y Navascu{\'e}s}, {Barros}, {Barstow}, {Becciani},
  {Bellazzini}, {Bellei}, {Bello Garc{\'\i}a}, {Belokurov}, {Bendjoya},
  {Berihuete}, {Bianchi}, {Bienaym{\'e}}, {Billebaud}, {Blagorodnova},
  {Blanco-Cuaresma}, {Boch}, {Bombrun}, {Borrachero}, {Bouquillon}, {Bourda},
  {Bouy}, {Bragaglia}, {Breddels}, {Brouillet}, {Br{\"u}semeister},
  {Bucciarelli}, {Budnik}, {Burgess}, {Burgon}, {Burlacu}, {Busonero}, {Buzzi},
  {Caffau}, {Cambras}, {Campbell}, {Cancelliere}, {Cantat-Gaudin}, {Carlucci},
  {Carrasco}, {Castellani}, {Charlot}, {Charnas}, {Charvet}, {Chassat},
  {Chiavassa}, {Clotet}, {Cocozza}, {Collins}, {Collins}, {Costigan}, {Crifo},
  {Cross}, {Crosta}, {Crowley}, {Dafonte}, {Damerdji}, {Dapergolas}, {David},
  {David}, {De Cat}, {de Felice}, {de Laverny}, {De Luise}, {De March}, {de
  Martino}, {de Souza}, {Debosscher}, {del Pozo}, {Delbo}, {Delgado},
  {Delgado}, {di Marco}, {Di Matteo}, {Diakite}, {Distefano}, {Dolding}, {Dos
  Anjos}, {Drazinos}, {Dur{\'a}n}, {Dzigan}, {Ecale}, {Edvardsson}, {Enke},
  {Erdmann}, {Escolar}, {Espina}, {Evans}, {Eynard Bontemps}, {Fabre},
  {Fabrizio}, {Faigler}, {Falc{\~a}o}, {Farr{\`a}s Casas}, {Faye}, {Federici},
  {Fedorets}, {Fern{\'a}ndez-Hern{\'a}ndez}, {Fernique}, {Fienga}, {Figueras},
  {Filippi}, {Findeisen}, {Fonti}, {Fouesneau}, {Fraile}, {Fraser}, {Fuchs},
  {Furnell}, {Gai}, {Galleti}, {Galluccio}, {Garabato}, {Garc{\'\i}a-Sedano},
  {Gar{\'e}}, {Garofalo}, {Garralda}, {Gavras}, {Gerssen}, {Geyer}, {Gilmore},
  {Girona}, {Giuffrida}, {Gomes}, {Gonz{\'a}lez-Marcos},
  {Gonz{\'a}lez-N{\'u}{\~n}ez}, {Gonz{\'a}lez-Vidal}, {Granvik}, {Guerrier},
  {Guillout}, {Guiraud}, {G{\'u}rpide}, {Guti{\'e}rrez-S{\'a}nchez}, {Guy},
  {Haigron}, {Hatzidimitriou}, {Haywood}, {Heiter}, {Helmi}, {Hobbs},
  {Hofmann}, {Holl}, {Holland }, {Hunt}, {Hypki}, {Icardi}, {Irwin}, {Jevardat
  de Fombelle}, {Jofr{\'e}}, {Jonker}, {Jorissen}, {Julbe}, {Karampelas},
  {Kochoska}, {Kohley}, {Kolenberg}, {Kontizas}, {Koposov}, {Kordopatis},
  {Koubsky}, {Kowalczyk}, {Krone-Martins}, {Kudryashova}, {Kull}, {Bachchan},
  {Lacoste-Seris}, {Lanza}, {Lavigne}, {Le Poncin-Lafitte}, {Lebreton},
  {Lebzelter}, {Leccia}, {Leclerc}, {Lecoeur-Taibi}, {Lemaitre}, {Lenhardt},
  {Leroux}, {Liao}, {Licata}, {Lindstr{\o}m}, {Lister}, {Livanou}, {Lobel},
  {L{\"o}ffler}, {L{\'o}pez}, {Lopez-Lozano}, {Lorenz}, {Loureiro},
  {MacDonald}, {Magalh{\~a}es Fernandes}, {Managau}, {Mann}, {Mantelet},
  {Marchal}, {Marchant}, {Marconi}, {Marie}, {Marinoni}, {Marrese},
  {Marschalk{\'o}}, {Marshall}, {Mart{\'\i}n-Fleitas}, {Martino}, {Mary},
  {Matijevi{\v{c}}}, {Mazeh}, {McMillan}, {Messina}, {Mestre}, {Michalik},
  {Millar}, {Miranda}, {Molina}, {Molinaro}, {Molinaro}, {Moln{\'a}r},
  {Moniez}, {Montegriffo}, {Monteiro}, {Mor}, {Mora}, {Morbidelli}, {Morel},
  {Morgenthaler}, {Morley}, {Morris}, {Mulone}, {Muraveva}, {Musella},
  {Narbonne}, {Nelemans}, {Nicastro}, {Noval}, {Ord{\'e}novic},
  {Ordieres-Mer{\'e}}, {Osborne}, {Pagani}, {Pagano}, {Pailler}, {Palacin},
  {Palaversa}, {Parsons}, {Paulsen}, {Pecoraro}, {Pedrosa}, {Pentik{\"a}inen},
  {Pereira}, {Pichon}, {Piersimoni}, {Pineau}, {Plachy}, {Plum}, {Poujoulet},
  {Pr{\v{s}}a}, {Pulone}, {Ragaini}, {Rago}, {Rambaux}, {Ramos-Lerate},
  {Ranalli}, {Rauw}, {Read}, {Regibo}, {Renk}, {Reyl{\'e}}, {Ribeiro},
  {Rimoldini}, {Ripepi}, {Riva}, {Rixon}, {Roelens}, {Romero-G{\'o}mez},
  {Rowell}, {Royer}, {Rudolph}, {Ruiz-Dern}, {Sadowski}, {Sagrist{\`a}
  Sell{\'e}s}, {Sahlmann}, {Salgado}, {Salguero}, {Sarasso}, {Savietto},
  {Schnorhk}, {Schultheis}, {Sciacca}, {Segol}, {Segovia}, {Segransan},
  {Serpell}, {Shih}, {Smareglia}, {Smart}, {Smith}, {Solano}, {Solitro},
  {Sordo}, {Soria Nieto}, {Souchay}, {Spagna}, {Spoto}, {Stampa}, {Steele},
  {Steidelm{\"u}ller}, {Stephenson}, {Stoev}, {Suess}, {S{\"u}veges}, {Surdej},
  {Szabados}, {Szegedi-Elek}, {Tapiador}, {Taris}, {Tauran}, {Taylor},
  {Teixeira}, {Terrett}, {Tingley}, {Trager}, {Turon}, {Ulla}, {Utrilla},
  {Valentini}, {van Elteren}, {Van Hemelryck}, {van Leeuwen}, {Varadi},
  {Vecchiato}, {Veljanoski}, {Via}, {Vicente}, {Vogt}, {Voss}, {Votruba},
  {Voutsinas}, {Walmsley}, {Weiler}, {Weingrill}, {Werner}, {Wevers},
  {Whitehead}, {Wyrzykowski}, {Yoldas}, {{\v{Z}}erjal}, {Zucker}, {Zurbach},
  {Zwitter}, {Alecu}, {Allen}, {Allende Prieto}, {Amorim},
  {Anglada-Escud{\'e}}, {Arsenijevic}, {Azaz}, {Balm}, {Beck}, {Bernstein},
  {Bigot}, {Bijaoui}, {Blasco}, {Bonfigli}, {Bono}, {Boudreault}, {Bressan},
  {Brown}, {Brunet}, {Bunclark}, {Buonanno}, {Butkevich}, {Carret}, {Carrion},
  {Chemin}, {Ch{\'e}reau}, {Corcione}, {Darmigny}, {de Boer}, {de Teodoro}, {de
  Zeeuw}, {Delle Luche}, {Domingues}, {Dubath}, {Fodor}, {Fr{\'e}zouls},
  {Fries}, {Fustes}, {Fyfe}, {Gallardo}, {Gallegos}, {Gardiol}, {Gebran},
  {Gomboc}, {G{\'o}mez}, {Grux}, {Gueguen}, {Heyrovsky}, {Hoar}, {Iannicola},
  {Isasi Parache}, {Janotto}, {Joliet}, {Jonckheere}, {Keil}, {Kim},
  {Klagyivik}, {Klar}, {Knude}, {Kochukhov}, {Kolka}, {Kos}, {Kutka}, {Lainey},
  {LeBouquin}, {Liu}, {Loreggia}, {Makarov}, {Marseille}, {Martayan},
  {Martinez-Rubi}, {Massart}, {Meynadier}, {Mignot}, {Munari}, {Nguyen},
  {Nordlander}, {Ocvirk}, {O'Flaherty}, {Olias Sanz}, {Ortiz}, {Osorio},
  {Oszkiewicz}, {Ouzounis}, {Palmer}, {Park}, {Pasquato}, {Peltzer}, {Peralta},
  {P{\'e}turaud}, {Pieniluoma}, {Pigozzi}, {Poels}, {Prat}, {Prod'homme},
  {Raison}, {Rebordao}, {Risquez}, {Rocca-Volmerange}, {Rosen}, {Ruiz-Fuertes},
  {Russo}, {Sembay}, {Serraller Vizcaino}, {Short}, {Siebert}, {Silva},
  {Sinachopoulos}, {Slezak}, {Soffel}, {Sosnowska}, {Strai{\v{z}}ys}, {ter
  Linden}, {Terrell}, {Theil}, {Tiede}, {Troisi}, {Tsalmantza}, {Tur},
  {Vaccari}, {Vachier}, {Valles}, {Van Hamme}, {Veltz}, {Virtanen}, {Wallut},
  {Wichmann}, {Wilkinson}, {Ziaeepour}, \& {Zschocke}}]{gaia2016}
{Gaia Collaboration}, {Prusti}, T., {de Bruijne}, J.~H.~J., {et~al.} 2016,
  \aap, 595, A1

\bibitem[{{Gillon} {et~al.}(2017){Gillon}, {Triaud}, {Demory}, {Jehin}, {Agol},
  {Deck}, {Lederer}, {de Wit}, {Burdanov}, {Ingalls}, {Bolmont}, {Leconte},
  {Raymond}, {Selsis}, {Turbet}, {Barkaoui}, {Burgasser}, {Burleigh}, {Carey},
  {Chaushev}, {Copperwheat}, {Delrez}, {Fernand es}, {Holdsworth}, {Kotze},
  {Van Grootel}, {Almleaky}, {Benkhaldoun}, {Magain}, \& {Queloz}}]{gillon2017}
{Gillon}, M., {Triaud}, A. H.~M.~J., {Demory}, B.-O., {et~al.} 2017, \nat, 542,
  456

\bibitem[{{Goldreich} \& {Tremaine}(1980)}]{goldreich1980}
{Goldreich}, P. \& {Tremaine}, S. 1980, \apj, 241, 425

\bibitem[{Goodman \& Weare(2010)}]{goodmanweare2010}
Goodman, J. \& Weare, J. 2010, Communications in applied mathematics and
  computational science, 5, 65

\bibitem[{{Hadden} \& {Lithwick}(2014)}]{hadden2014}
{Hadden}, S. \& {Lithwick}, Y. 2014, \apj, 787, 80

\bibitem[{{H{\'e}brard} {et~al.}(2020){H{\'e}brard}, {D{\'\i}az}, {Correia},
  {Collier Cameron}, {Laskar}, {Pollacco}, {Almenara}, {Anderson}, {Barros},
  {Boisse}, {Bonomo}, {Bouchy}, {Bou{\'e}}, {Boumis}, {Brown}, {Dalal},
  {Deleuil}, {Demangeon}, {Doyle}, {Haswell}, {Hellier}, {Osborn}, {Kiefer},
  {Kolb}, {Lam}, {Lecavelier des {\'E}tangs}, {Lopez}, {Martin-Lagarde},
  {Maxted}, {McCormac}, {Nielsen}, {Pall{\'e}}, {Prieto-Arranz}, {Queloz},
  {Santerne}, {Smalley}, {Turner}, {Udry}, {Verilhac}, {West}, {Wheatley}, \&
  {Wilson}}]{hebrard2020}
{H{\'e}brard}, G., {D{\'\i}az}, R.~F., {Correia}, A.~C.~M., {et~al.} 2020,
  \aap, 640, A32

\bibitem[{{Holman} {et~al.}(2010){Holman}, {Fabrycky}, {Ragozzine}, {Ford},
  {Steffen}, {Welsh}, {Lissauer}, {Latham}, {Marcy}, {Walkowicz}, {Batalha},
  {Jenkins}, {Rowe}, {Cochran}, {Fressin}, {Torres}, {Buchhave}, {Sasselov},
  {Borucki}, {Koch}, {Basri}, {Brown}, {Caldwell}, {Charbonneau}, {Dunham},
  {Gautier}, {Geary}, {Gilliland}, {Haas}, {Howell}, {Ciardi}, {Endl},
  {Fischer}, {F{\"u}r{\'e}sz}, {Hartman}, {Isaacson}, {Johnson}, {MacQueen},
  {Moorhead}, {Morehead}, \& {Orosz}}]{holman2010}
{Holman}, M.~J., {Fabrycky}, D.~C., {Ragozzine}, D., {et~al.} 2010, Science,
  330, 51

\bibitem[{{Holman} \& {Murray}(2005)}]{holman2005}
{Holman}, M.~J. \& {Murray}, N.~W. 2005, Science, 307, 1288

\bibitem[{{Hut}(1981)}]{Hut_1981}
{Hut}, P. 1981, \aap, 99, 126

\bibitem[{{Irwin}(1952)}]{irwin1952}
{Irwin}, J.~B. 1952, \apj, 116, 211

\bibitem[{{Jenkins} {et~al.}(2016){Jenkins}, {Twicken}, {McCauliff},
  {Campbell}, {Sanderfer}, {Lung}, {Mansouri-Samani}, {Girouard}, {Tenenbaum},
  {Klaus}, {Smith}, {Caldwell}, {Chacon}, {Henze}, {Heiges}, {Latham},
  {Morgan}, {Swade}, {Rinehart}, \& {Vanderspek}}]{jenkins2016}
{Jenkins}, J.~M., {Twicken}, J.~D., {McCauliff}, S., {et~al.} 2016, in Society
  of Photo-Optical Instrumentation Engineers (SPIE) Conference Series, Vol.
  9913, Software and Cyberinfrastructure for Astronomy IV, ed. G.~{Chiozzi} \&
  J.~C. {Guzman}, 99133E

\bibitem[{{Jontof-Hutter} {et~al.}(2021){Jontof-Hutter}, {Wolfgang}, {Ford},
  {Lissauer}, {Fabrycky}, \& {Rowe}}]{Jontof-Hutter2021}
{Jontof-Hutter}, D., {Wolfgang}, A., {Ford}, E.~B., {et~al.} 2021, \aj, 161,
  246

\bibitem[{Kass \& Raftery(1995)}]{kass1995}
Kass, R.~E. \& Raftery, A.~E. 1995, Journal of the American Statistical
  Association, 90, 773

\bibitem[{{Kipping}(2010{\natexlab{a}})}]{kipping2010}
{Kipping}, D.~M. 2010{\natexlab{a}}, \mnras, 408, 1758

\bibitem[{{Kipping}(2010{\natexlab{b}})}]{kipping2010b}
{Kipping}, D.~M. 2010{\natexlab{b}}, \mnras, 407, 301

\bibitem[{{Kipping}(2013)}]{kipping2013}
{Kipping}, D.~M. 2013, \mnras, 435, 2152

\bibitem[{{Kozai}(1962)}]{kozai1962}
{Kozai}, Y. 1962, \aj, 67, 591

\bibitem[{Kraft(1988)}]{kraft1988}
Kraft, D. 1988, A software package for sequential quadratic programming, Tech.
  Rep. DFVLR-FB 88-28, DFVLR Deutsche Forschungs- und Versuchsanstalt f\"{u}r
  Luft- und Raumfahrt

\bibitem[{{Kreidberg}(2015)}]{kreidberg2015}
{Kreidberg}, L. 2015, \pasp, 127, 1161

\bibitem[{{Lainey} {et~al.}(2009){Lainey}, {Arlot}, {Karatekin}, \& {van
  Hoolst}}]{Lainey_etal_2009}
{Lainey}, V., {Arlot}, J.-E., {Karatekin}, {\"O}., \& {van Hoolst}, T. 2009,
  Nature, 459, 957

\bibitem[{{Laskar}(1990)}]{laskar1990}
{Laskar}, J. 1990, \icarus, 88, 266

\bibitem[{Laskar(1993)}]{laskar1993}
Laskar, J. 1993, Physica D: Nonlinear Phenomena, 67, 257

\bibitem[{{Lecar} {et~al.}(2006){Lecar}, {Podolak}, {Sasselov}, \&
  {Chiang}}]{lecar2006}
{Lecar}, M., {Podolak}, M., {Sasselov}, D., \& {Chiang}, E. 2006, \apj, 640,
  1115

\bibitem[{{Lee} \& {Peale}(2002)}]{lee2002}
{Lee}, M.~H. \& {Peale}, S.~J. 2002, \apj, 567, 596

\bibitem[{{Libert} \& {Tsiganis}(2009)}]{libert2009}
{Libert}, A.~S. \& {Tsiganis}, K. 2009, \mnras, 400, 1373

\bibitem[{{Lidov}(1962)}]{lidov1962}
{Lidov}, M.~L. 1962, \planss, 9, 719

\bibitem[{{Lindegren} {et~al.}(2021){Lindegren}, {Klioner}, {Hern{\'a}ndez},
  {Bombrun}, {Ramos-Lerate}, {Steidelm{\"u}ller}, {Bastian}, {Biermann}, {de
  Torres}, {Gerlach}, {Geyer}, {Hilger}, {Hobbs}, {Lammers}, {McMillan},
  {Stephenson}, {Casta{\~n}eda}, {Davidson}, {Fabricius}, {Gracia-Abril},
  {Portell}, {Rowell}, {Teyssier}, {Torra}, {Bartolom{\'e}}, {Clotet},
  {Garralda}, {Gonz{\'a}lez-Vidal}, {Torra}, {Abbas}, {Altmann}, {Anglada
  Varela}, {Balaguer-N{\'u}{\~n}ez}, {Balog}, {Barache}, {Becciani}, {Bernet},
  {Bertone}, {Bianchi}, {Bouquillon}, {Brown}, {Bucciarelli}, {Busonero},
  {Butkevich}, {Buzzi}, {Cancelliere}, {Carlucci}, {Charlot}, {Cioni},
  {Crosta}, {Crowley}, {del Peloso}, {del Pozo}, {Drimmel}, {Esquej}, {Fienga},
  {Fraile}, {Gai}, {Garcia-Reinaldos}, {Guerra}, {Hambly}, {Hauser},
  {Jan{\ss}en}, {Jordan}, {Kostrzewa-Rutkowska}, {Lattanzi}, {Liao}, {Licata},
  {Lister}, {L{\"o}ffler}, {Marchant}, {Masip}, {Mignard}, {Mints}, {Molina},
  {Mora}, {Morbidelli}, {Murphy}, {Pagani}, {Panuzzo}, {Pe{\~n}alosa Esteller},
  {Poggio}, {Re Fiorentin}, {Riva}, {Sagrist{\`a} Sell{\'e}s}, {Sanchez
  Gimenez}, {Sarasso}, {Sciacca}, {Siddiqui}, {Smart}, {Souami}, {Spagna},
  {Steele}, {Taris}, {Utrilla}, {van Reeven}, \& {Vecchiato}}]{lindegren2021}
{Lindegren}, L., {Klioner}, S.~A., {Hern{\'a}ndez}, J., {et~al.} 2021, \aap,
  649, A2

\bibitem[{{Lissauer} {et~al.}(2011){Lissauer}, {Fabrycky}, {Ford}, {Borucki},
  {Fressin}, {Marcy}, {Orosz}, {Rowe}, {Torres}, {Welsh}, {Batalha}, {Bryson},
  {Buchhave}, {Caldwell}, {Carter}, {Charbonneau}, {Christiansen}, {Cochran},
  {Desert}, {Dunham}, {Fanelli}, {Fortney}, {Gautier}, {Geary}, {Gilliland},
  {Haas}, {Hall}, {Holman}, {Koch}, {Latham}, {Lopez}, {McCauliff}, {Miller},
  {Morehead}, {Quintana}, {Ragozzine}, {Sasselov}, {Short}, \&
  {Steffen}}]{lissauer2011}
{Lissauer}, J.~J., {Fabrycky}, D.~C., {Ford}, E.~B., {et~al.} 2011, \nat, 470,
  53

\bibitem[{{Lithwick} {et~al.}(2012){Lithwick}, {Xie}, \& {Wu}}]{lithwick2012}
{Lithwick}, Y., {Xie}, J., \& {Wu}, Y. 2012, \apj, 761, 122

\bibitem[{{Maciejewski} {et~al.}(2020){Maciejewski}, {Fern{\'a}ndez}, {Sota},
  \& {Garc{\'\i}a Segura}}]{Maciejewski2020}
{Maciejewski}, G., {Fern{\'a}ndez}, M., {Sota}, A., \& {Garc{\'\i}a Segura},
  A.~J. 2020, \actaa, 70, 203

\bibitem[{{Mandel} \& {Agol}(2002)}]{mandelagol2002}
{Mandel}, K. \& {Agol}, E. 2002, \apjl, 580, L171

\bibitem[{{Manduca} {et~al.}(1977){Manduca}, {Bell}, \&
  {Gustafsson}}]{manduca1977}
{Manduca}, A., {Bell}, R.~A., \& {Gustafsson}, B. 1977, \aap, 61, 809

\bibitem[{{McArthur} {et~al.}(2010){McArthur}, {Benedict}, {Barnes},
  {Martioli}, {Korzennik}, {Nelan}, \& {Butler}}]{mcarthur2010}
{McArthur}, B.~E., {Benedict}, G.~F., {Barnes}, R., {et~al.} 2010, \apj, 715,
  1203

\bibitem[{{Meibom} {et~al.}(2015){Meibom}, {Barnes}, {Platais}, {Gilliland},
  {Latham}, \& {Mathieu}}]{meibom2015}
{Meibom}, S., {Barnes}, S.~A., {Platais}, I., {et~al.} 2015, \nat, 517, 589

\bibitem[{{Mills} \& {Fabrycky}(2017)}]{mills2017}
{Mills}, S.~M. \& {Fabrycky}, D.~C. 2017, \aj, 153, 45

\bibitem[{{Nelson} {et~al.}(2020){Nelson}, {Ford}, {Buchner}, {Cloutier},
  {D{\'\i}az}, {Faria}, {Hara}, {Rajpaul}, \& {Rukdee}}]{nelson2020}
{Nelson}, B.~E., {Ford}, E.~B., {Buchner}, J., {et~al.} 2020, \aj, 159, 73

\bibitem[{{Nesvorn{\'y}} {et~al.}(2013){Nesvorn{\'y}}, {Kipping}, {Terrell},
  {Hartman}, {Bakos}, \& {Buchhave}}]{nesvorny2013}
{Nesvorn{\'y}}, D., {Kipping}, D., {Terrell}, D., {et~al.} 2013, \apj, 777, 3

\bibitem[{Pedregosa {et~al.}(2011)Pedregosa, Varoquaux, Gramfort, Michel,
  Thirion, Grisel, Blondel, Prettenhofer, Weiss, Dubourg, Vanderplas, Passos,
  Cournapeau, Brucher, Perrot, \& Duchesnay}]{scikit-learn}
Pedregosa, F., Varoquaux, G., Gramfort, A., {et~al.} 2011, Journal of Machine
  Learning Research, 12, 2825

\bibitem[{{Petit} {et~al.}(2020){Petit}, {Petigura}, {Davies}, \&
  {Johansen}}]{petit2020}
{Petit}, A.~C., {Petigura}, E.~A., {Davies}, M.~B., \& {Johansen}, A. 2020,
  \mnras, 496, 3101

\bibitem[{{Rasio} \& {Ford}(1996)}]{rasio1996}
{Rasio}, F.~A. \& {Ford}, E.~B. 1996, Science, 274, 954

\bibitem[{Rasmussen \& Williams(2005)}]{rasmussenwilliams2005}
Rasmussen, C.~E. \& Williams, C. K.~I. 2005, Gaussian Processes for Machine
  Learning (Adaptive Computation and Machine Learning) (The MIT Press)

\bibitem[{{Rein} \& {Liu}(2012)}]{rein2012}
{Rein}, H. \& {Liu}, S.-F. 2012, \aap, 537, A128

\bibitem[{{Rein} \& {Papaloizou}(2010)}]{rein2010}
{Rein}, H. \& {Papaloizou}, J.~C.~B. 2010, in EAS Publications Series, Vol.~42,
  EAS Publications Series, ed. K.~{Go{\.z}dziewski}, A.~{Niedzielski}, \&
  J.~{Schneider}, 299--302

\bibitem[{{Rein} \& {Tamayo}(2015)}]{rein2015}
{Rein}, H. \& {Tamayo}, D. 2015, \mnras, 452, 376

\bibitem[{{Ricker} {et~al.}(2015){Ricker}, {Winn}, {Vanderspek}, {Latham},
  {Bakos}, {Bean}, {Berta-Thompson}, {Brown}, {Buchhave}, {Butler}, {Butler},
  {Chaplin}, {Charbonneau}, {Christensen-Dalsgaard}, {Clampin}, {Deming},
  {Doty}, {De Lee}, {Dressing}, {Dunham}, {Endl}, {Fressin}, {Ge}, {Henning},
  {Holman}, {Howard}, {Ida}, {Jenkins}, {Jernigan}, {Johnson}, {Kaltenegger},
  {Kawai}, {Kjeldsen}, {Laughlin}, {Levine}, {Lin}, {Lissauer}, {MacQueen},
  {Marcy}, {McCullough}, {Morton}, {Narita}, {Paegert}, {Palle}, {Pepe},
  {Pepper}, {Quirrenbach}, {Rinehart}, {Sasselov}, {Sato}, {Seager},
  {Sozzetti}, {Stassun}, {Sullivan}, {Szentgyorgyi}, {Torres}, {Udry}, \&
  {Villasenor}}]{ricker2015}
{Ricker}, G.~R., {Winn}, J.~N., {Vanderspek}, R., {et~al.} 2015, Journal of
  Astronomical Telescopes, Instruments, and Systems, 1, 014003

\bibitem[{{Riello} {et~al.}(2021){Riello}, {De Angeli}, {Evans}, {Montegriffo},
  {Carrasco}, {Busso}, {Palaversa}, {Burgess}, {Diener}, {Davidson}, {Rowell},
  {Fabricius}, {Jordi}, {Bellazzini}, {Pancino}, {Harrison}, {Cacciari}, {van
  Leeuwen}, {Hambly}, {Hodgkin}, {Osborne}, {Altavilla}, {Barstow}, {Brown},
  {Castellani}, {Cowell}, {De Luise}, {Gilmore}, {Giuffrida}, {Hidalgo},
  {Holland}, {Marinoni}, {Pagani}, {Piersimoni}, {Pulone}, {Ragaini}, {Rainer},
  {Richards}, {Sanna}, {Walton}, {Weiler}, \& {Yoldas}}]{riello2021}
{Riello}, M., {De Angeli}, F., {Evans}, D.~W., {et~al.} 2021, \aap, 649, A3

\bibitem[{{Skrutskie} {et~al.}(2006){Skrutskie}, {Cutri}, {Stiening},
  {Weinberg}, {Schneider}, {Carpenter}, {Beichman}, {Capps}, {Chester},
  {Elias}, {Huchra}, {Liebert}, {Lonsdale}, {Monet}, {Price}, {Seitzer},
  {Jarrett}, {Kirkpatrick}, {Gizis}, {Howard}, {Evans}, {Fowler}, {Fullmer},
  {Hurt}, {Light}, {Kopan}, {Marsh}, {McCallon}, {Tam}, {Van Dyk}, \&
  {Wheelock}}]{2mass}
{Skrutskie}, M.~F., {Cutri}, R.~M., {Stiening}, R., {et~al.} 2006, \aj, 131,
  1163

\bibitem[{{Speagle}(2020)}]{speagle2020}
{Speagle}, J.~S. 2020, \mnras, 493, 3132

\bibitem[{{Stassun} {et~al.}(2019){Stassun}, {Oelkers}, {Paegert}, {Torres},
  {Pepper}, {De Lee}, {Collins}, {Latham}, {Muirhead}, {Chittidi},
  {Rojas-Ayala}, {Fleming}, {Rose}, {Tenenbaum}, {Ting}, {Kane}, {Barclay},
  {Bean}, {Brassuer}, {Charbonneau}, {Ge}, {Lissauer}, {Mann}, {McLean},
  {Mullally}, {Narita}, {Plavchan}, {Ricker}, {Sasselov}, {Seager}, {Sharma},
  {Shiao}, {Sozzetti}, {Stello}, {Vanderspek}, {Wallace}, \&
  {Winn}}]{stassun2019}
{Stassun}, K.~G., {Oelkers}, R.~J., {Paegert}, M., {et~al.} 2019, \aj, 158, 138

\bibitem[{{Thommes} \& {Lissauer}(2003)}]{Thommes2003}
{Thommes}, E.~W. \& {Lissauer}, J.~J. 2003, \apj, 597, 566

\bibitem[{{Tingley} \& {Sackett}(2005)}]{tingley2005}
{Tingley}, B. \& {Sackett}, P.~D. 2005, \apj, 627, 1011

\bibitem[{{Udry} {et~al.}(2003){Udry}, {Mayor}, \& {Santos}}]{udry2003}
{Udry}, S., {Mayor}, M., \& {Santos}, N.~C. 2003, \aap, 407, 369

\bibitem[{Virtanen {et~al.}(2020)Virtanen, Gommers, Oliphant, Haberland, Reddy,
  Cournapeau, Burovski, Peterson, Weckesser, Bright, {van der Walt}, Brett,
  Wilson, Millman, Mayorov, Nelson, Jones, Kern, Larson, Carey, Polat, Feng,
  Moore, {VanderPlas}, Laxalde, Perktold, Cimrman, Henriksen, Quintero, Harris,
  Archibald, Ribeiro, Pedregosa, {van Mulbregt}, \& {SciPy 1.0
  Contributors}}]{scipy2020}
Virtanen, P., Gommers, R., Oliphant, T.~E., {et~al.} 2020, Nature Methods, 17,
  261

\bibitem[{{Wang} {et~al.}(2014){Wang}, {Xie}, {Barclay}, \&
  {Fischer}}]{wang2014}
{Wang}, J., {Xie}, J.-W., {Barclay}, T., \& {Fischer}, D.~A. 2014, \apj, 783, 4

\bibitem[{{Wang} {et~al.}(2022){Wang}, {Rice}, {Wang}, {Pu}, {Stef{\'a}nsson},
  {Mahadevan}, {Radzom}, {Giacalone}, {Wu}, {Esposito}, {Dalba}, {Avsar},
  {Holden}, {Skiff}, {Polakis}, {Voeller}, {Logsdon}, {Klusmeyer}, {Schweiker},
  {Wu}, {Beard}, {Dai}, {Lubin}, {Weiss}, {Bender}, {Blake}, {Dressing},
  {Halverson}, {Hearty}, {Howard}, {Huber}, {Isaacson}, {Jackman}, {Llama},
  {McElwain}, {Rajagopal}, {Roy}, {Robertson}, {Schwab}, {Shkolnik}, {Wright},
  \& {Laughlin}}]{wang2022}
{Wang}, X.-Y., {Rice}, M., {Wang}, S., {et~al.} 2022, \apjl, 926, L8

\bibitem[{{Wright} {et~al.}(2010){Wright}, {Eisenhardt}, {Mainzer}, {Ressler},
  {Cutri}, {Jarrett}, {Kirkpatrick}, {Padgett}, {McMillan}, {Skrutskie},
  {Stanford}, {Cohen}, {Walker}, {Mather}, {Leisawitz}, {Gautier}, {McLean},
  {Benford}, {Lonsdale}, {Blain}, {Mendez}, {Irace}, {Duval}, {Liu}, {Royer},
  {Heinrichsen}, {Howard}, {Shannon}, {Kendall}, {Walsh}, {Larsen}, {Cardon},
  {Schick}, {Schwalm}, {Abid}, {Fabinsky}, {Naes}, \& {Tsai}}]{wise}
{Wright}, E.~L., {Eisenhardt}, P.~R.~M., {Mainzer}, A.~K., {et~al.} 2010, \aj,
  140, 1868

\bibitem[{{Wu} \& {Lithwick}(2011)}]{wu2011}
{Wu}, Y. \& {Lithwick}, Y. 2011, \apj, 735, 109

\bibitem[{Zhu {et~al.}(1997)Zhu, Byrd, Lu, \& Nocedal}]{zhu1997}
Zhu, C., Byrd, R.~H., Lu, P., \& Nocedal, J. 1997, ACM Transactions on
  Mathematical Software (TOMS), 23, 550

\end{thebibliography}

\begin{appendix}
\section{Additional figures and tables}

\begin{table}[!ht]
\small
  \caption{Photometric measurements used for the SED analysis of WASP-148.}
  \centering
\begin{tabular}{lcc}
\hline
\hline
Filter & Magnitude & $\pm1\sigma$  \\
\hline
Gaia-G    &  12.0790 &  0.0028 \\
Gaia-BP   &  12.4636 &  0.0030 \\
Gaia-RP   &  11.5315 &  0.0029 \\
2MASS-J   &  10.938  &  0.024  \\
2MASS-H   &  10.585  &  0.018  \\ 
2MASS-Ks  &  10.506  &  0.017  \\
WISE-W1   &  10.466  &  0.022  \\
WISE-W2   &  10.519  &  0.020  \\
WISE-W3$^{(\dagger)}$ &  10.476  &  0.064  \\
\hline
\end{tabular}
\tablefoot{$^{(\dagger)}$ Not covered by the Castelli \& Kurucz models.}
\label{table.sedmag}
\end{table}

\newpage
\clearpage

\begin{table*}
  \tiny
\renewcommand{\arraystretch}{1.25}
\centering
\caption{Inferred system parameters for the transit-only analysis.}\label{table.juliet}
\begin{tabular}{lccccccc}
\hline
\multicolumn{3}{l}{Parameter} & Units & Prior & Posterior median and 68.3\% CI\\
\hline
\multicolumn{3}{l}{\emph{\bf Star}} \\
\multicolumn{3}{l}{Stellar mean density}    & [$\rm{g\;cm^{-3}}$]  &  $N$(1.75, 0.15)    & 1.77 $\pm$ 0.13 \\
\multicolumn{3}{l}{\citet{kipping2013} $q_1$ for Clear, Johnson-R, RISE, TESS} & & $U$(0, 1) & 0.38$^{+0.25}_{-0.18}$, 0.204$^{+0.13}_{-0.090}$, 0.78 $\pm$ 0.16, 0.35$^{+0.22}_{-0.15}$ \\
\multicolumn{3}{l}{\citet{kipping2013} $q_2$ for Clear, Johnson-R, RISE, TESS} & & $U$(0, 1) & 0.29$^{+0.26}_{-0.19}$, 0.57 $\pm$ 0.28, 0.40$^{+0.14}_{-0.12}$, 0.39$^{+0.29}_{-0.23}$ \\

\multicolumn{3}{l}{\emph{\bf Planet b}} \\
\multicolumn{3}{l}{\citet{espinoza2018} $r_1$} &       & $U$(0, 1) & 0.432$^{+0.080}_{-0.065}$ \\
\multicolumn{3}{l}{\citet{espinoza2018} $r_2$} &       & $U$(0, 1) & 0.0854 $\pm$ 0.0013 \\
\multicolumn{3}{l}{$\sqrt{e}\cos{\omega}$}    &       & $N$(0.24, 0.09) & 0.243 $\pm$ 0.080 \\
\multicolumn{3}{l}{$\sqrt{e}\sin{\omega}$}    &       & $N$(0.40, 0.12) & 0.384$^{+0.043}_{-0.049}$ \\

\multicolumn{3}{l}{\emph{\bf Transit timings (instrument, band, epoch)}} \\
WASP1 & Clear & -377   & [BJD$_{\rm TDB}$] & $U$(2454638.265, 2454638.773) & 2454638.461$^{+0.21}_{-0.097}$ \\
WASP2 & Clear & -296   & [BJD$_{\rm TDB}$] & $U$(2455351.288, 2455351.835) & 2455351.517$^{+0.033}_{-0.041}$ \\
WASP3 & Clear & -255   & [BJD$_{\rm TDB}$] & $U$(2455712.42, 2455712.65)   & 2455712.5208$^{+0.0080}_{-0.0067}$ \\
NITES1 & Clear & -122   & [BJD$_{\rm TDB}$] & $U$(2456883.440, 2456883.449) & 2456883.4442 $\pm$ 0.0010 \\
SANCHEZ & Johnson-R & -87 & [BJD$_{\rm TDB}$] & $U$(2457191.55, 2457191.56) & 2457191.5555 $\pm$ 0.0010 \\
NITES2 & Johnson-R & -46  & [BJD$_{\rm TDB}$] & $U$(2457552.5005, 2457552.5082) & 2457552.50421 $\pm$ 0.00088 \\
MARS & Clear & -11        & [BJD$_{\rm TDB}$] & $U$(2457860.648, 2457860.660) & 2457860.6550 $\pm$ 0.0012 \\
RISE & RISE & 0       & [BJD$_{\rm TDB}$] & $U$(2457957.480, 2457957.484) & 2457957.48202 $\pm$ 0.00042 \\
TESS sector 24 & TESS & 114     & [BJD$_{\rm TDB}$] & $U$(2458961.1150, 2458961.1331) & 2458961.1226 $\pm$ 0.0019 \\
TESS sector 24 & TESS & 115     & [BJD$_{\rm TDB}$] & $U$(2458969.9220, 2458969.9358) & 2458969.9285 $\pm$ 0.0015 \\
TESS sector 24 & TESS & 116     & [BJD$_{\rm TDB}$] & $U$(2458978.7245, 2458978.7438) & 2458978.7346 $\pm$ 0.0022 \\
TESS sector 25 & TESS & 117     & [BJD$_{\rm TDB}$] & $U$(2458987.531, 2458987.5476) & 2458987.5403 $\pm$ 0.0019 \\
TESS sector 25 & TESS & 119     & [BJD$_{\rm TDB}$] & $U$(2459005.1466, 2459005.1581) & 2459005.1521 $\pm$ 0.0016 \\
TESS sector 26 & TESS & 120     & [BJD$_{\rm TDB}$] & $U$(2459013.9516, 2459013.970) & 2459013.9600 $\pm$ 0.0021 \\
TESS sector 26 & TESS & 122     & [BJD$_{\rm TDB}$] & $U$(2459031.5623, 2459031.5792) & 2459031.5701 $\pm$ 0.0018 \\

\multicolumn{3}{l}{\emph{\bf Instruments}} \\
\multicolumn{3}{l}{Dilution factor for WASP}                   &       & $TN$(1.0, 0.1, 0.0, 1.0) & 0.923$^{+0.047}_{-0.069}$ \\
\multicolumn{3}{l}{Dilution factor for TESS sector 24, 25, 26} &       & $TN$(1.0, 0.1, 0.0, 1.0) & 0.949 $\pm$ 0.033, 0.936 $\pm$ 0.035, 0.945$^{+0.032}_{-0.041}$ \\
\multicolumn{3}{l}{Offset relative flux for WASP1, WASP2, WASP3}        & [Relative flux] & $N$(0.0, 0.1) & 0.0030$^{+0.0038}_{-0.0033}$, -0.0006$^{+0.0037}_{-0.0030}$, 0.0008 $\pm$ 0.0022 \\
\multicolumn{3}{l}{Offset relative flux for NITES1, NITES2}             & [Relative flux] & $N$(0.0, 0.1) & 0.00061$^{+0.00058}_{-0.00086}$, -0.00015$^{+0.0016}_{-0.00068}$ \\
\multicolumn{3}{l}{Offset relative flux for MARS}                       & [Relative flux] & $N$(0.0, 0.1) & -0.0019$^{+0.0062}_{-0.0060}$ \\
\multicolumn{3}{l}{Offset relative flux for SANCHEZ}                    & [Relative flux] & $N$(0.0, 0.1) & -0.00201$^{+0.00081}_{-0.00090}$ \\
\multicolumn{3}{l}{Offset relative flux for RISE}                       & [Relative flux] & $N$(0.0, 0.1) & 0.0034$^{+0.035}_{-0.0059}$ \\
\multicolumn{3}{l}{Offset relative flux for TESS sector 24, 25, 26}     & [Relative flux] & $N$(0.0, 0.1) & 0.0004 $\pm$ 0.0013, 0.0005 $\pm$ 0.0019, -0.0003 $\pm$ 0.0014 \\
\multicolumn{3}{l}{Additive jitter for WASP1, WASP2, WASP3}      & [ppm] & $J$(1, 10000) & 82$^{+980}_{-76}$, 61$^{+630}_{-56}$, 210$^{+2800}_{-200}$ \\
\multicolumn{3}{l}{Additive jitter for NITES1, NITES2}           & [ppm] & $J$(1, 4000)  & 18$^{+100}_{-15}$, 3290 $\pm$ 130 \\
\multicolumn{3}{l}{Additive jitter for MARS}                     & [ppm] & $J$(1, 1600)  & 860 $\pm$ 120 \\
\multicolumn{3}{l}{Additive jitter for SANCHEZ}                  & [ppm] & $J$(1, 1600)  & 27$^{+150}_{-24}$ \\
\multicolumn{3}{l}{Additive jitter for RISE}                     & [ppm] & $J$(1, 300)   & 10.5$^{+36}_{-8.1}$ \\
\multicolumn{3}{l}{Additive jitter for TESS sector 24, 25, 26}   & [ppm] & $J$(1, 900)   & 27$^{+110}_{-23}$, 94$^{+120}_{-85}$, 744 $\pm$ 27 \\
\multicolumn{3}{l}{Timescale of the GP for WASP1, WASP2, WASP3}       & [days] & $J$(0.001, 1000) & 1.0$^{+91}_{-1.0}$, 1.8$^{+75}_{-1.8}$, 0.021$^{+0.034}_{-0.012}$ \\
\multicolumn{3}{l}{Timescale of the GP for NITES1, NITES2}            & [days] & $J$(0.001, 1000) & 6.0$^{+160}_{-5.9}$, 12$^{+220}_{-12}$ \\
\multicolumn{3}{l}{Timescale of the GP for MARS}                      & [days] & $J$(0.001, 1000) & 1.3$^{+15}_{-1.1}$ \\
\multicolumn{3}{l}{Timescale of the GP for SANCHEZ}                  & [days] & $J$(0.001, 1000) & 0.0233$^{+0.016}_{-0.0098}$ \\
\multicolumn{3}{l}{Timescale of the GP for RISE}                      & [days] & $J$(0.001, 1000) & 0.34$^{+0.75}_{-0.22}$ \\
\multicolumn{3}{l}{Timescale of the GP for TESS sector 24, 25, 26}    & [days] & $J$(0.001, 1000) & 1.81$^{+0.38}_{-0.30}$, 5.1$^{+2.4}_{-1.4}$, 1.04$^{+0.16}_{-0.13}$ \\
\multicolumn{3}{l}{Amplitude of the GP for WASP1, WASP2, WASP3}    & [Relative flux] & $J$(10$^{-6}$, 1) & 0.00022$^{+0.0046}_{-0.00021}$, 0.00072$^{+0.013}_{-0.00070}$, 0.0040$^{+0.0020}_{-0.0013}$ \\
\multicolumn{3}{l}{Amplitude of the GP for NITES1, NITES2}          & [Relative flux] & $J$(10$^{-6}$, 1) & 0.00020$^{+0.0044}_{-0.00019}$, 0.00040$^{+0.0088}_{-0.00038}$ \\
\multicolumn{3}{l}{Amplitude of the GP for MARS}                    & [Relative flux] & $J$(10$^{-6}$, 1) & 0.014$^{+0.10}_{-0.012}$ \\
\multicolumn{3}{l}{Amplitude of the GP for SANCHEZ}                 & [Relative flux] & $J$(10$^{-6}$, 1) & 0.00177$^{+0.00085}_{-0.00045}$ \\
\multicolumn{3}{l}{Amplitude of the GP for RISE}                    & [Relative flux] & $J$(10$^{-6}$, 1) & 0.0104$^{+0.041}_{-0.0074}$ \\
\multicolumn{3}{l}{Amplitude of the GP for TESS sector 24, 25, 26}  & [Relative flux] & $J$(10$^{-6}$, 1) & 0.00312$^{+0.00077}_{-0.00050}$, 0.00285$^{+0.0019}_{-0.00090}$, 0.00430$^{+0.00077}_{-0.00058}$ \\

\multicolumn{3}{l}{\emph{\bf Derived}} \\
\multicolumn{3}{l}{Radius ratio, $R_{\mathrm{p}}/R_\star$} &       &       &  0.0854 $\pm$ 0.0013 \\
\multicolumn{3}{l}{Impact parameter} &       &       &  0.148$^{+0.12}_{-0.098}$ \\
\multicolumn{3}{l}{Eccentricity, $e$} &       &       &  0.213 $\pm$ 0.035 \\
\multicolumn{3}{l}{Argument of pericentre, $\omega$} &  [\degree]     &       &  58 $\pm$ 11 \smallskip\\
\hline
\end{tabular}
\tablefoot{The table lists: Priors and posterior median as well as 68.3\% CI for the transit-only analysis with \juliet (Section~\ref{section.juliet}). $N(\mu, \sigma)$: Normal distribution with mean $\mu$ and standard deviation $\sigma$. $TN(\mu, \sigma, a, b)$: Normal distribution with mean $\mu$ and standard deviation $\sigma$, truncated between a lower $a$ and upper $b$ limit. $U(a, b)$: A uniform distribution defined between a lower $a$ and upper $b$ limit. $J(a, b)$: Jeffreys (or log-uniform) distribution defined between a lower $a$ and upper $b$ limit.}
\end{table*}

\newpage
\clearpage

\begin{table*}
  \tiny
\renewcommand{\arraystretch}{1.25}
\centering
\caption{Idem.\ as Figure~\ref{fig.juliet}, but for the analysis of OSN150 transits.}\label{table.OSN}
\begin{tabular}{lccccccc}
\hline
\multicolumn{3}{l}{Parameter} & Units & Prior & Posterior median and 68.3\% CI\\
\hline
\multicolumn{3}{l}{\emph{\bf Star}} \\
\multicolumn{3}{l}{Stellar mean density}    & [$\rm{g\;cm^{-3}}$]  &  $N$(1.75, 0.15)    & 1.73 $\pm$ 0.15 \\
\multicolumn{3}{l}{\citet{kipping2013} $q_1$ for Clear} & & $U$(0, 1) & 0.45 $^{+0.25}_{-0.16}$\\
\multicolumn{3}{l}{\citet{kipping2013} $q_2$ for Clear} & & $U$(0, 1) & 0.17 $^{+0.17}_{-0.11}$\\

\multicolumn{3}{l}{\emph{\bf Planet b}} \\
\multicolumn{3}{l}{\citet{espinoza2018} $r_1$} &       & $U$(0, 1) & 0.550 $^{+0.064}_{-0.11}$ \\
\multicolumn{3}{l}{\citet{espinoza2018} $r_2$} &       & $U$(0, 1) & 0.0851 $^{+0.0020}_{-0.0016}$ \\
\multicolumn{3}{l}{$\sqrt{e}\cos{\omega}$}    &       & $N$(0.24, 0.09) & 0.238 $\pm$ 0.091 \\
\multicolumn{3}{l}{$\sqrt{e}\sin{\omega}$}    &       & $N$(0.40, 0.12) & 0.366$^{+0.053}_{-0.062}$ \\

\multicolumn{3}{l}{\emph{\bf Transit timings (instrument, band, epoch)}} \\
OSN1 & Clear & 123   & [BJD$_{\rm TDB}$] & $U$(2459040.365914, 2459040.49543)  & 2459040.37337 $^{+0.00052}_{-0.00047}$ \\
OSN2 & Clear & 128   & [BJD$_{\rm TDB}$] & $U$(2459084.327292, 2459084.50194)  & 2459084.39721 $^{+0.00058}_{-0.00067}$ \\
OSN3 & Clear & 152   & [BJD$_{\rm TDB}$] & $U$(2459295.515766, 2459295.733722) & 2459295.65861 $^{+0.00053}_{-0.00053}$ \\
OSN4 & Clear & 163   & [BJD$_{\rm TDB}$] & $U$(2459392.390943, 2459392.628101) & 2459392.51001 $^{+0.00036}_{-0.00036}$ \\

\multicolumn{3}{l}{\emph{\bf Instruments}} \\
\multicolumn{3}{l}{Offset relative flux for OSN1}        & [Relative flux] & $U$(-0.001, 0.001) & -0.00035 $^{+0.00067}_{-0.00041}$ \\
\multicolumn{3}{l}{Offset relative flux for OSN2}        & [Relative flux] & $U$(-0.001, 0.001) &  0.00037 $^{+0.00043}_{-0.00068}$ \\
\multicolumn{3}{l}{Offset relative flux for OSN3}        & [Relative flux] & $U$(-0.001, 0.001) &  0.00040 $^{+0.00042}_{-0.00066}$ \\
\multicolumn{3}{l}{Offset relative flux for OSN4}        & [Relative flux] & $U$(-0.001, 0.001) &  0.00006 $^{+0.00047}_{-0.00055}$ \\
\multicolumn{3}{l}{Additive jitter for OSN1}      & [ppm] & $J$(1, 2000) & 979 $\pm$ 63 \\
\multicolumn{3}{l}{Additive jitter for OSN2}      & [ppm] & $J$(1, 2000) & 1374 $\pm$ 77 \\
\multicolumn{3}{l}{Additive jitter for OSN3}      & [ppm] & $J$(1, 2000) & 1051 $\pm$ 56 \\
\multicolumn{3}{l}{Additive jitter for OSN4}      & [ppm] & $J$(1, 2000) & 19 $^{+130}_{-17}$ \\
\multicolumn{3}{l}{Timescale of the GP for OSN1}       & [days] & $J$(0.001, 1000) & 0.118 $^{+0.33}_{-0.075}$\\
\multicolumn{3}{l}{Timescale of the GP for OSN2}       & [days] & $J$(0.001, 1000) & 0.063 $^{+0.068}_{-0.032}$\\
\multicolumn{3}{l}{Timescale of the GP for OSN3}       & [days] & $J$(0.001, 1000) & 0.048 $^{+0.060}_{-0.024}$\\
\multicolumn{3}{l}{Timescale of the GP for OSN4}       & [days] & $J$(0.001, 1000) & 0.066 $^{+0.076}_{-0.035}$\\
\multicolumn{3}{l}{Amplitude of the GP for OSN1}    & [Relative flux] & $J$(10$^{-6}$, 0.05) & 0.00106 $^{+0.0023}_{-0.00056}$ \\
\multicolumn{3}{l}{Amplitude of the GP for OSN2}    & [Relative flux] & $J$(10$^{-6}$, 0.05) & 0.00120 $^{+0.0010}_{-0.00056}$ \\
\multicolumn{3}{l}{Amplitude of the GP for OSN3}    & [Relative flux] & $J$(10$^{-6}$, 0.05) & 0.00117 $^{+0.0010}_{-0.00044}$ \\
\multicolumn{3}{l}{Amplitude of the GP for OSN4}    & [Relative flux] & $J$(10$^{-6}$, 0.05) & 0.00090 $^{+0.00095}_{-0.00034}$ \\

\multicolumn{3}{l}{\emph{\bf Derived}} \\
\multicolumn{3}{l}{Radius ratio, $R_{\mathrm{p}}/R_\star$} &       &       &  0.0851 $^{+0.0020}_{-0.0016}$ \\
\multicolumn{3}{l}{Impact parameter} &       &       &  0.324 $^{+0.096}_{-0.17}$ \\
\multicolumn{3}{l}{Eccentricity, $e$} &       &       &  0.197 $\pm$ 0.046 \\
\multicolumn{3}{l}{Argument of pericentre, $\omega$} &  [\degree]     &       &  57 $\pm$ 13 \smallskip\\
\hline
\end{tabular}
\tablefoot{$N(\mu, \sigma)$: Normal distribution with mean $\mu$ and standard deviation $\sigma$. $TN(\mu, \sigma, a, b)$: Normal distribution with mean $\mu$ and standard deviation $\sigma$, truncated between a lower $a$ and upper $b$ limit. $U(a, b)$: A uniform distribution defined between a lower $a$ and upper $b$ limit. $J(a, b)$: Jeffreys (or log-uniform) distribution defined between a lower $a$ and upper $b$ limit.}
\end{table*}

\newpage
\clearpage

\begin{table*}
  \tiny
\renewcommand{\arraystretch}{1.2}
\centering
\caption{Continuation of Table~\ref{table.results} for the photodynamical modelling.}\label{table.results2}
\begin{tabular}{lcccc}
\hline
Parameter & Units & Prior & Median and 68.3\% CI \\
\hline
SOPHIE offset                           & [\kms]          & $U$(-100, 100)    &    5.6179 $\pm$ 0.0014 \\
SOPHIE multiplicative jitter            &                 & $U$(0, 10)        &   1.53 $^{+0.12}_{-0.10}$ \\
Dilution factor for WASP                &                 & $U$(0, 1)         &    0.53 $^{+0.25}_{-0.31}$ \\
Dilution factor for TESS sector 24      &                 & $U$(0, 1)         &   0.104 $^{+0.035}_{-0.041}$ \\
Dilution factor for TESS sector 25      &                 & $U$(0, 1)         &   0.122 $^{+0.037}_{-0.044}$ \\
Dilution factor for TESS sector 26      &                 & $U$(0, 1)         &   0.116 $\pm$ 0.053 \\
Relative flux for WASP1                 & [Relative flux] & $U$(0.9, 1.1)     &   1.0002 $^{+0.0023}_{-0.0027}$ \\
Relative flux for WASP2                 & [Relative flux] & $U$(0.9, 1.1)     &    1.0022 $\pm$ 0.0015 \\
Relative flux for WASP3                 & [Relative flux] & $U$(0.9, 1.1)     &    0.9995 $\pm$ 0.0011 \\
Relative flux for NITES1                & [Relative flux] & $U$(0.9, 1.1)     &    1.00014 $\pm$ 0.00018 \\
Relative flux for NITES2                & [Relative flux] & $U$(0.9, 1.1)     &    0.99994 $\pm$ 0.00016 \\
Relative flux for MARS                  & [Relative flux] & $U$(0.9, 1.1)     &    0.99994 $\pm$ 0.00013 \\
Relative flux for SANCHEZ               & [Relative flux] & $U$(0.9, 1.1)     &   1.00006 $^{+0.00015}_{-0.00013}$ \\
Relative flux for RISE                  & [Relative flux] & $U$(0.9, 1.1)     &   1.000018 $^{+0.000065}_{-0.000057}$ \\
Relative flux for TESS sector 24        & [Relative flux] & $U$(0.9, 1.1)     &    0.999921 $\pm$ 0.00011  \\
Relative flux for TESS sector 25        & [Relative flux] & $U$(0.9, 1.1)     &    1.00001 $\pm$ 0.00012 \\
Relative flux for TESS sector 26        & [Relative flux] & $U$(0.9, 1.1)     &   0.99991 $\pm$ 0.00014 \\
Relative flux for OSN1                  & [Relative flux] & $U$(0.9, 1.1)     &   1.000054 $\pm$ 0.000086 \\
Relative flux for OSN2                  & [Relative flux] & $U$(0.9, 1.1)     &   1.000009 $\pm$ 0.000097 \\
Relative flux for OSN3                  & [Relative flux] & $U$(0.9, 1.1)     &   0.999996 $\pm$ 0.000072 \\
Relative flux for OSN4                  & [Relative flux] & $U$(0.9, 1.1)     &   0.999998 $\pm$ 0.000052 \\
Additive jitter for WASP1               & [ppm]           & $J$(1, 10000)     &   69 $^{+1000}_{-65}$ \\
Additive jitter for WASP2               & [ppm]           & $J$(1, 10000)     &   53 $^{+540}_{-49}$ \\
Additive jitter for WASP3               & [ppm]           & $J$(1, 10000)     &   320 $^{+2500}_{-310}$ \\
Additive jitter for NITES1              & [ppm]           & $J$(1, 10000)     &   25 $^{+130}_{-22}$ \\
Additive jitter for NITES2              & [ppm]           & $J$(1, 10000)     &  3300 $\pm$ 140  \\
Additive jitter for MARS                & [ppm]           & $J$(1, 10000)     &   870 $\pm$ 140 \\
Additive jitter for SANCHEZ             & [ppm]           & $J$(1, 10000)     &  31 $^{+180}_{-29}$ \\
Additive jitter for RISE                & [ppm]           & $J$(1, 10000)     &    15 $^{+43}_{-12}$ \\
Additive jitter for TESS sector 24      & [ppm]           & $J$(1, 10000)     &    26 $^{+100}_{-23}$ \\
Additive jitter for TESS sector 25      & [ppm]           & $J$(1, 10000)     &  110 $^{+120}_{-100}$ \\
Additive jitter for TESS sector 26      & [ppm]           & $J$(1, 10000)     &  745 $\pm$ 27 \\
Additive jitter for OSN1                & [ppm]           & $J$(1, 10000)     &   972 $\pm$ 64 \\
Additive jitter for OSN2                & [ppm]           & $J$(1, 10000)     &   1372 $\pm$ 77 \\
Additive jitter for OSN3                & [ppm]           & $J$(1, 10000)     &   1052 $\pm$ 61 \\
Additive jitter for OSN4                & [ppm]           & $J$(1, 10000)     &   20 $^{+110}_{-18}$ \\
Timescale of the GP for WASP1          & [days]          & $J$(0.001, 1000)  &  3.0 $^{+100}_{-3.0}$ \\
Timescale of the GP for WASP2          & [days]          & $J$(0.001, 1000)  &  2.0 $^{+63}_{-1.9}$ \\
Timescale of the GP for WASP3          & [days]          & $J$(0.001, 1000)  &  0.040 $^{+0.061}_{-0.024}$ \\
Timescale of the GP for NITES1         & [days]          & $J$(0.001, 1000)  &  2.7 $^{+72}_{-2.7}$ \\
Timescale of the GP for NITES2         & [days]          & $J$(0.001, 1000)  &  13 $^{+160}_{-13}$ \\
Timescale of the GP for MARS           & [days]          & $J$(0.001, 1000)  &  1.2 $^{+12}_{-1.1}$ \\
Timescale of the GP for SANCHEZ        & [days]          & $J$(0.001, 1000)  &  0.0230 $^{+0.017}_{-0.0076}$ \\
Timescale of the GP for RISE           & [days]          & $J$(0.001, 1000)  &  1.03 $^{+1.5}_{-0.91}$ \\
Timescale of the GP for TESS sector 24 & [days]          & $J$(0.001, 1000)  &  1.79 $^{+0.40}_{-0.30}$ \\
Timescale of the GP for TESS sector 25 & [days]          & $J$(0.001, 1000)  &  4.7 $^{+2.4}_{-1.3}$ \\
Timescale of the GP for TESS sector 26 & [days]          & $J$(0.001, 1000)  &  1.05 $^{+0.17}_{-0.14}$ \\
Timescale of the GP for OSN1           & [days]          & $J$(0.001, 1000)  &  0.130 $^{+0.16}_{-0.083}$ \\
Timescale of the GP for OSN2           & [days]          & $J$(0.001, 1000)  &  0.081 $^{+3.6}_{-0.041}$ \\
Timescale of the GP for OSN3           & [days]          & $J$(0.001, 1000)  &  0.062 $^{+0.080}_{-0.028}$ \\
Timescale of the GP for OSN4           & [days]          & $J$(0.001, 1000)  &  0.046 $^{+0.040}_{-0.018}$ \\
Amplitude of the GP for WASP1           & [Relative flux] & $J$(10$^{-6}$, 1) &  0.00036 $^{+0.0050}_{-0.00035}$ \\
Amplitude of the GP for WASP2           & [Relative flux] & $J$(10$^{-6}$, 1) &  0.0019 $^{+0.0053}_{-0.0019}$ \\
Amplitude of the GP for WASP3           & [Relative flux] & $J$(10$^{-6}$, 1) &  0.0058 $^{+0.0038}_{-0.0024}$ \\
Amplitude of the GP for NITES1          & [Relative flux] & $J$(10$^{-6}$, 1) &  0.00051 $^{+0.0011}_{-0.00049}$ \\
Amplitude of the GP for NITES2          & [Relative flux] & $J$(10$^{-6}$, 1) &  0.00023 $^{+0.0049}_{-0.00022}$ \\
Amplitude of the GP for MARS            & [Relative flux] & $J$(10$^{-6}$, 1) &  0.0052 $^{+0.0053}_{-0.0040}$ \\
Amplitude of the GP for SANCHEZ         & [Relative flux] & $J$(10$^{-6}$, 1) &  0.00181 $^{+0.00082}_{-0.00040}$ \\
Amplitude of the GP for RISE            & [Relative flux] & $J$(10$^{-6}$, 1) &  0.051 $^{+0.11}_{-0.049}$ \\
Amplitude of the GP for TESS sector 24  & [Relative flux] & $J$(10$^{-6}$, 1) &  0.00312 $^{+0.00084}_{-0.00056}$ \\
Amplitude of the GP for TESS sector 25  & [Relative flux] & $J$(10$^{-6}$, 1) &  0.00266 $^{+0.0018}_{-0.00087}$ \\
Amplitude of the GP for TESS sector 26  & [Relative flux] & $J$(10$^{-6}$, 1) &   0.00436 $^{+0.00076}_{-0.00062}$ \\
Amplitude of the GP for OSN1            & [Relative flux] & $J$(10$^{-6}$, 1) &  0.0019 $^{+0.0039}_{-0.0013}$ \\
Amplitude of the GP for OSN2            & [Relative flux] & $J$(10$^{-6}$, 1) &  0.00108 $^{+0.0013}_{-0.00037}$ \\
Amplitude of the GP for OSN3            & [Relative flux] & $J$(10$^{-6}$, 1) &  0.00102 $^{+0.00076}_{-0.00022}$ \\
Amplitude of the GP for OSN4            & [Relative flux] & $J$(10$^{-6}$, 1) &  0.00076 $^{+0.00058}_{-0.00022}$ \\
\hline
\end{tabular}
\tablefoot{$U(a, b)$: A uniform distribution defined between a lower $a$ and upper $b$ limit. $J(a, b)$: Jeffreys (or log-uniform) distribution defined between a lower $a$ and upper $b$ limit.}
\end{table*}

\newpage
\clearpage
\begin{figure*}
  \hspace{-2cm}\includegraphics[width=1.2\textwidth]{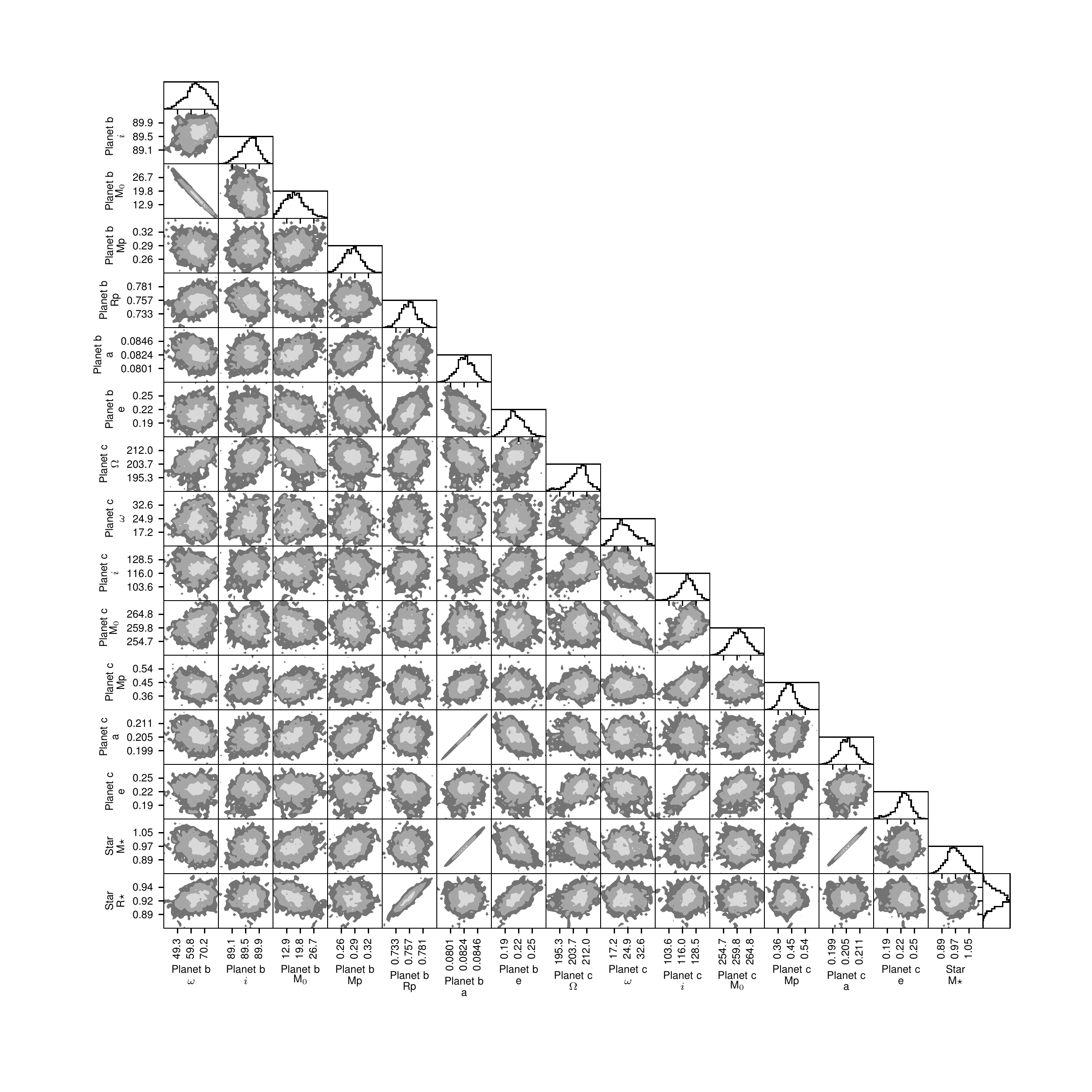}
  \vspace{-2cm}\caption{Two-parameter joint posterior distributions for the most relevant model parameters from the photodynamical modelling (Section~\ref{section.photodynamical}). The 39.3, 86.5, and 98.9\% two-variable joint confidence regions are denoted by three different grey levels; in the case of a Gaussian posterior, these regions project on to the one-dimensional 1, 2, and 3~$\sigma$ intervals. The histogram of the marginal distribution for each parameter is shown at the top of each column, except for the parameter on the last line, which is shown at the end of the line. Units are the same as in Table~\ref{table.results}.} \label{fig.pyramid}
\end{figure*}

\newpage
\clearpage
\begin{figure*}
  \centering
  \includegraphics[width=0.88\textwidth]{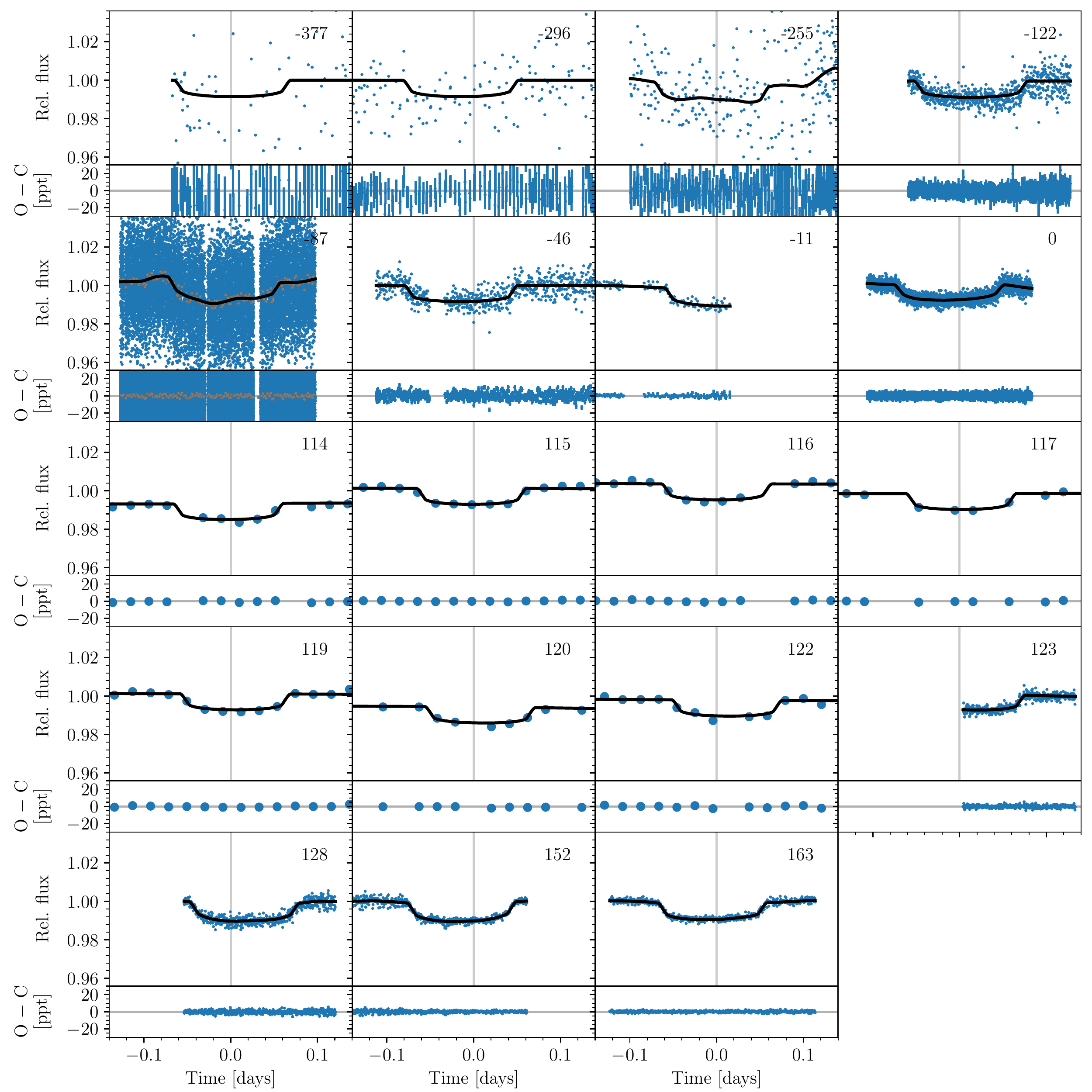}
  \caption{Transits of WASP-148~b (blue points) and the MAP model (black line) from the photodynamical modelling (Section~\ref{section.photodynamical}). Each panel is centred at the linear ephemeris (indicated by the vertical grey lines, and reported in the caption of Fig.~\ref{fig.TTVs}). For the SANCHEZ transit, 90-second binned data are shown in grey in addition to the observed data points. Each panel is labelled with the epoch; zero is the transit at $t_{\mathrm{ref}}$. In the lower part of each panel, the residuals after subtracting the MAP model are shown.} \label{fig.PH}
\end{figure*}
\begin{figure*}
  \centering
  \includegraphics[width=0.9\textwidth]{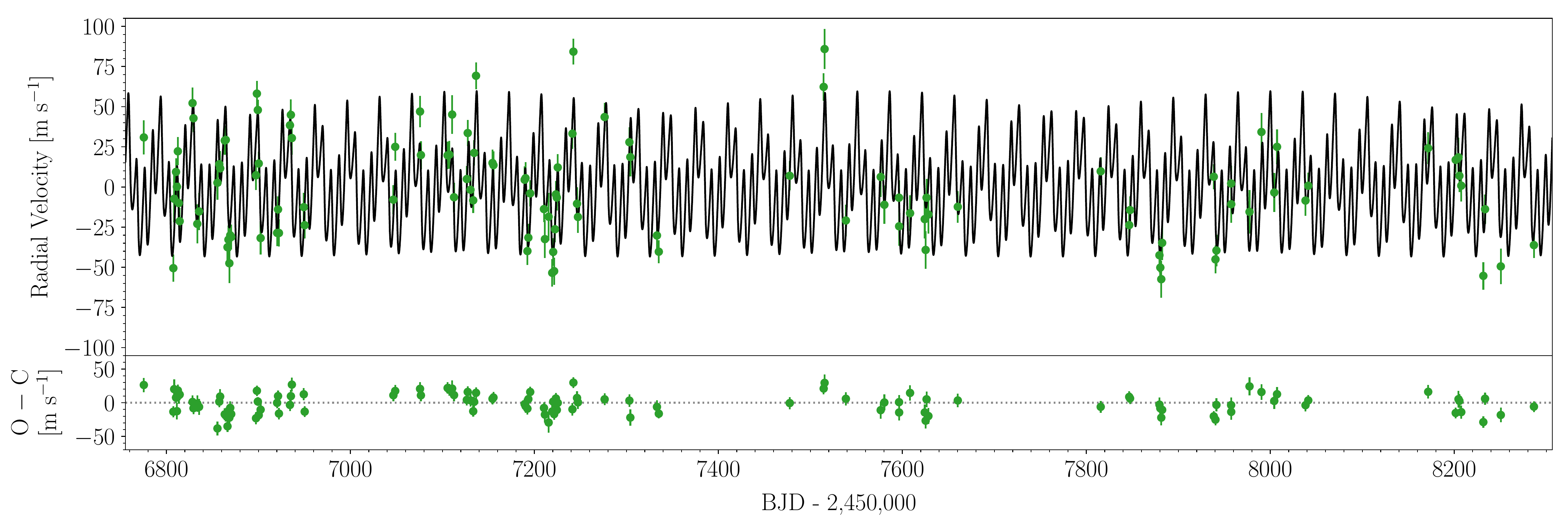}
  \caption{SOPHIE radial velocities of WASP-148 (green error bars) and the MAP model (black line) from the photodynamical modelling (Section~\ref{section.photodynamical}). In the lower panel, the residuals after subtracting the MAP model are shown.} \label{fig.RV}
\end{figure*}

\newpage
\clearpage

\begin{table*}
    \scriptsize
\caption{Planet~b transit-time predictions from 2021 to 2026.} \label{table.transits}
\begin{tabular}{lccc}
\hline
Date & Mid-transit time & \multicolumn{2}{c}{68.3\% CI} \\
 &  [BJD$_{\rm TDB}$] & [days] & [days] \\
\hline
2021-01-01 & 2459216.4351 & -0.0014 & +0.0012 \\
2021-01-10 & 2459225.2363 & -0.0012 & +0.0011 \\
2021-01-19 & 2459234.0382 & -0.0011 & +0.0011 \\
2021-01-28 & 2459242.8399 & -0.0012 & +0.0011 \\
2021-02-06 & 2459251.6440 & -0.0010 & +0.0009 \\
2021-02-14 & 2459260.4460 & -0.0008 & +0.0007 \\
2021-02-23 & 2459269.2488 & -0.0006 & +0.0007 \\
2021-03-04 & 2459278.0509 & -0.0006 & +0.0006 \\
2021-03-13 & 2459286.8557 & -0.0005 & +0.0005 \\
2021-03-22 & 2459295.6588 & -0.0004 & +0.0004 \\
2021-03-30 & 2459304.4627 & -0.0004 & +0.0004 \\
2021-04-08 & 2459313.2654 & -0.0004 & +0.0005 \\
2021-04-17 & 2459322.0709 & -0.0004 & +0.0004 \\
2021-04-26 & 2459330.8749 & -0.0004 & +0.0005 \\
2021-05-05 & 2459339.6799 & -0.0004 & +0.0005 \\
2021-05-13 & 2459348.4830 & -0.0004 & +0.0005 \\
2021-05-22 & 2459357.2891 & -0.0005 & +0.0004 \\
2021-05-31 & 2459366.0938 & -0.0004 & +0.0004 \\
2021-06-09 & 2459374.8997 & -0.0004 & +0.0004 \\
2021-06-18 & 2459383.7032 & -0.0004 & +0.0004 \\
2021-06-27 & 2459392.5095 & -0.0003 & +0.0003 \\
2021-07-05 & 2459401.3150 & -0.0003 & +0.0003 \\
2021-07-14 & 2459410.1215 & -0.0003 & +0.0004 \\
2021-07-23 & 2459418.9253 & -0.0004 & +0.0004 \\
2021-08-01 & 2459427.7316 & -0.0004 & +0.0004 \\
2021-08-10 & 2459436.5372 & -0.0004 & +0.0004 \\
2021-08-18 & 2459445.3440 & -0.0004 & +0.0005 \\
2021-08-27 & 2459454.1481 & -0.0004 & +0.0005 \\
2021-09-05 & 2459462.9537 & -0.0004 & +0.0004 \\
2021-09-14 & 2459471.7591 & -0.0004 & +0.0005 \\
2021-09-23 & 2459480.5656 & -0.0004 & +0.0004 \\
2021-10-01 & 2459489.3695 & -0.0004 & +0.0005 \\
2021-10-10 & 2459498.1741 & -0.0004 & +0.0005 \\
2021-10-19 & 2459506.9785 & -0.0005 & +0.0005 \\
2021-10-28 & 2459515.7840 & -0.0005 & +0.0005 \\
2021-11-06 & 2459524.5876 & -0.0005 & +0.0005 \\
2021-11-14 & 2459533.3907 & -0.0005 & +0.0005 \\
2021-11-23 & 2459542.1940 & -0.0006 & +0.0006 \\
2021-12-02 & 2459550.9982 & -0.0008 & +0.0008 \\
2021-12-11 & 2459559.8014 & -0.0008 & +0.0008 \\
2021-12-20 & 2459568.6032 & -0.0009 & +0.0009 \\
2021-12-28 & 2459577.4054 & -0.0011 & +0.0010 \\
2022-01-06 & 2459586.2083 & -0.0013 & +0.0012 \\
2022-01-15 & 2459595.0113 & -0.0014 & +0.0012 \\
2022-01-24 & 2459603.8124 & -0.0014 & +0.0013 \\
2022-02-02 & 2459612.6139 & -0.0015 & +0.0014 \\
2022-02-10 & 2459621.4159 & -0.0018 & +0.0016 \\
2022-02-19 & 2459630.2190 & -0.0017 & +0.0015 \\
2022-02-28 & 2459639.0200 & -0.0017 & +0.0015 \\
2022-03-09 & 2459647.8215 & -0.0017 & +0.0015 \\
2022-03-18 & 2459656.6231 & -0.0018 & +0.0017 \\
2022-03-26 & 2459665.4267 & -0.0016 & +0.0015 \\
2022-04-04 & 2459674.2281 & -0.0014 & +0.0014 \\
2022-04-13 & 2459683.0302 & -0.0012 & +0.0013 \\
2022-04-22 & 2459691.8320 & -0.0013 & +0.0013 \\
2022-05-01 & 2459700.6363 & -0.0010 & +0.0012 \\
2022-05-09 & 2459709.4387 & -0.0008 & +0.0009 \\
2022-05-18 & 2459718.2418 & -0.0007 & +0.0008 \\
2022-05-27 & 2459727.0441 & -0.0007 & +0.0008 \\
2022-06-05 & 2459735.8492 & -0.0006 & +0.0006 \\
2022-06-14 & 2459744.6525 & -0.0005 & +0.0005 \\
2022-06-22 & 2459753.4568 & -0.0005 & +0.0005 \\
2022-07-01 & 2459762.2596 & -0.0005 & +0.0006 \\
2022-07-10 & 2459771.0653 & -0.0005 & +0.0005 \\
2022-07-19 & 2459779.8695 & -0.0005 & +0.0005 \\
2022-07-28 & 2459788.6748 & -0.0005 & +0.0005 \\
2022-08-05 & 2459797.4781 & -0.0005 & +0.0005 \\
2022-08-14 & 2459806.2842 & -0.0005 & +0.0004 \\
2022-08-23 & 2459815.0892 & -0.0005 & +0.0005 \\
2022-09-01 & 2459823.8953 & -0.0004 & +0.0004 \\
2022-09-10 & 2459832.6990 & -0.0004 & +0.0005 \\
2022-09-19 & 2459841.5053 & -0.0004 & +0.0004 \\
2022-09-27 & 2459850.3109 & -0.0004 & +0.0004 \\
2022-10-06 & 2459859.1176 & -0.0004 & +0.0005 \\
2022-10-15 & 2459867.9216 & -0.0005 & +0.0006 \\
2022-10-24 & 2459876.7277 & -0.0005 & +0.0005 \\
2022-11-02 & 2459885.5334 & -0.0005 & +0.0005 \\
2022-11-10 & 2459894.3402 & -0.0005 & +0.0006 \\
2022-11-19 & 2459903.1443 & -0.0006 & +0.0006 \\
2022-11-28 & 2459911.9496 & -0.0005 & +0.0006 \\
2022-12-07 & 2459920.7548 & -0.0005 & +0.0006 \\
2022-12-16 & 2459929.5610 & -0.0005 & +0.0005 \\
2022-12-24 & 2459938.3649 & -0.0005 & +0.0006 \\
\hline
\end{tabular}
\begin{tabular}{lccc}
\hline
Date & Mid-transit time & \multicolumn{2}{c}{68.3\% CI} \\
 &  [BJD$_{\rm TDB}$] & [days] & [days] \\
\hline
2023-01-02 & 2459947.1690 & -0.0005 & +0.0005 \\
2023-01-11 & 2459955.9730 & -0.0006 & +0.0006 \\
2023-01-20 & 2459964.7781 & -0.0006 & +0.0006 \\
2023-01-29 & 2459973.5816 & -0.0007 & +0.0006 \\
2023-02-06 & 2459982.3843 & -0.0008 & +0.0007 \\
2023-02-15 & 2459991.1872 & -0.0009 & +0.0009 \\
2023-02-24 & 2459999.9909 & -0.0011 & +0.0011 \\
2023-03-05 & 2460008.7940 & -0.0012 & +0.0011 \\
2023-03-14 & 2460017.5955 & -0.0014 & +0.0012 \\
2023-03-22 & 2460026.3974 & -0.0016 & +0.0014 \\
2023-03-31 & 2460035.1999 & -0.0018 & +0.0017 \\
2023-04-09 & 2460044.0029 & -0.0018 & +0.0017 \\
2023-04-18 & 2460052.8039 & -0.0019 & +0.0017 \\
2023-04-27 & 2460061.6053 & -0.0020 & +0.0018 \\
2023-05-05 & 2460070.4070 & -0.0022 & +0.0020 \\
2023-05-14 & 2460079.2103 & -0.0021 & +0.0019 \\
2023-05-23 & 2460088.0113 & -0.0020 & +0.0019 \\
2023-06-01 & 2460096.8130 & -0.0019 & +0.0019 \\
2023-06-10 & 2460105.6146 & -0.0020 & +0.0020 \\
2023-06-18 & 2460114.4184 & -0.0017 & +0.0018 \\
2023-06-27 & 2460123.2200 & -0.0015 & +0.0016 \\
2023-07-06 & 2460132.0224 & -0.0013 & +0.0016 \\
2023-07-15 & 2460140.8243 & -0.0013 & +0.0015 \\
2023-07-24 & 2460149.6289 & -0.0011 & +0.0013 \\
2023-08-01 & 2460158.4316 & -0.0009 & +0.0010 \\
2023-08-10 & 2460167.2351 & -0.0009 & +0.0009 \\
2023-08-19 & 2460176.0375 & -0.0009 & +0.0009 \\
2023-08-28 & 2460184.8428 & -0.0007 & +0.0007 \\
2023-09-06 & 2460193.6464 & -0.0006 & +0.0006 \\
2023-09-14 & 2460202.4510 & -0.0006 & +0.0006 \\
2023-09-23 & 2460211.2539 & -0.0007 & +0.0007 \\
2023-10-02 & 2460220.0598 & -0.0006 & +0.0006 \\
2023-10-11 & 2460228.8643 & -0.0006 & +0.0006 \\
2023-10-20 & 2460237.6699 & -0.0006 & +0.0006 \\
2023-10-28 & 2460246.4732 & -0.0007 & +0.0007 \\
2023-11-06 & 2460255.2795 & -0.0006 & +0.0006 \\
2023-11-15 & 2460264.0848 & -0.0006 & +0.0006 \\
2023-11-24 & 2460272.8911 & -0.0006 & +0.0006 \\
2023-12-03 & 2460281.6949 & -0.0006 & +0.0007 \\
2023-12-12 & 2460290.5013 & -0.0006 & +0.0006 \\
2023-12-20 & 2460299.3070 & -0.0006 & +0.0007 \\
2023-12-29 & 2460308.1138 & -0.0006 & +0.0007 \\
2024-01-07 & 2460316.9178 & -0.0006 & +0.0008 \\
2024-01-16 & 2460325.7238 & -0.0006 & +0.0007 \\
2024-01-25 & 2460334.5294 & -0.0007 & +0.0007 \\
2024-02-02 & 2460343.3361 & -0.0007 & +0.0007 \\
2024-02-11 & 2460352.1402 & -0.0007 & +0.0008 \\
2024-02-20 & 2460360.9452 & -0.0007 & +0.0007 \\
2024-02-29 & 2460369.7500 & -0.0007 & +0.0007 \\
2024-03-09 & 2460378.5560 & -0.0006 & +0.0006 \\
2024-03-17 & 2460387.3598 & -0.0007 & +0.0007 \\
2024-03-26 & 2460396.1633 & -0.0007 & +0.0007 \\
2024-04-04 & 2460404.9670 & -0.0008 & +0.0007 \\
2024-04-13 & 2460413.7717 & -0.0009 & +0.0009 \\
2024-04-22 & 2460422.5750 & -0.0010 & +0.0009 \\
2024-04-30 & 2460431.3772 & -0.0011 & +0.0011 \\
2024-05-09 & 2460440.1798 & -0.0014 & +0.0013 \\
2024-05-18 & 2460448.9831 & -0.0017 & +0.0016 \\
2024-05-27 & 2460457.7861 & -0.0017 & +0.0016 \\
2024-06-05 & 2460466.5873 & -0.0019 & +0.0017 \\
2024-06-13 & 2460475.3890 & -0.0021 & +0.0020 \\
2024-06-22 & 2460484.1912 & -0.0024 & +0.0022 \\
2024-07-01 & 2460492.9942 & -0.0023 & +0.0021 \\
2024-07-10 & 2460501.7951 & -0.0023 & +0.0022 \\
2024-07-19 & 2460510.5965 & -0.0024 & +0.0023 \\
2024-07-27 & 2460519.3981 & -0.0025 & +0.0026 \\
2024-08-05 & 2460528.2014 & -0.0023 & +0.0024 \\
2024-08-14 & 2460537.0027 & -0.0022 & +0.0023 \\
2024-08-23 & 2460545.8045 & -0.0020 & +0.0022 \\
2024-09-01 & 2460554.6061 & -0.0021 & +0.0023 \\
2024-09-09 & 2460563.4101 & -0.0018 & +0.0021 \\
2024-09-18 & 2460572.2121 & -0.0017 & +0.0018 \\
2024-09-27 & 2460581.0148 & -0.0014 & +0.0017 \\
2024-10-06 & 2460589.8169 & -0.0014 & +0.0016 \\
2024-10-15 & 2460598.6217 & -0.0013 & +0.0014 \\
2024-10-23 & 2460607.4246 & -0.0011 & +0.0011 \\
2024-11-01 & 2460616.2285 & -0.0010 & +0.0010 \\
2024-11-10 & 2460625.0311 & -0.0010 & +0.0010 \\
2024-11-19 & 2460633.8365 & -0.0009 & +0.0009 \\
2024-11-28 & 2460642.6405 & -0.0008 & +0.0008 \\
2024-12-06 & 2460651.4453 & -0.0008 & +0.0008 \\
2024-12-15 & 2460660.2484 & -0.0009 & +0.0009 \\
2024-12-24 & 2460669.0544 & -0.0008 & +0.0007 \\
\hline
\end{tabular}
\begin{tabular}{lccc}
\hline
Date & Mid-transit time & \multicolumn{2}{c}{68.3\% CI} \\
 &  [BJD$_{\rm TDB}$] & [days] & [days] \\
\hline
2025-01-02 & 2460677.8592 & -0.0008 & +0.0007 \\
2025-01-11 & 2460686.6651 & -0.0007 & +0.0008 \\
2025-01-19 & 2460695.4686 & -0.0008 & +0.0008 \\
2025-01-28 & 2460704.2750 & -0.0008 & +0.0008 \\
2025-02-06 & 2460713.0804 & -0.0008 & +0.0008 \\
2025-02-15 & 2460721.8870 & -0.0008 & +0.0008 \\
2025-02-24 & 2460730.6909 & -0.0008 & +0.0009 \\
2025-03-04 & 2460739.4972 & -0.0008 & +0.0008 \\
2025-03-13 & 2460748.3030 & -0.0008 & +0.0009 \\
2025-03-22 & 2460757.1099 & -0.0008 & +0.0009 \\
2025-03-31 & 2460765.9140 & -0.0009 & +0.0010 \\
2025-04-09 & 2460774.7198 & -0.0008 & +0.0009 \\
2025-04-18 & 2460783.5252 & -0.0008 & +0.0009 \\
2025-04-26 & 2460792.3318 & -0.0008 & +0.0008 \\
2025-05-05 & 2460801.1358 & -0.0008 & +0.0009 \\
2025-05-14 & 2460809.9403 & -0.0008 & +0.0008 \\
2025-05-23 & 2460818.7448 & -0.0008 & +0.0008 \\
2025-06-01 & 2460827.5504 & -0.0007 & +0.0008 \\
2025-06-09 & 2460836.3541 & -0.0008 & +0.0008 \\
2025-06-18 & 2460845.1572 & -0.0009 & +0.0009 \\
2025-06-27 & 2460853.9604 & -0.0011 & +0.0010 \\
2025-07-06 & 2460862.7646 & -0.0013 & +0.0013 \\
2025-07-15 & 2460871.5678 & -0.0014 & +0.0014 \\
2025-07-23 & 2460880.3697 & -0.0016 & +0.0016 \\
2025-08-01 & 2460889.1719 & -0.0019 & +0.0018 \\
2025-08-10 & 2460897.9747 & -0.0023 & +0.0021 \\
2025-08-19 & 2460906.7777 & -0.0023 & +0.0022 \\
2025-08-28 & 2460915.5787 & -0.0024 & +0.0023 \\
2025-09-05 & 2460924.3803 & -0.0026 & +0.0025 \\
2025-09-14 & 2460933.1821 & -0.0029 & +0.0029 \\
2025-09-23 & 2460941.9852 & -0.0028 & +0.0028 \\
2025-10-02 & 2460950.7861 & -0.0027 & +0.0028 \\
2025-10-11 & 2460959.5876 & -0.0027 & +0.0028 \\
2025-10-19 & 2460968.3892 & -0.0028 & +0.0030 \\
2025-10-28 & 2460977.1927 & -0.0026 & +0.0029 \\
2025-11-06 & 2460985.9941 & -0.0024 & +0.0026 \\
2025-11-15 & 2460994.7961 & -0.0021 & +0.0025 \\
2025-11-24 & 2461003.5978 & -0.0022 & +0.0026 \\
2025-12-02 & 2461012.4021 & -0.0020 & +0.0023 \\
2025-12-11 & 2461021.2043 & -0.0018 & +0.0020 \\
2025-12-20 & 2461030.0074 & -0.0016 & +0.0018 \\
2025-12-29 & 2461038.8096 & -0.0015 & +0.0018 \\
2026-01-07 & 2461047.6146 & -0.0014 & +0.0015 \\
2026-01-15 & 2461056.4179 & -0.0013 & +0.0013 \\
2026-01-24 & 2461065.2221 & -0.0012 & +0.0012 \\
2026-02-02 & 2461074.0248 & -0.0012 & +0.0012 \\
2026-02-11 & 2461082.8305 & -0.0011 & +0.0010 \\
2026-02-20 & 2461091.6347 & -0.0010 & +0.0010 \\
2026-02-28 & 2461100.4399 & -0.0010 & +0.0010 \\
2026-03-09 & 2461109.2431 & -0.0011 & +0.0010 \\
2026-03-18 & 2461118.0493 & -0.0010 & +0.0010 \\
2026-03-27 & 2461126.8543 & -0.0010 & +0.0010 \\
2026-04-05 & 2461135.6604 & -0.0009 & +0.0010 \\
2026-04-13 & 2461144.4641 & -0.0010 & +0.0011 \\
2026-04-22 & 2461153.2706 & -0.0010 & +0.0010 \\
2026-05-01 & 2461162.0762 & -0.0010 & +0.0010 \\
2026-05-10 & 2461170.8830 & -0.0010 & +0.0010 \\
2026-05-19 & 2461179.6870 & -0.0011 & +0.0012 \\
2026-05-27 & 2461188.4932 & -0.0010 & +0.0011 \\
2026-06-05 & 2461197.2990 & -0.0011 & +0.0011 \\
2026-06-14 & 2461206.1059 & -0.0010 & +0.0011 \\
2026-06-23 & 2461214.9101 & -0.0011 & +0.0012 \\
2026-07-02 & 2461223.7155 & -0.0010 & +0.0010 \\
2026-07-11 & 2461232.5207 & -0.0010 & +0.0010 \\
2026-07-19 & 2461241.3270 & -0.0009 & +0.0009 \\
2026-07-28 & 2461250.1309 & -0.0010 & +0.0009 \\
2026-08-06 & 2461258.9350 & -0.0009 & +0.0009 \\
2026-08-15 & 2461267.7391 & -0.0009 & +0.0009 \\
2026-08-24 & 2461276.5442 & -0.0011 & +0.0010 \\
2026-09-01 & 2461285.3478 & -0.0012 & +0.0010 \\
2026-09-10 & 2461294.1504 & -0.0013 & +0.0013 \\
2026-09-19 & 2461302.9533 & -0.0016 & +0.0015 \\
2026-09-28 & 2461311.7569 & -0.0019 & +0.0019 \\
2026-10-07 & 2461320.5600 & -0.0020 & +0.0020 \\
2026-10-15 & 2461329.3616 & -0.0022 & +0.0022 \\
2026-10-24 & 2461338.1634 & -0.0025 & +0.0025 \\
2026-11-02 & 2461346.9659 & -0.0029 & +0.0029 \\
2026-11-11 & 2461355.7689 & -0.0029 & +0.0028 \\
2026-11-20 & 2461364.5698 & -0.0030 & +0.0029 \\
2026-11-28 & 2461373.3712 & -0.0030 & +0.0032 \\
2026-12-07 & 2461382.1729 & -0.0033 & +0.0035 \\
2026-12-16 & 2461390.9761 & -0.0032 & +0.0034 \\
2026-12-25 & 2461399.7771 & -0.0030 & +0.0033 \\
\hline
\end{tabular}
\tablefoot{The table lists: Calendar UT date (YYYY-MM-DD), mid-transit time, and 68.3\% CI.}
\end{table*}

\newpage
\clearpage

\begin{figure*}
  \includegraphics[width=0.49\textwidth]{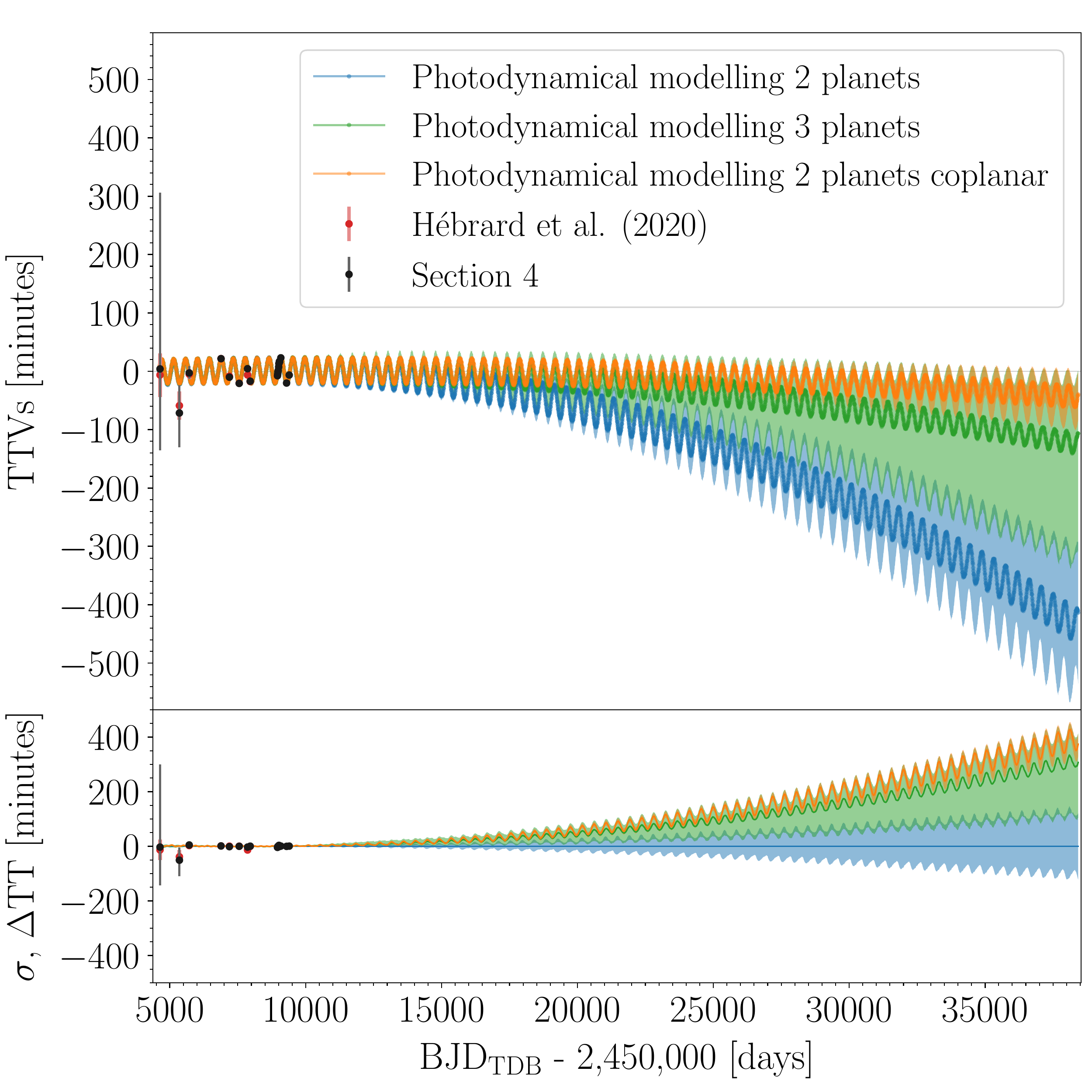}\hspace{0.1cm}\includegraphics[width=0.49\textwidth]{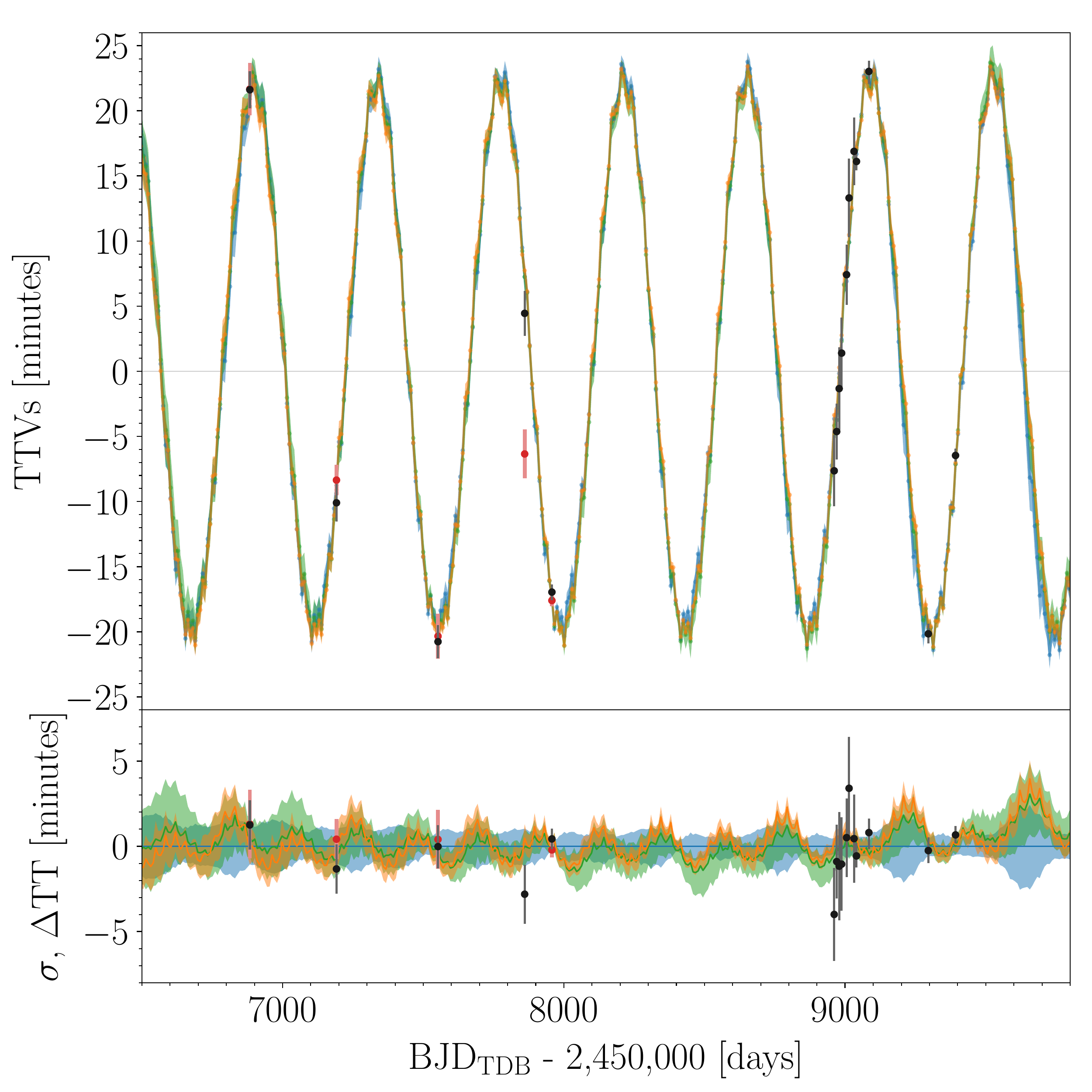}
  \caption{Comparison of TTVs for different models. Left: Idem.\ as Figure~\ref{fig.TTVs}, but up to the year 2100 with the two-planet photodynamical modelling (Section~\ref{section.photodynamical}), three-planet photodynamical modelling (Section~\ref{sec.3planets}), and coplanar two-planet photodynamical modelling (Section~\ref{section.mutualInclination}). Right: Zoom of the plot on the left.} \label{fig.TTVs2100}
\end{figure*}

\begin{figure*}
  \includegraphics[width=0.49\textwidth]{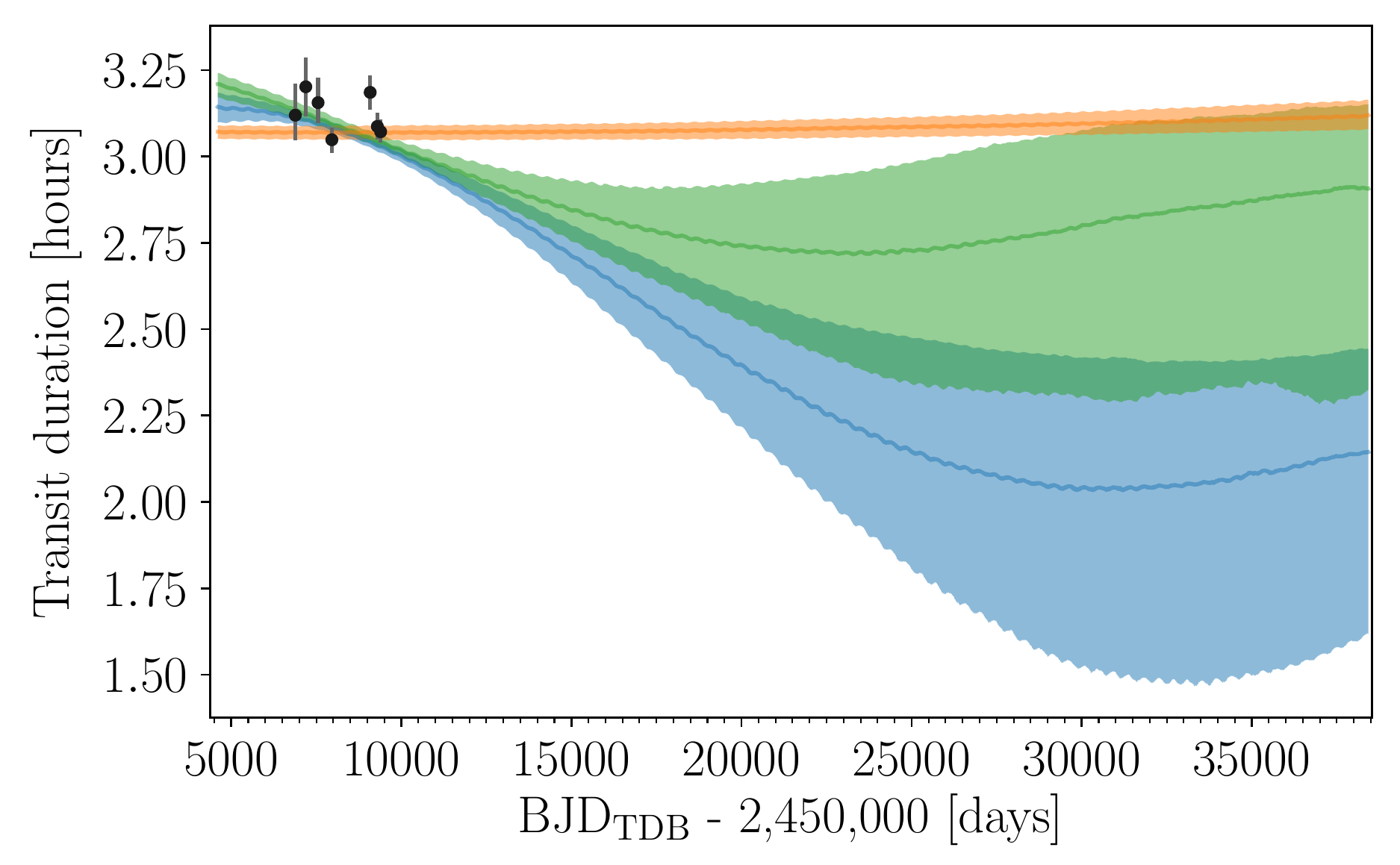}\hspace{0.1cm}\includegraphics[width=0.49\textwidth]{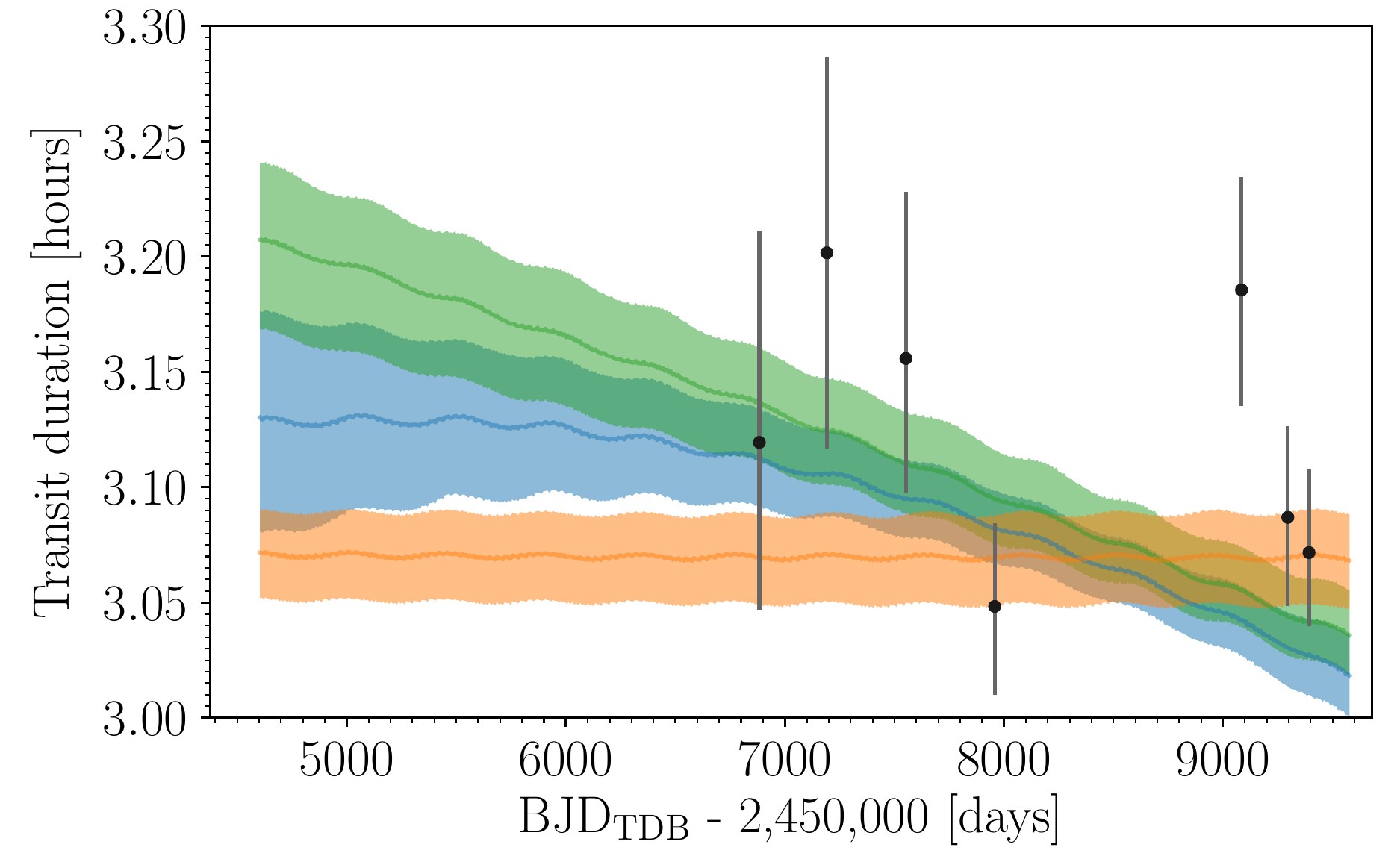}
  \caption{Comparison of transit duration variations for different models. Left: Posterior transit duration variations of planet~b up to the end of the year 2100 from 1,000 random draws from the posterior distribution of the two-planet photodynamical modelling (Section~\ref{section.photodynamical}), three-planet photodynamical modelling (Section~\ref{sec.3planets}), and coplanar two-planet photodynamical modelling (Section~\ref{section.mutualInclination}). The colours correspond with the ones used in Figure~\ref{fig.TTVs2100}. The points with error bars are the durations measured on individual transits (Section~\ref{section.mutualInclination}). Right: Zoom of the plot on the left.} \label{fig.TDVs}
\end{figure*}

\begin{table}
\tiny
\renewcommand{\arraystretch}{1.24}
\centering
\caption{Inferred system parameters with the photodynamical modelling assuming coplanar orbits.}\label{table.coplanar}
\begin{tabular}{lccc}
\hline
Parameter & & Median   \\
&  & and 68.3\% CI \\
\hline

\emph{\bf Star} \\
Mass, $M_\star$                           & (\Msun)                 & 0.968 $\pm$ 0.050 \\
Radius, $R_\star$                         & (\Rnom)                 & 0.918 $\pm$ 0.017 \\
Stellar mean density, $\rho_{\star}$      & [$\mathrm{g\;cm^{-3}}$] & 1.76 $^{+0.14}_{-0.12}$ \smallskip\\

\emph{\bf Planet~b} \\
Semi-major axis, $a$                      & [au]                    & 0.0825 $\pm$ 0.0015 \\
Eccentricity, $e$                         &                         & 0.194 $\pm$ 0.035 \\
Argument of pericentre, $\omega$          & [\degree]               & 61.4 $\pm$ 7.2 \\
Inclination, $i$                          & [\degree]               & 89.06 $^{+0.43}_{-0.31}$ \\
Longitude of the ascending node, $\Omega$ & [\degree]               & 180 (fixed at $t_{\mathrm{ref}}$) \\
Mean anomaly, $M_0$                       & [\degree]               & 19.0 $^{+6.1}_{-5.0}$ \\
Impact parameter, $b$                     &                         & 0.259 $^{+0.096}_{-0.12}$ \smallskip\\

Mass, $M_{\mathrm{p}}$                    &(\Mjup)                  & 0.289 $\pm$ 0.023 \\
Radius, $R_{\mathrm{p}}$                  &(\RJnom)                 & 0.756 $\pm$ 0.017 \\
Planet mean density, $\rho_{\mathrm{p}}$  &[$\mathrm{g\;cm^{-3}}$]  & 0.825 $^{+0.094}_{-0.078}$ \smallskip\\

\emph{\bf Planet~c} \smallskip\\
Semi-major axis, $a$                      & [au]                    & 0.2054 $\pm$ 0.0036 \\
Eccentricity, $e$                         &                         & 0.168 $^{+0.016}_{-0.010}$ \\
Argument of pericentre, $\omega$          & [\degree]               & 26.3 $\pm$ 6.0 \\
Inclination, $i$                          & [\degree]               & $\equiv$ Planet~b inclination\\
Longitude of the ascending node, $\Omega$ & [\degree]               & 180 (fixed at $t_{\mathrm{ref}}$) \\
Mean anomaly, $M_0$                       & [\degree]               & 258.1 $\pm$ 3.6 \\
Impact parameter, $b$                     &                         & 0.71 $^{+0.23}_{-0.33}$ \smallskip\\

Mass, $M_{\mathrm{p}}$                    &(\Mjup)                  & 0.375 $\pm$ 0.040 \smallskip\\

\hline
\end{tabular}
\tablefoot{The table lists: Posterior median and 68.3\% CI for the photodynamical modelling assuming coplanar orbits (Section~\ref{section.mutualInclination}). Only masses, radii, densities, and orbital parameters (at $t_{\mathrm{ref}}~=~2\;457\;957.48167$~BJD$_{\mathrm{TDB}}$) are listed.}
\end{table}

\begin{table}
\tiny
\renewcommand{\arraystretch}{1.24}
\centering
\caption{Inferred system parameters with the photodynamical modelling without stellar priors.} \label{table.absolute}
\begin{tabular}{lccc}
\hline
Parameter & & Median   \\
&  & and 68.3\% CI \\
\hline

\emph{\bf Star} \\
Mass, $M_\star$                           & (\Msun)                 & 9.7 $^{+12}_{-5.3}$ \\
Radius, $R_\star$                         & (\Rnom)                 & 2.14 $^{+0.69}_{-0.54}$ \\
Stellar mean density, $\rho_{\star}$      & [$\mathrm{g\;cm^{-3}}$] & 1.41 $^{+0.29}_{-0.22}$ \smallskip\\

\emph{\bf Planet~b} \\
Semi-major axis, $a$                      & [au]                    & 0.178 $^{+0.053}_{-0.041}$ \\
Eccentricity, $e$                         &                         & 0.254 $\pm$ 0.045 \\
Argument of pericentre, $\omega$          & [\degree]               & 65.6 $\pm$ 9.6 \\
Inclination, $i$                          & [\degree]               & 88.90 $^{+0.56}_{-0.45}$ \\
Longitude of the ascending node, $\Omega$ & [\degree]               & 180 (fixed at $t_{\mathrm{ref}}$) \\
Mean anomaly, $M_0$                       & [\degree]               & 13.9 $^{+7.0}_{-5.6}$ \smallskip\\

Mass, $M_{\mathrm{p}}$                    &(\Mjup)                  & 1.32 $^{+0.88}_{-0.53}$ \\
Radius, $R_{\mathrm{p}}$                  &(\RJnom)                 & 1.77 $^{+0.58}_{-0.45}$ \\
Planet mean density, $\rho_{\mathrm{p}}$  &[$\mathrm{g\;cm^{-3}}$]  & 0.293 $^{+0.14}_{-0.089}$ \smallskip\\

\emph{\bf Planet~c} \smallskip\\
Semi-major axis, $a$                      & [au]                    & 0.44 $^{+0.13}_{-0.10}$ \\
Eccentricity, $e$                         &                         & 0.280 $\pm$ 0.023 \\
Argument of pericentre, $\omega$          & [\degree]               & 23.6 $\pm$ 7.0 \\
Inclination, $i$                          & [\degree]               & 129 $\pm$ 13 \\
Longitude of the ascending node, $\Omega$ & [\degree]               & 214.4 $\pm$ 8.2 \\
Mean anomaly, $M_0$                       & [\degree]               & 262.1 $\pm$ 4.4 \smallskip\\

Mass, $M_{\mathrm{p}}$                    &(\Mjup)                  & 2.4 $^{+2.1}_{-1.2}$ \medskip\\

Mutual inclination, $I$                   & [\degree]               & 52 $^{+10}_{-13}$ \smallskip\\
\hline
\end{tabular}
\tablefoot{The table lists: Posterior median and 68.3\% CI for the photodynamical modelling without stellar priors (Section~\ref{section.noprior}). Only masses, radii, densities, and orbital parameters (at $t_{\mathrm{ref}}~=~2\;457\;957.48167$~BJD$_{\mathrm{TDB}}$) are listed.}
\end{table}

\begin{table}
\tiny
\renewcommand{\arraystretch}{1.25}
\centering
\caption{Inferred system parameters with the photodynamical modelling with three planets.} \label{table.3planets}
\begin{tabular}{lccc}
\hline
Parameter & & Median   \\
&  & and 68.3\% CI \\
\hline

\emph{\bf Star} \\
Mass, $M_\star$                           & (\Msun)                 & 0.976 $\pm$ 0.053 \\
Radius, $R_\star$                         & (\Rnom)                 & 0.918 $\pm$ 0.017 \\
Stellar mean density, $\rho_{\star}$      & [$\mathrm{g\;cm^{-3}}$] & 1.79 $\pm$ 0.13 \smallskip\\

\emph{\bf Planet~b} \\
Semi-major axis, $a$                      & [au]                    & 0.0828 $\pm$ 0.0015 \\
Eccentricity, $e$                         &                         & 0.183 $^{+0.028}_{-0.031}$ \\
Argument of pericentre, $\omega$          & [\degree]               & 50.7 $^{+9.8}_{-8.6}$ \\
Inclination, $i$                          & [\degree]               & 88.87 $\pm$ 0.21 \\
Longitude of the ascending node, $\Omega$ & [\degree]               & 180 (fixed at $t_{\mathrm{ref}}$) \\
Mean anomaly, $M_0$                       & [\degree]               & 27.3 $^{+6.4}_{-7.1}$ \\
Impact parameter, $b$                     &                         & 0.323 $\pm$ 0.064 \\
$T_0'$\;-\;2\;450\;000                    & [BJD$_{\mathrm{TDB}}$]  & 7957.48176 $\pm$ 0.00025 \\
$P'$                                      & [d]                     & 8.80358 $\pm$ 0.00018 \\ 
$K'$                                      & [\ms]                   & 29.1 $\pm$ 1.9 \smallskip\\

Mass, $M_{\mathrm{p}}$                    &(\Mjup)                  & 0.284 $^{+0.021}_{-0.018}$ \\
Radius, $R_{\mathrm{p}}$                  &(\RJnom)                 & 0.765 $\pm$ 0.016 \\
Planet mean density, $\rho_{\mathrm{p}}$  &[$\mathrm{g\;cm^{-3}}$]  & 0.788 $\pm$ 0.071 \smallskip\\

\emph{\bf Planet~c} \smallskip\\
Semi-major axis, $a$                      & [au]                    & 0.2059 $\pm$ 0.0037 \\
Eccentricity, $e$                         &                         & 0.196 $^{+0.023}_{-0.019}$ \\
Argument of pericentre, $\omega$          & [\degree]               & 25.4 $\pm$ 5.6 \\
Inclination, $i$                          & [\degree]               & 108 $^{+11}_{-14}$ \\
Longitude of the ascending node, $\Omega$ & [\degree]               & 188.4 $^{+4.5}_{-2.9}$ \\
Mean anomaly, $M_0$                       & [\degree]               & 259.0 $\pm$ 3.8 \\
Impact parameter, $b$                     &                         & 13.3 $^{+7.5}_{-9.6}$ \\
$T_0'$\;-\;2\;450\;000                    & [BJD$_{\mathrm{TDB}}$]  & 7971.56 $\pm$ 0.31 \\
$P'$                                      & [d]                     & 34.5441 $\pm$ 0.0026 \\ 
$K'$                                      & [\ms]                   & 25.0 $^{+2.5}_{-2.2}$ \smallskip\\

Mass, $M_{\mathrm{p}}$                    &(\Mjup)                  & 0.411 $^{+0.054}_{-0.044}$ \smallskip\\

\emph{\bf Planet~(d)} \smallskip\\
Semi-major axis, $a$                      & [au]                    & 0.551 $^{+0.010}_{-0.012}$ \\
Eccentricity, $e$                         &                         & 0.19 $^{+0.19}_{-0.13}$ \\
Argument of pericentre, $\omega$          & [\degree]               & 148 $\pm$ 96 \\
Inclination, $i$                          & [\degree]               & 82 $^{+39}_{-33}$ \\
Longitude of the ascending node, $\Omega$ & [\degree]               & 199 $^{+110}_{-150}$ \\
Mean anomaly, $M_0$                       & [\degree]               & 134 $^{+110}_{-85}$ \\
Impact parameter, $b$                     &                         & 48 $^{+40}_{-34}$ \\
$T_0'$\;-\;2\;450\;000                    & [BJD$_{\mathrm{TDB}}$]  & 7891 $^{+20}_{-11}$ \\
$P'$                                      & [d]                     & 151.2 $^{+2.1}_{-1.7}$ \\ 
$K'$                                      & [\ms]                   & 8.9 $\pm$ 2.2 \smallskip\\

Mass, $M_{\mathrm{p}}$                    &(\Mjup)                  & 0.262 $^{+0.10}_{-0.076}$ \medskip\\

Mutual inclination $b,c$, $I_{b,c}$       & [\degree]               & 21 $\pm$ 12 \\ 
Mutual inclination $b,d$, $I_{b,d}$       & [\degree]               & 93 $\pm$ 46 \\
Mutual inclination $c,d$, $I_{c,d}$       & [\degree]               & 95 $\pm$ 43 \smallskip\\
\hline
\end{tabular}
\tablefoot{The table lists: Posterior median and 68.3\% CI for the photodynamical modelling with three planets (Section~\ref{sec.3planets}). The orbital parameters are given for the reference time $t_{\mathrm{ref}}~=~2\;457\;957.48167$~BJD$_{\mathrm{TDB}}$.}
\end{table}

\newpage
\clearpage

\begin{figure*}
\includegraphics[height=3.2cm]{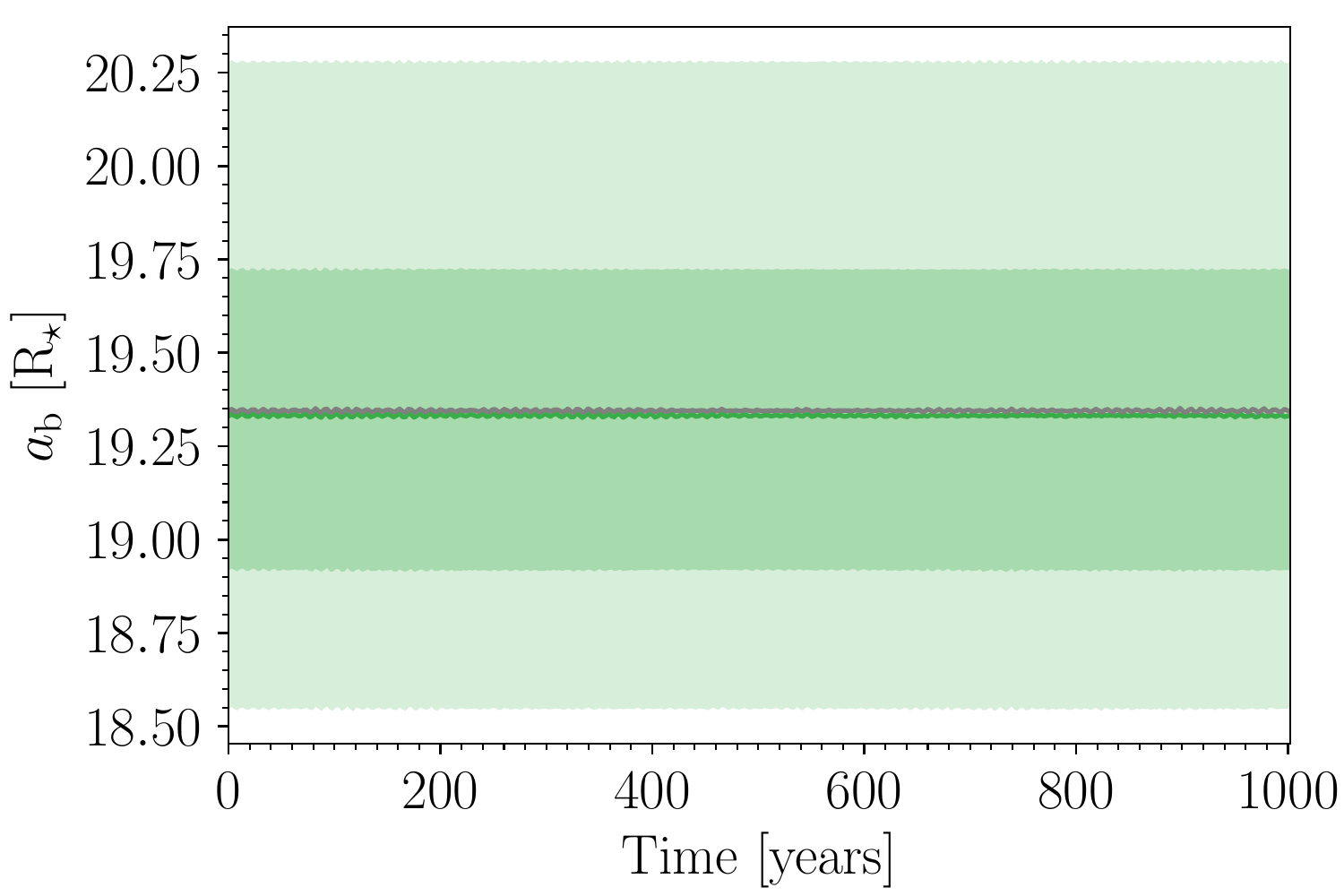}\includegraphics[height=3.2cm]{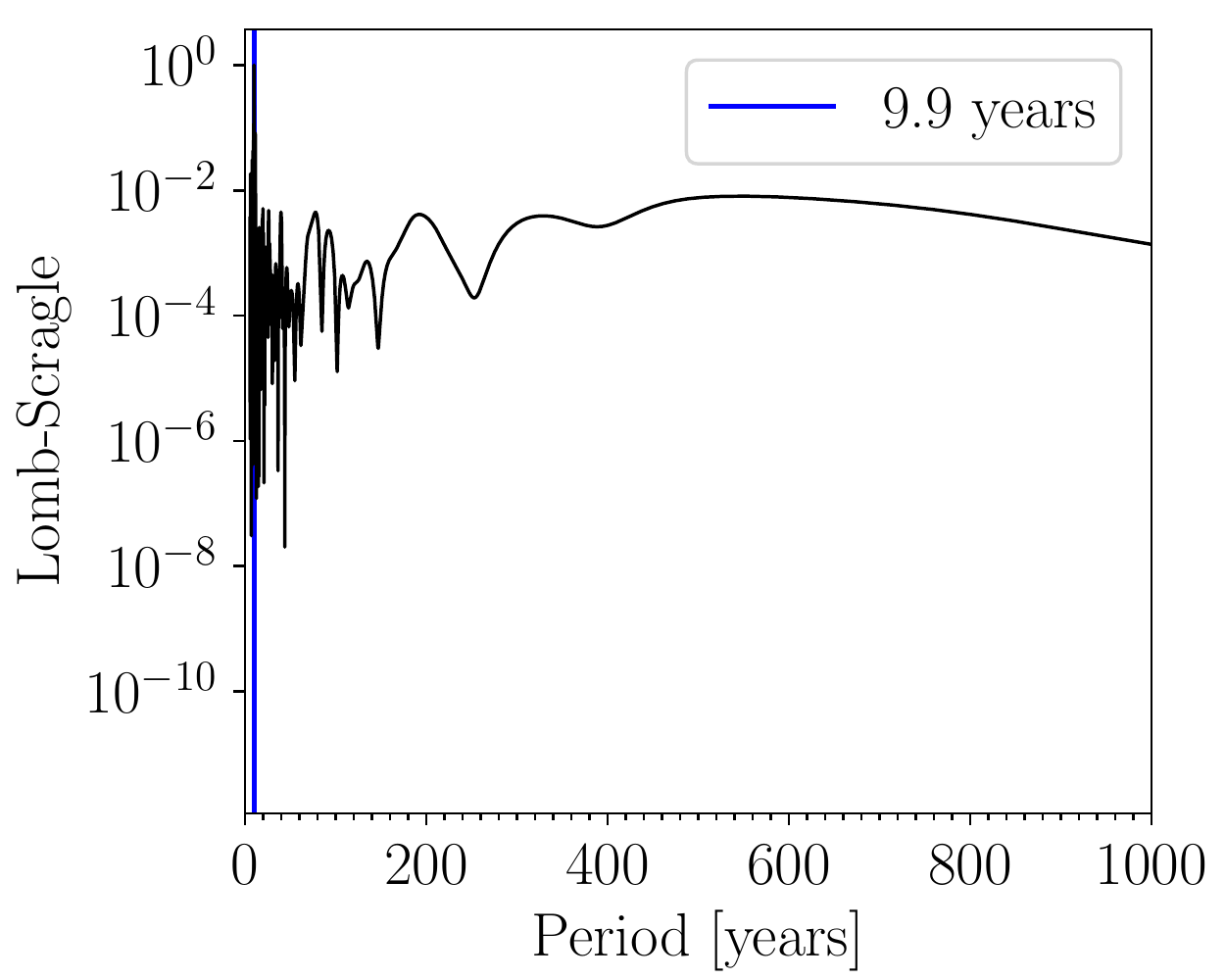} 
\includegraphics[height=3.2cm]{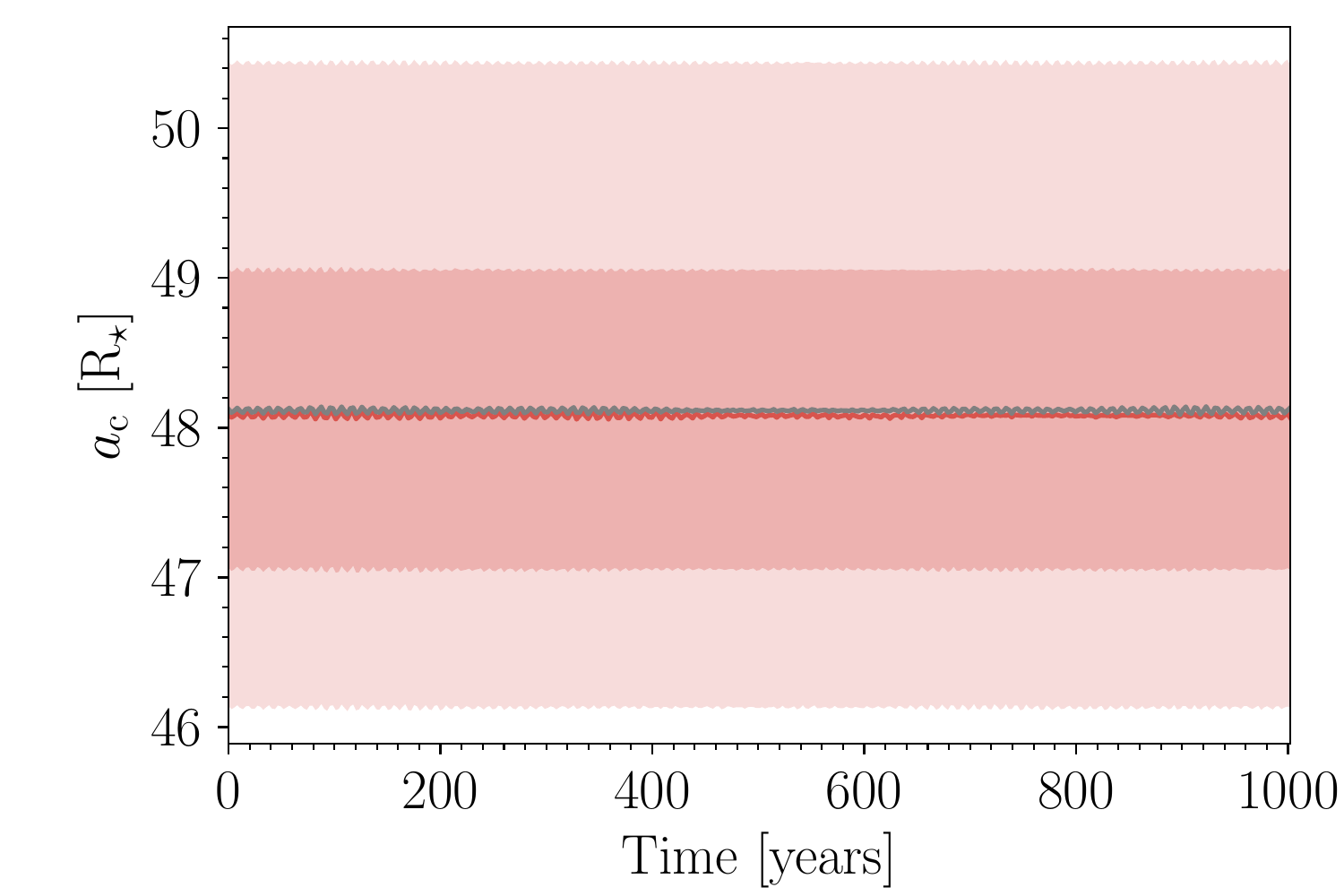}\includegraphics[height=3.2cm]{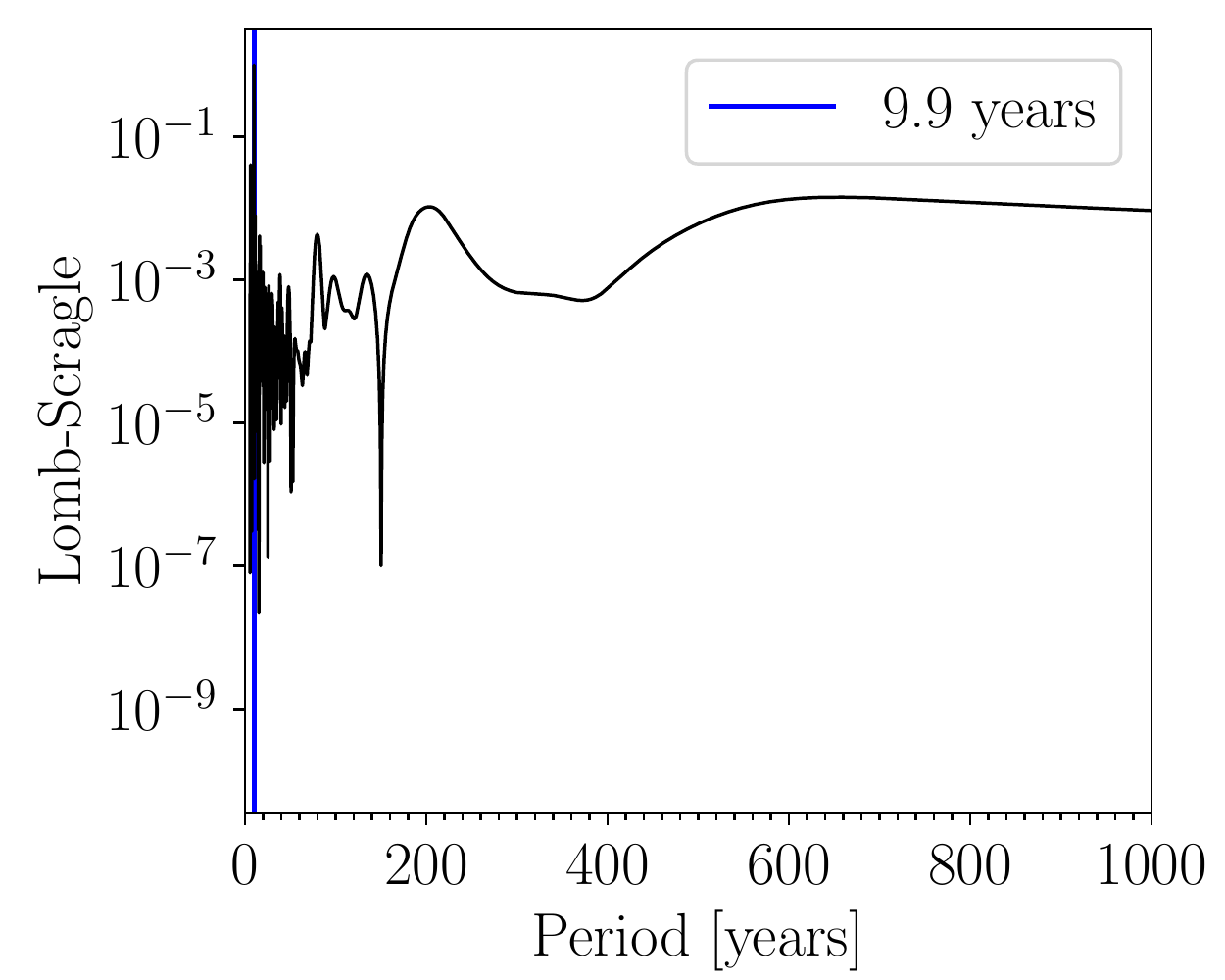}
 
\includegraphics[height=3.2cm]{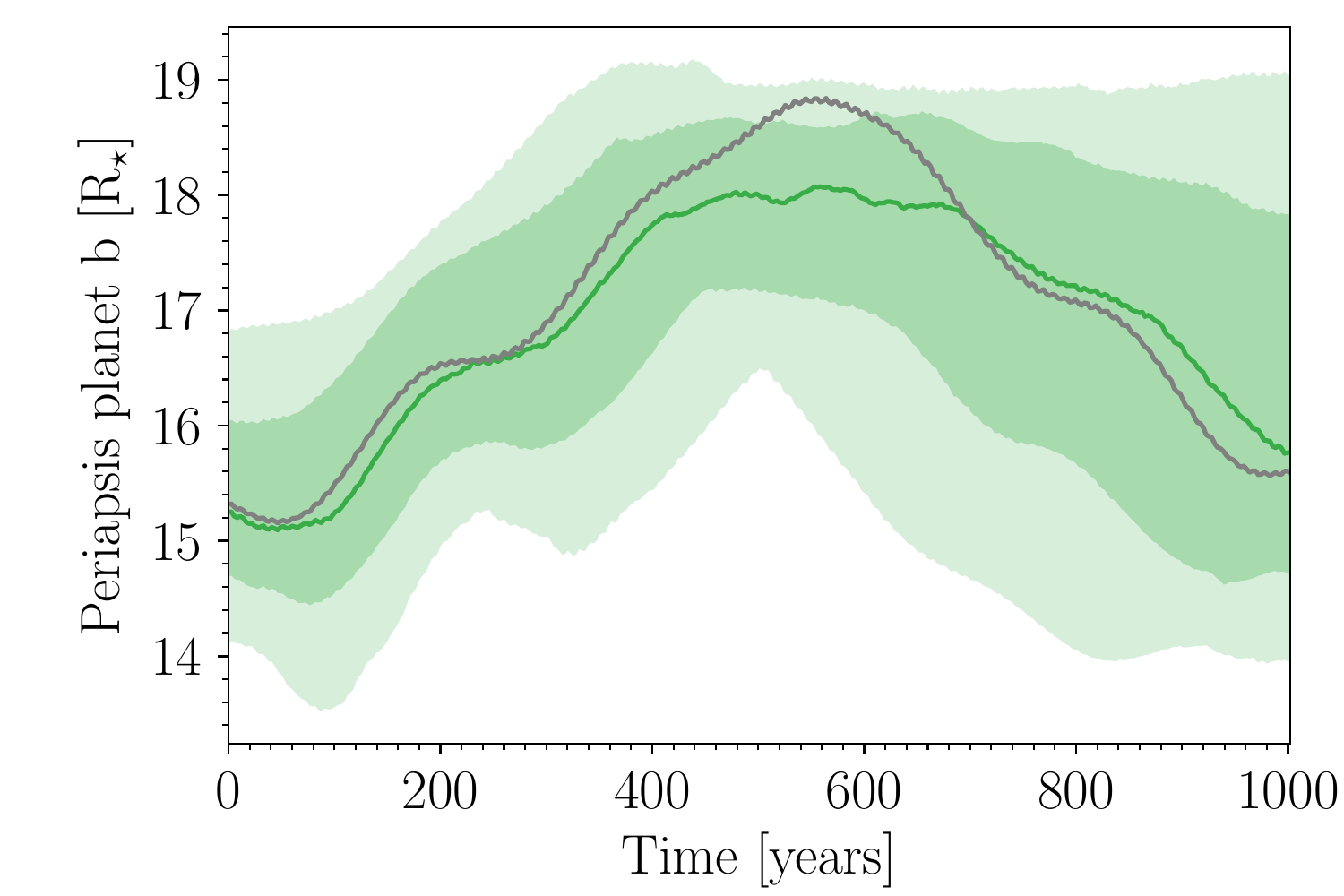}\includegraphics[height=3.2cm]{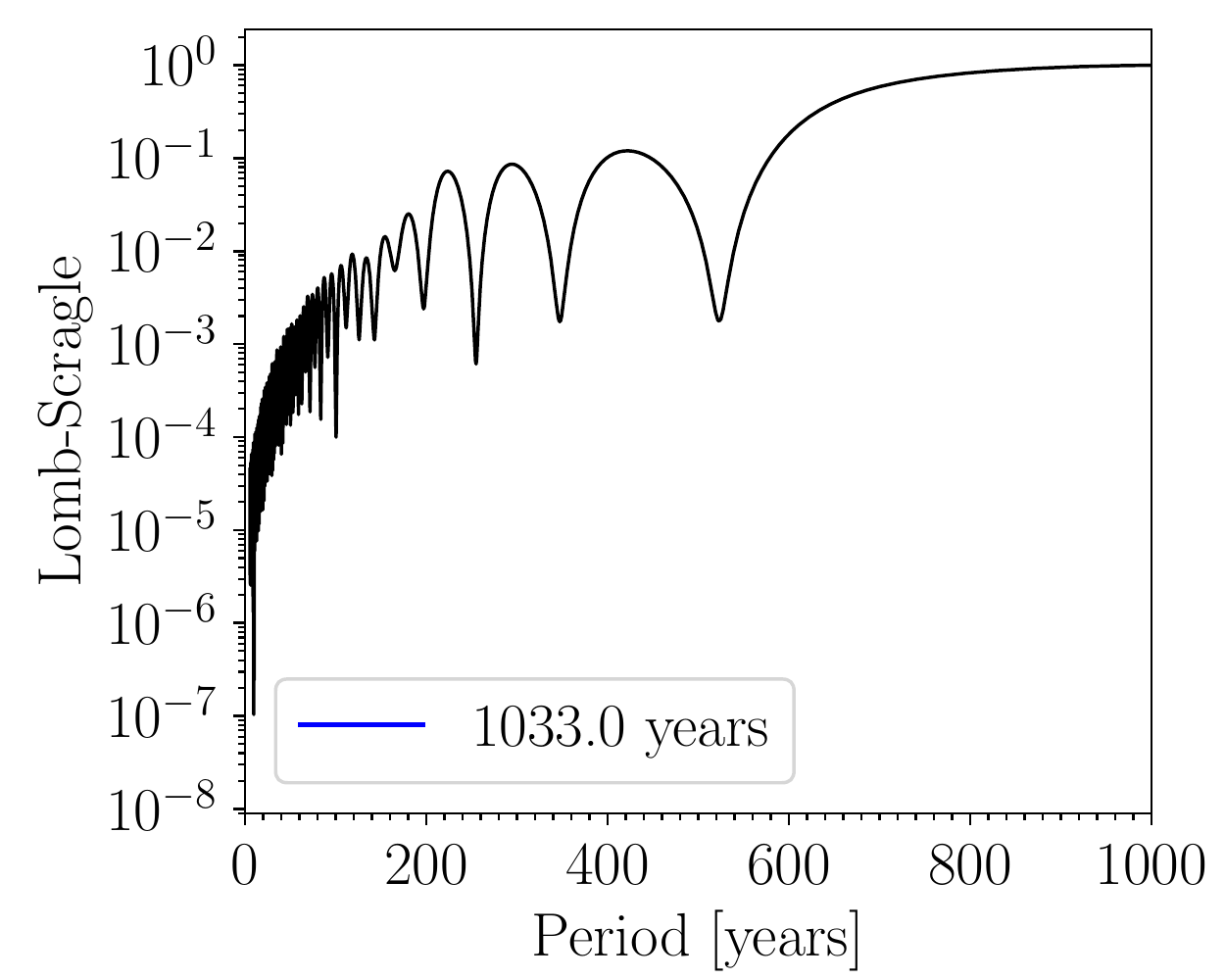}

\includegraphics[height=3.2cm]{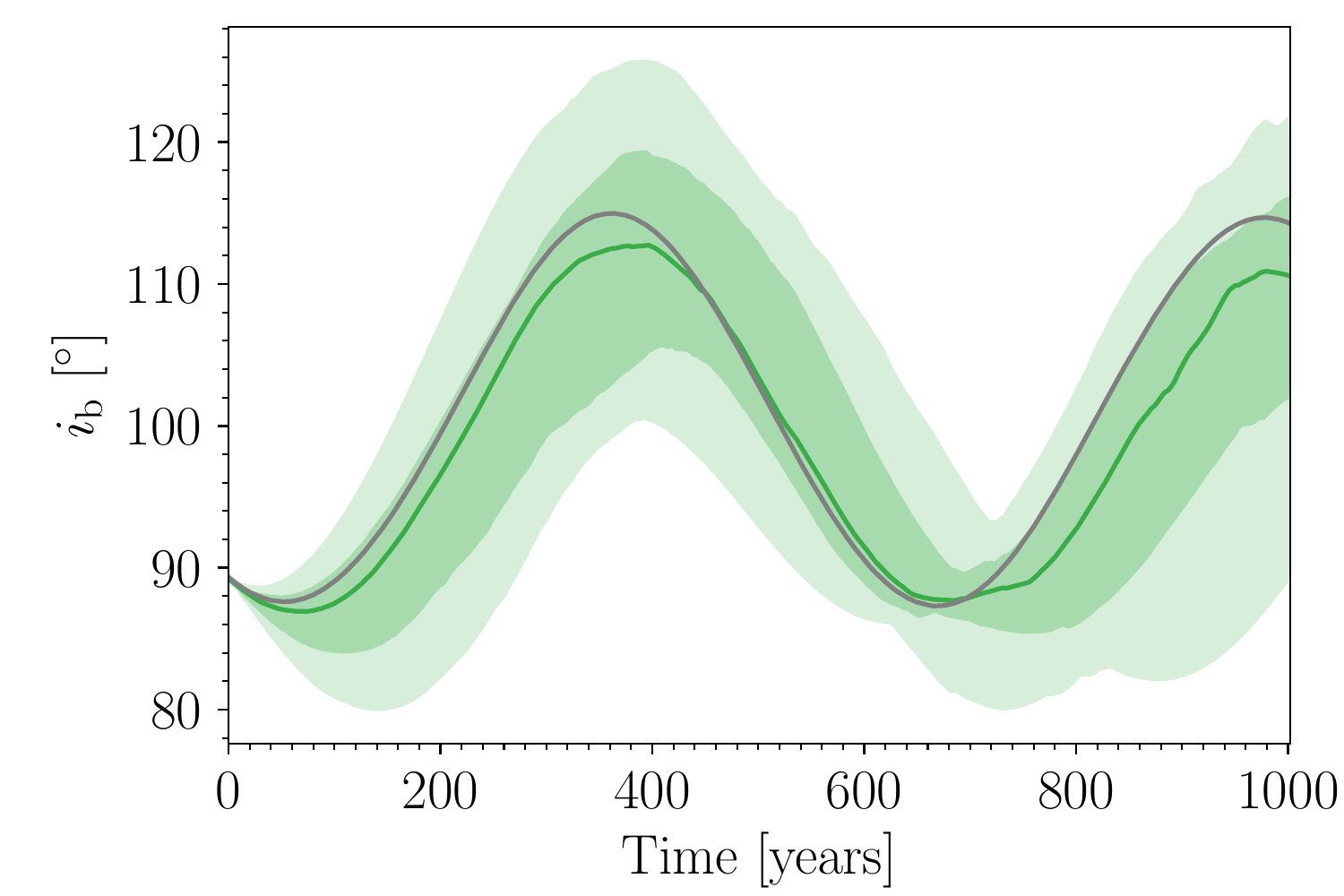}\includegraphics[height=3.2cm]{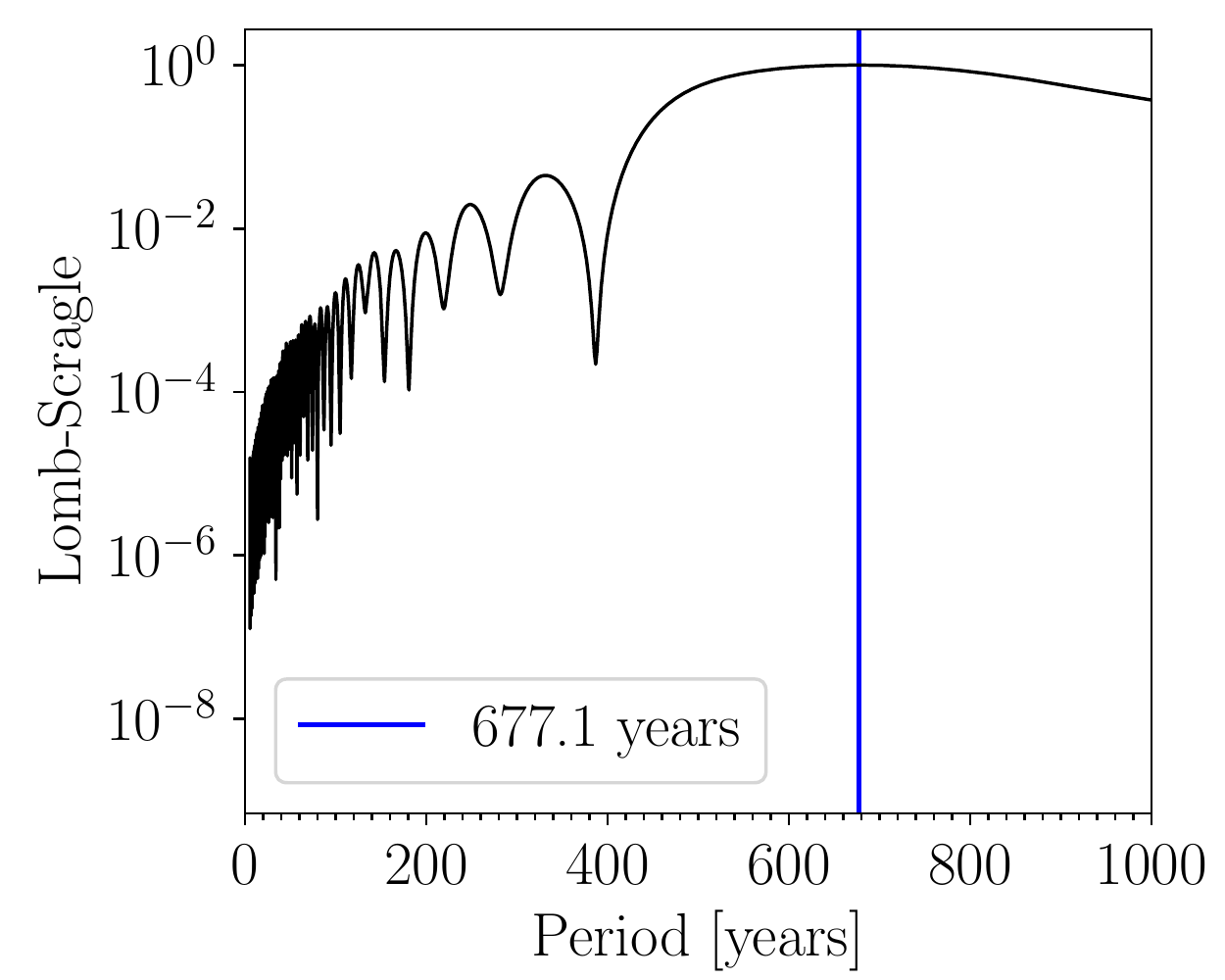}
\includegraphics[height=3.2cm]{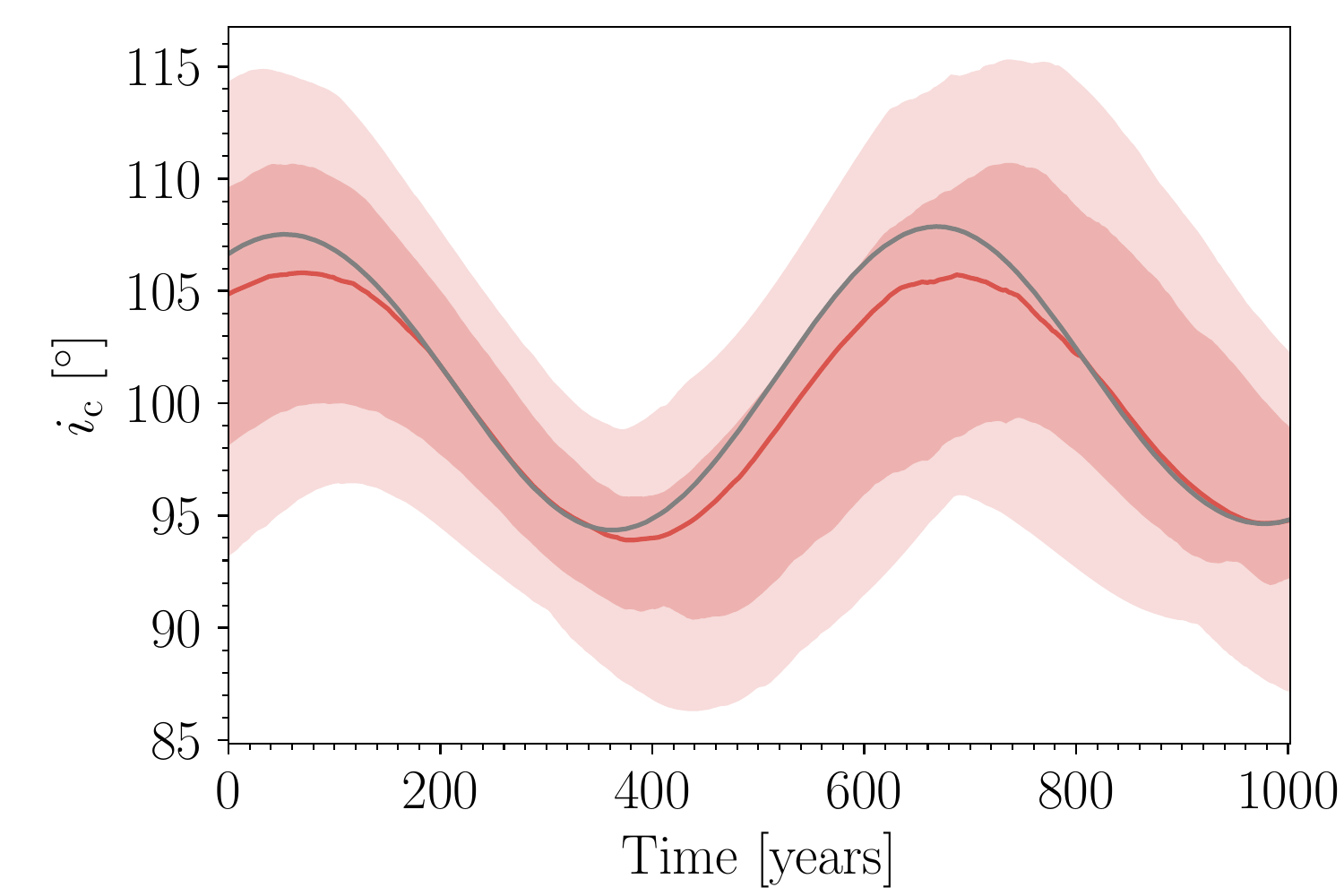}\includegraphics[height=3.2cm]{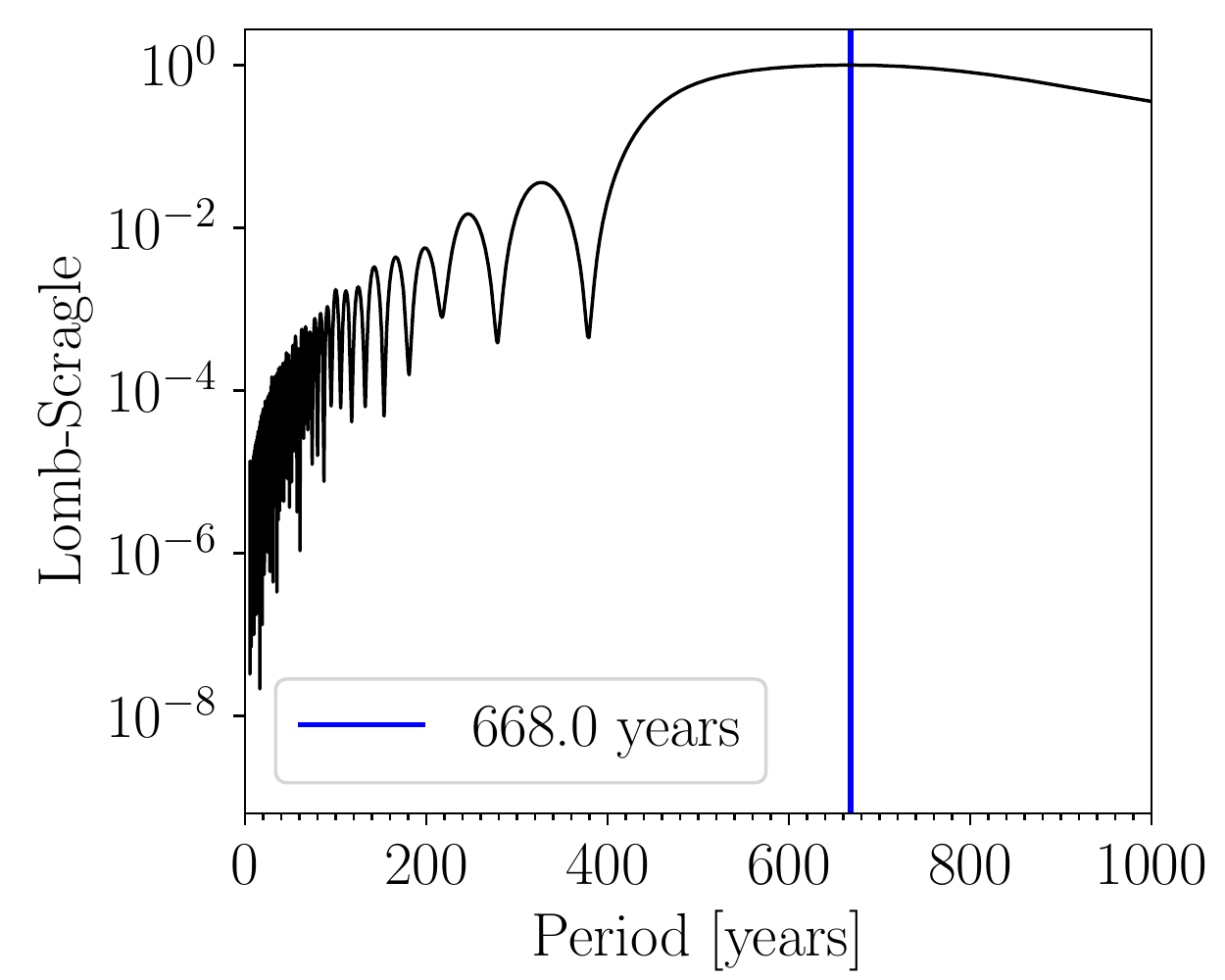}

\includegraphics[height=3.2cm]{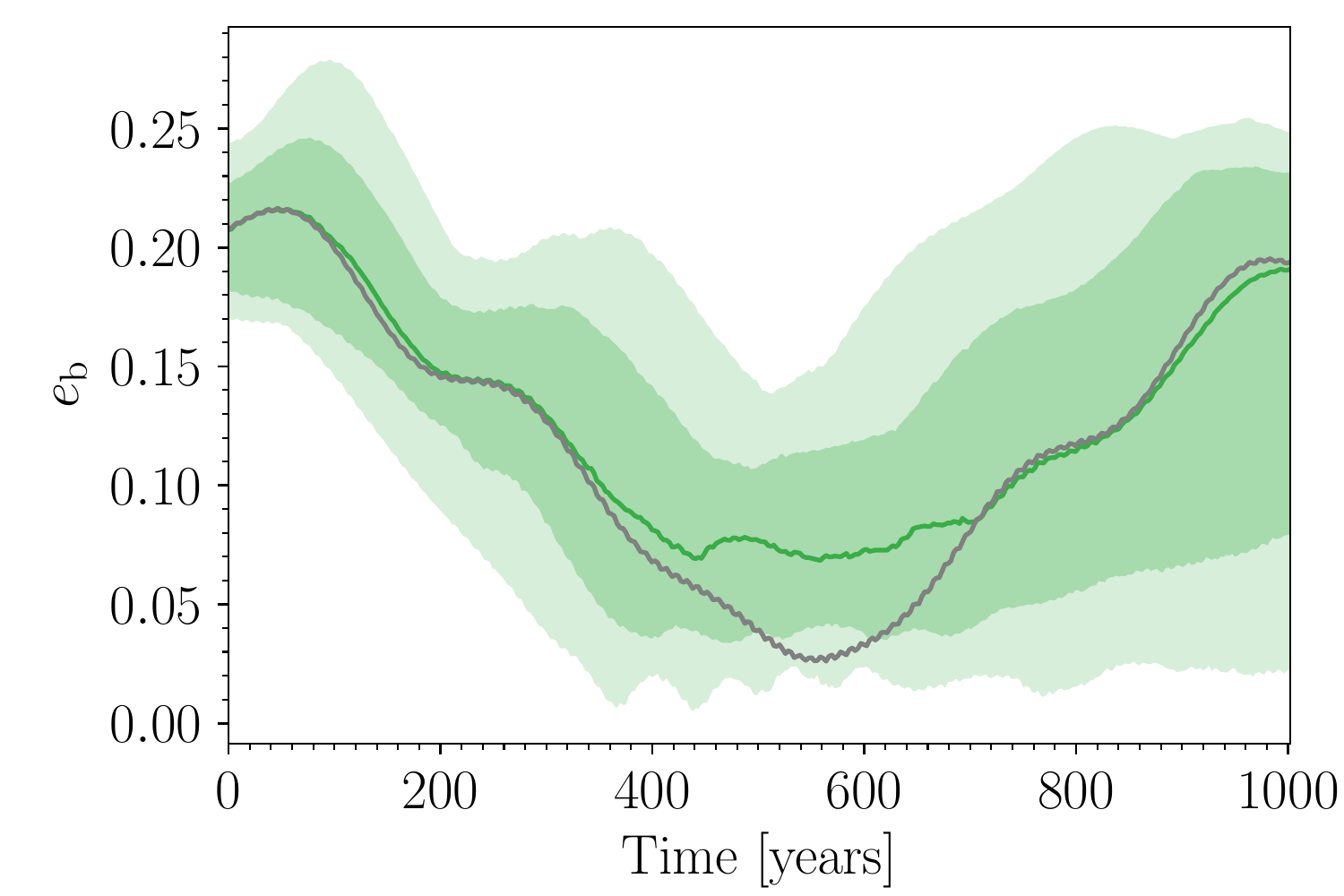}\includegraphics[height=3.2cm]{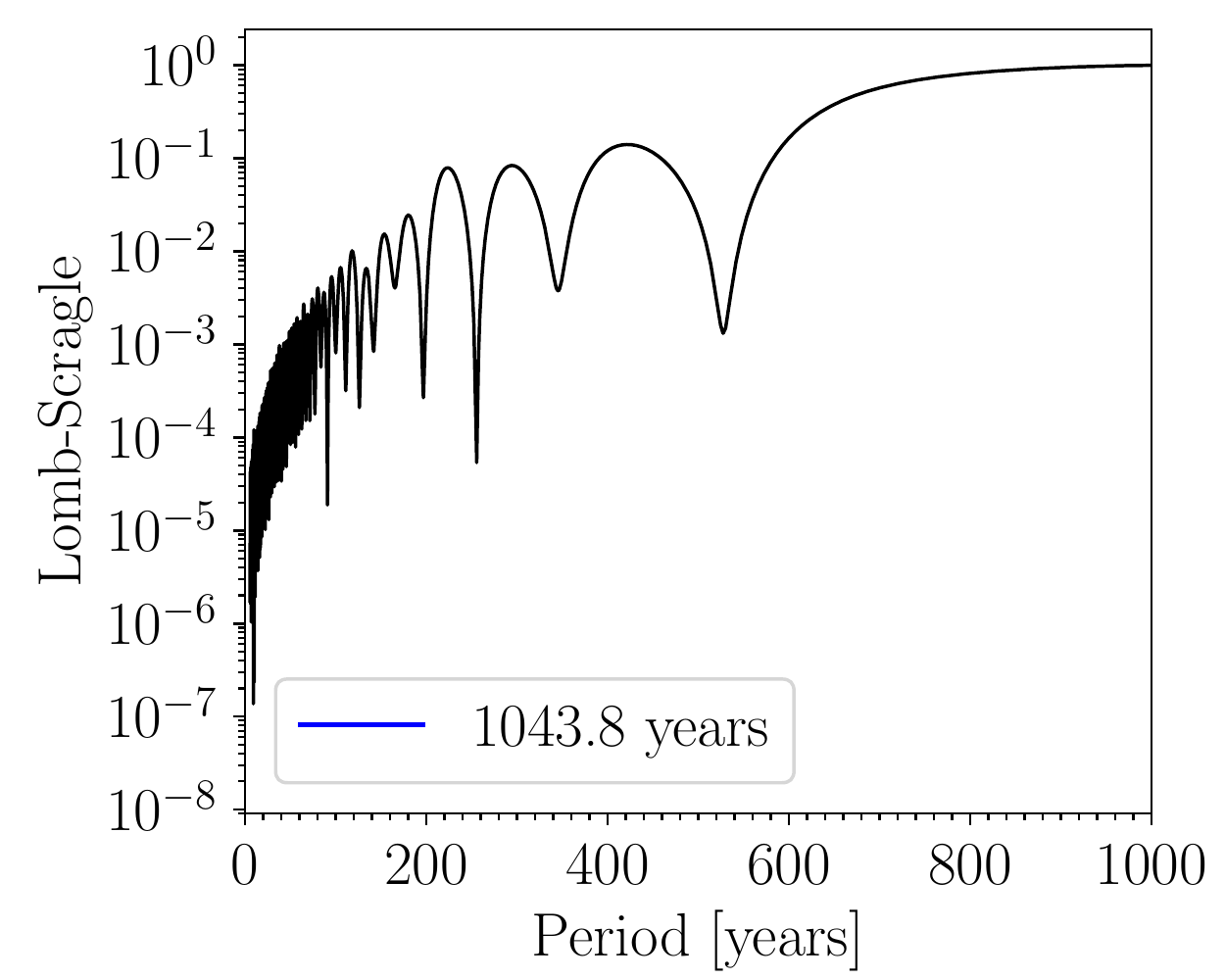}
\includegraphics[height=3.2cm]{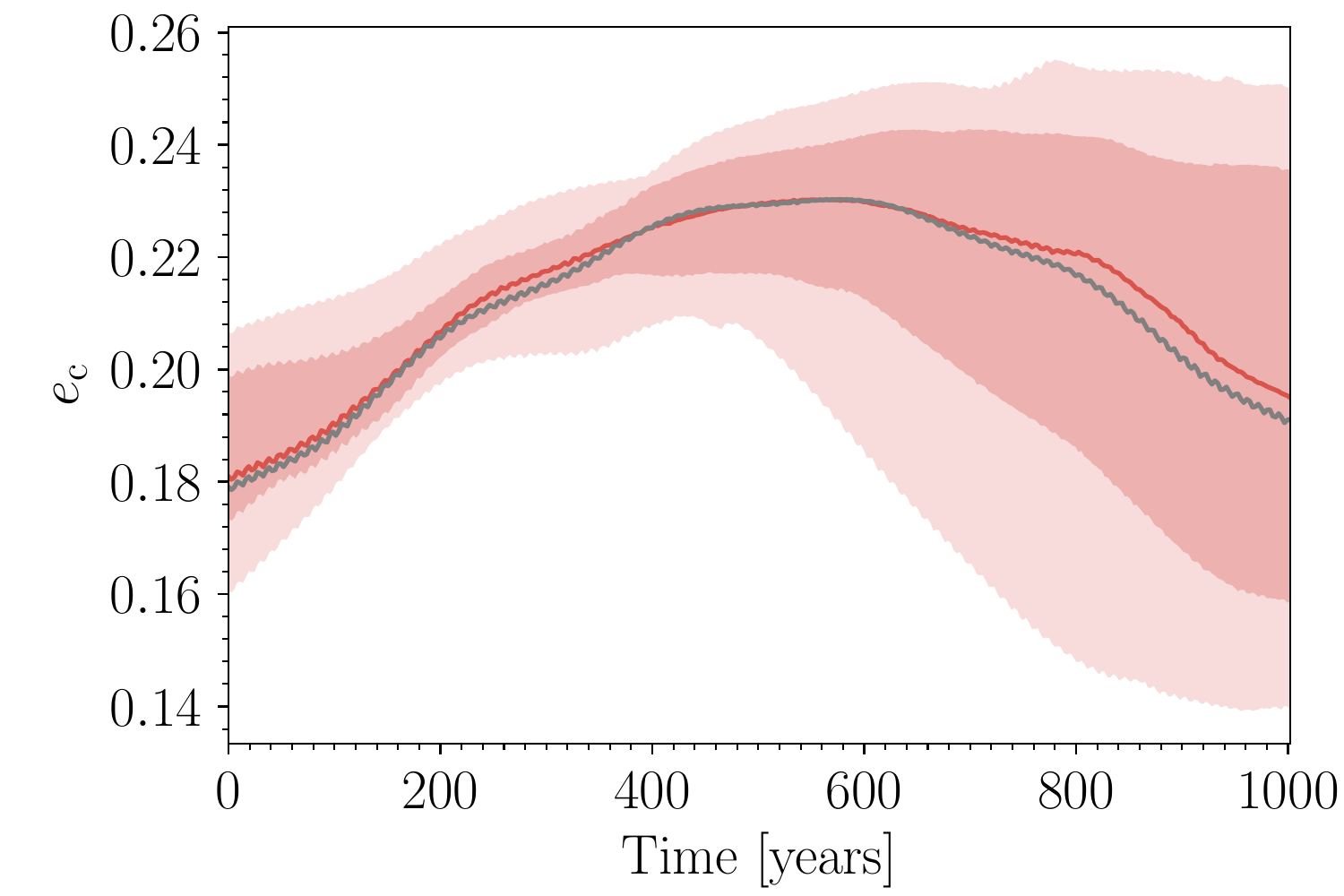}\includegraphics[height=3.2cm]{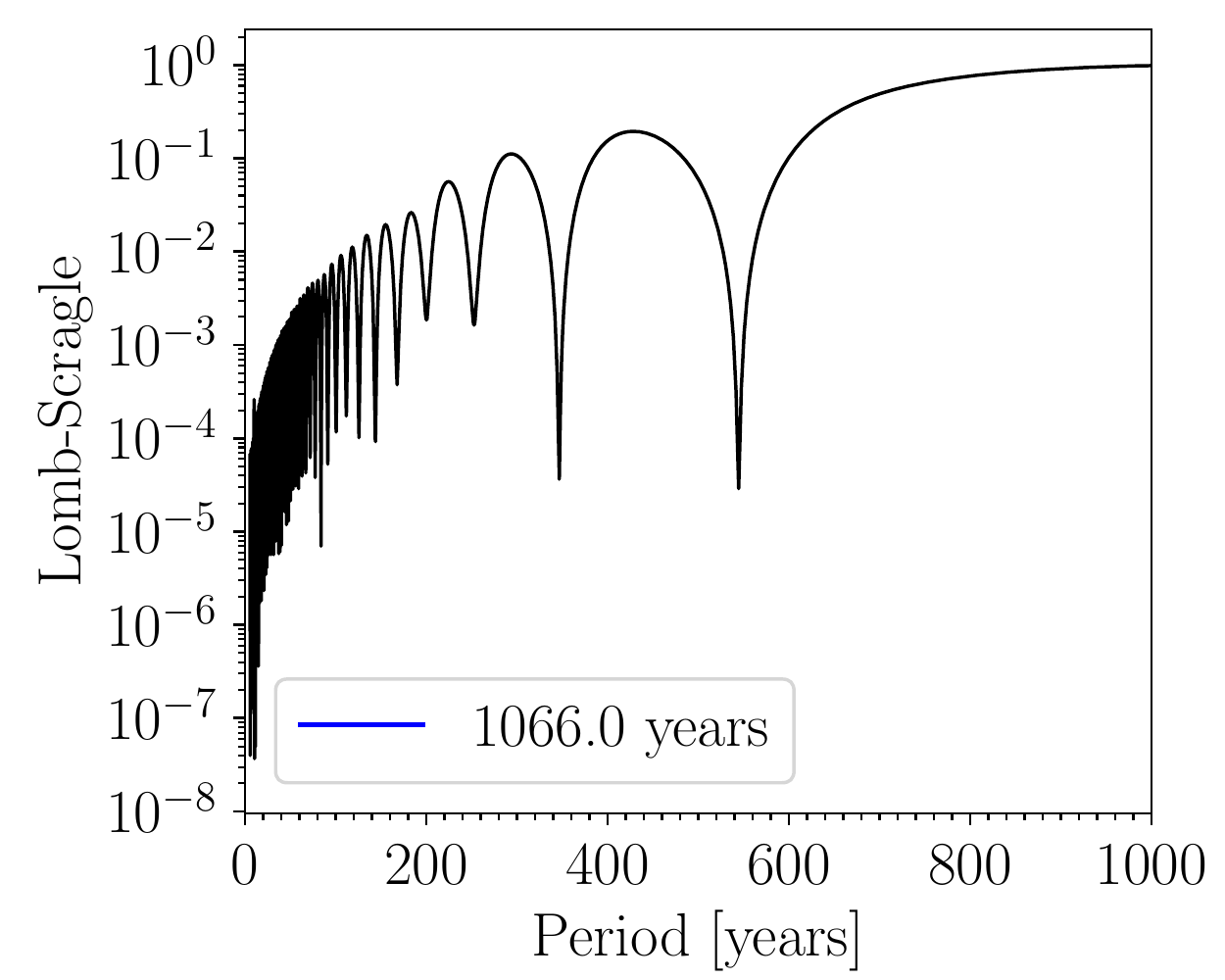}

\includegraphics[height=3.2cm]{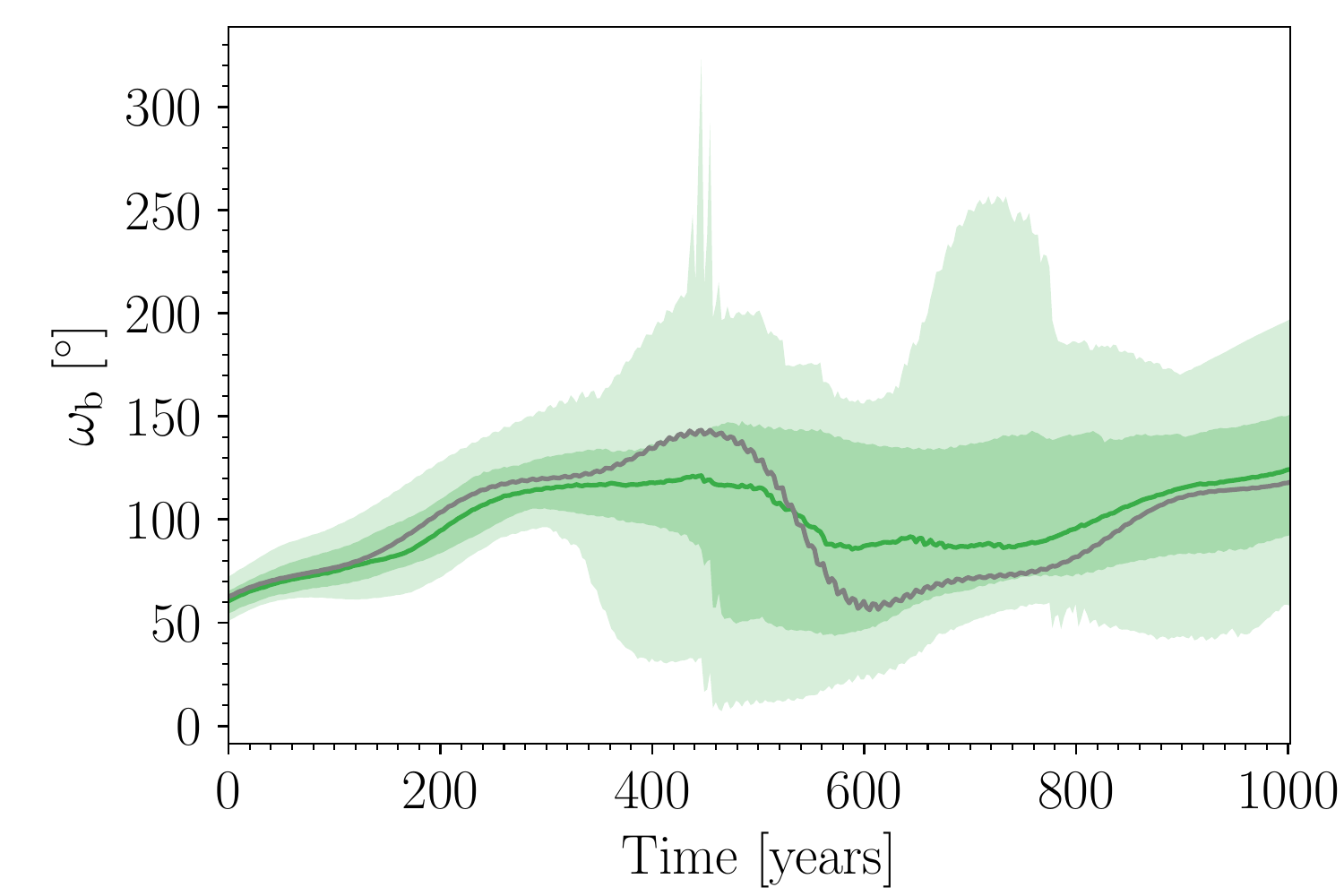}\includegraphics[height=3.2cm]{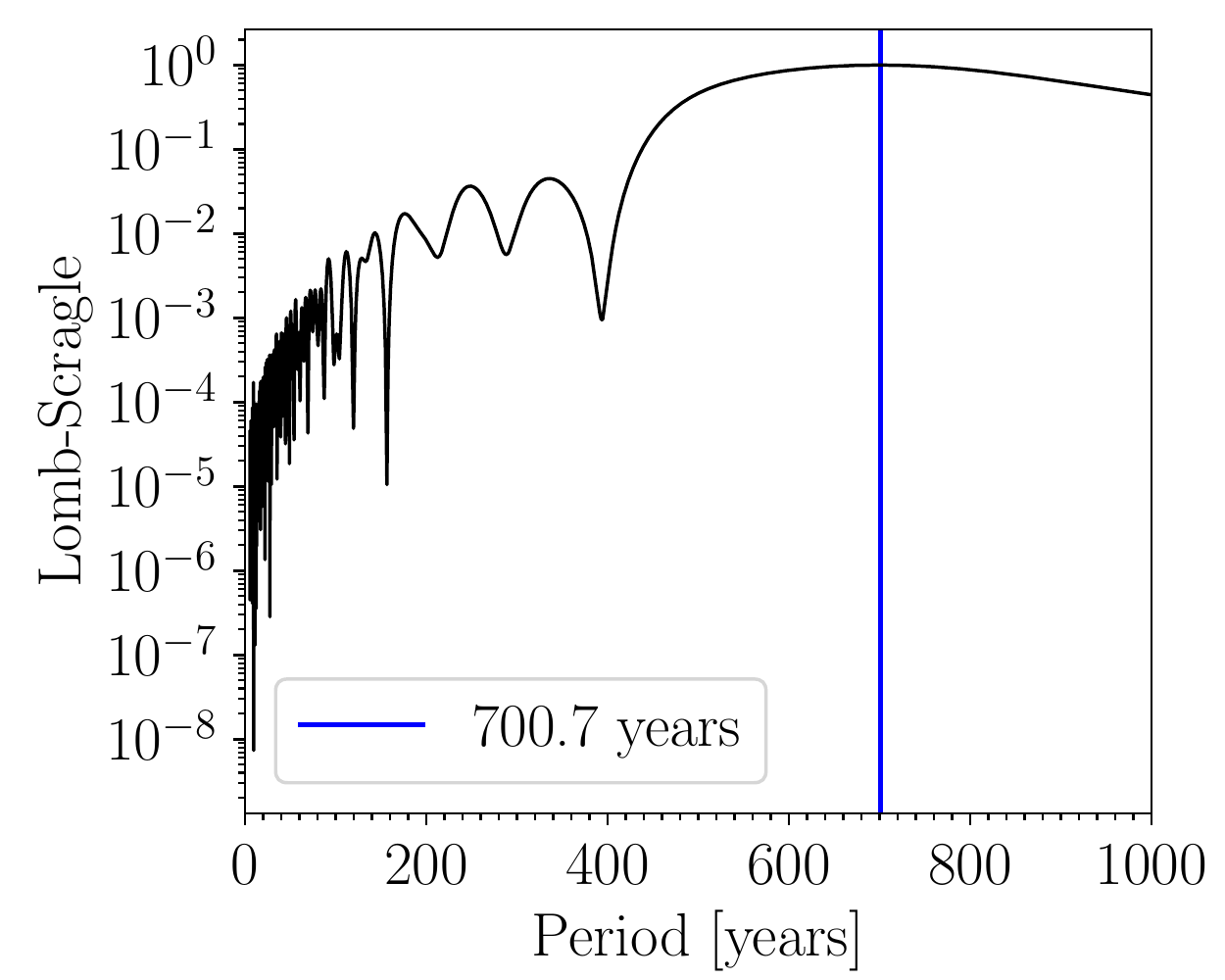}
\includegraphics[height=3.2cm]{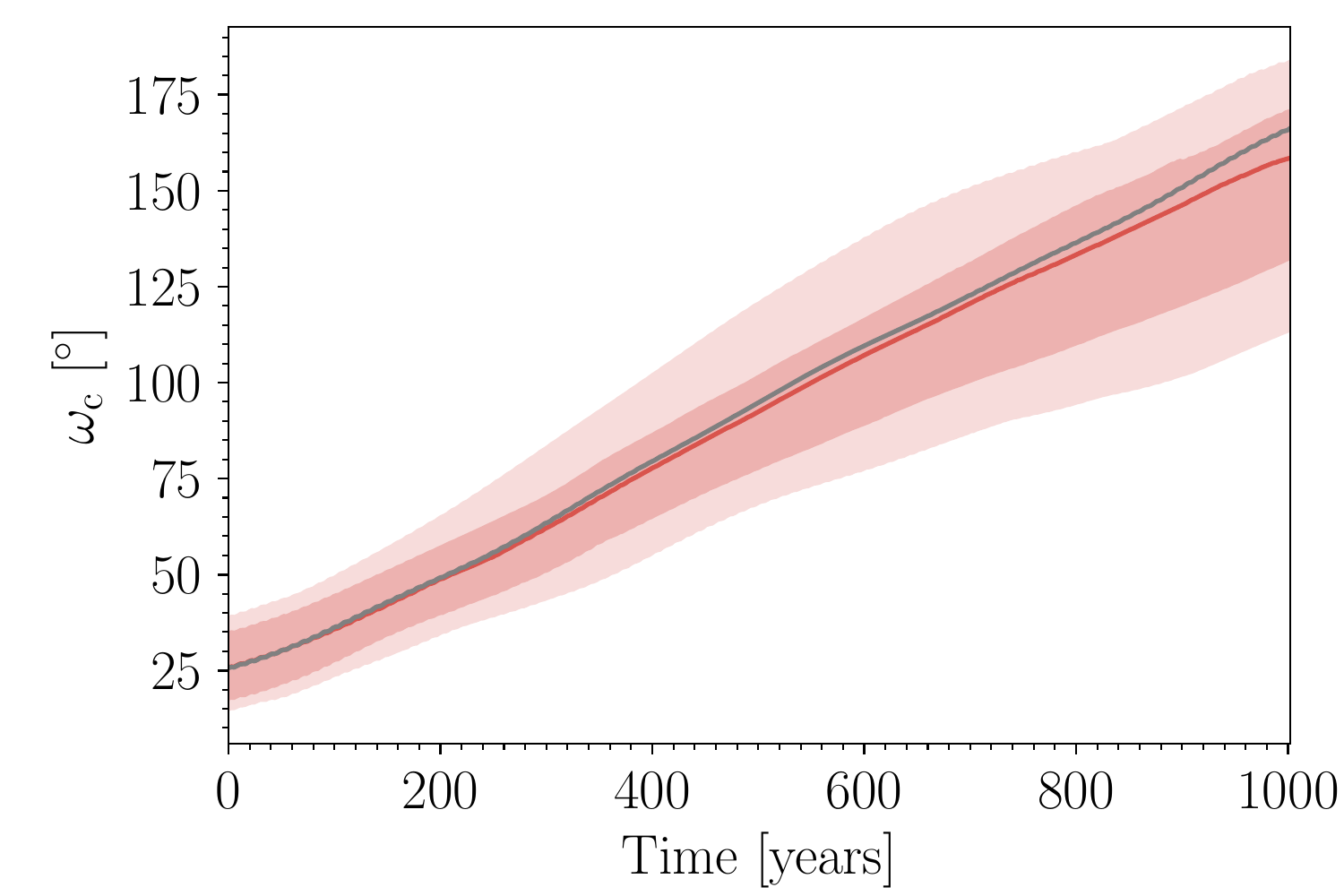}\includegraphics[height=3.2cm]{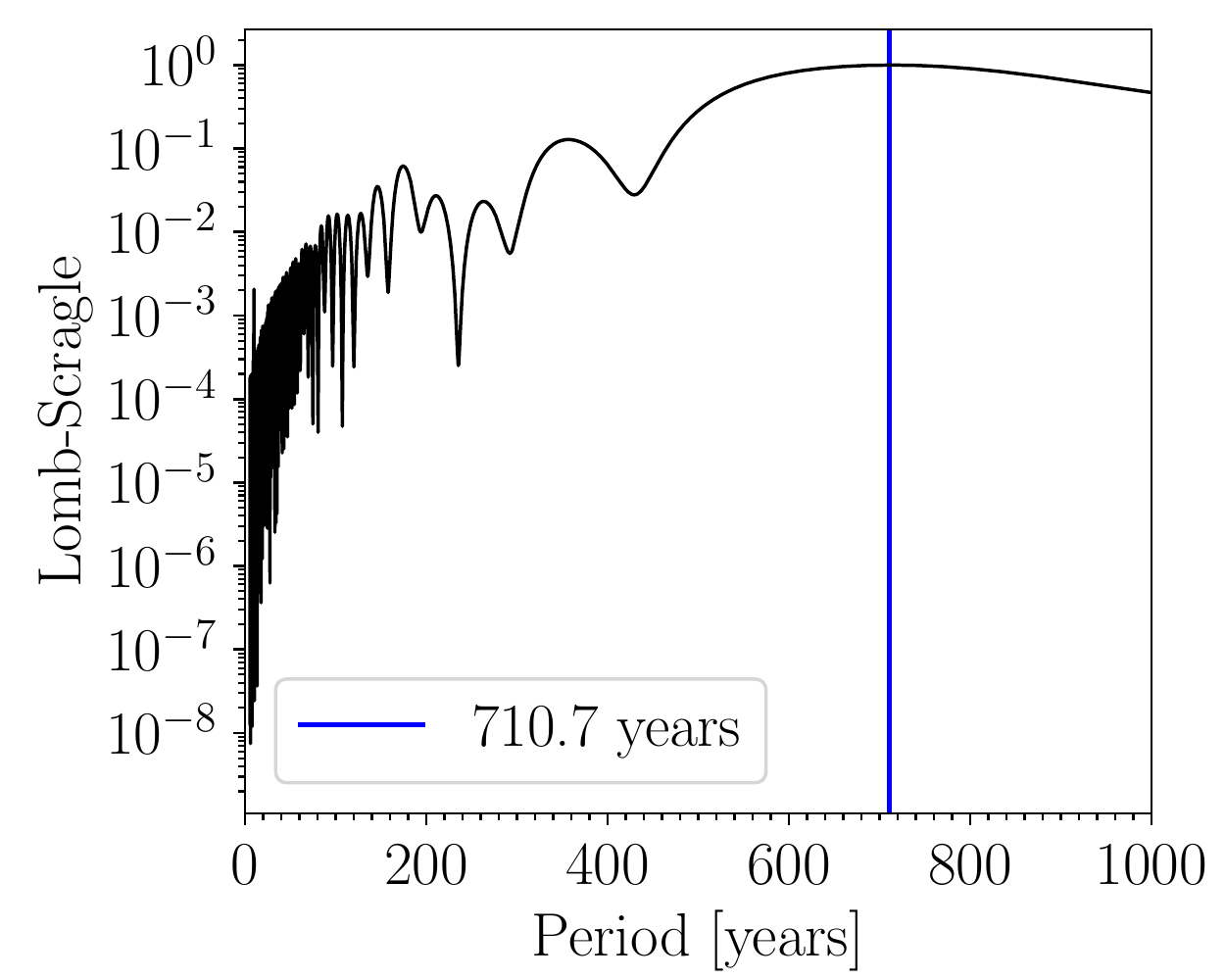}

\includegraphics[height=3.2cm]{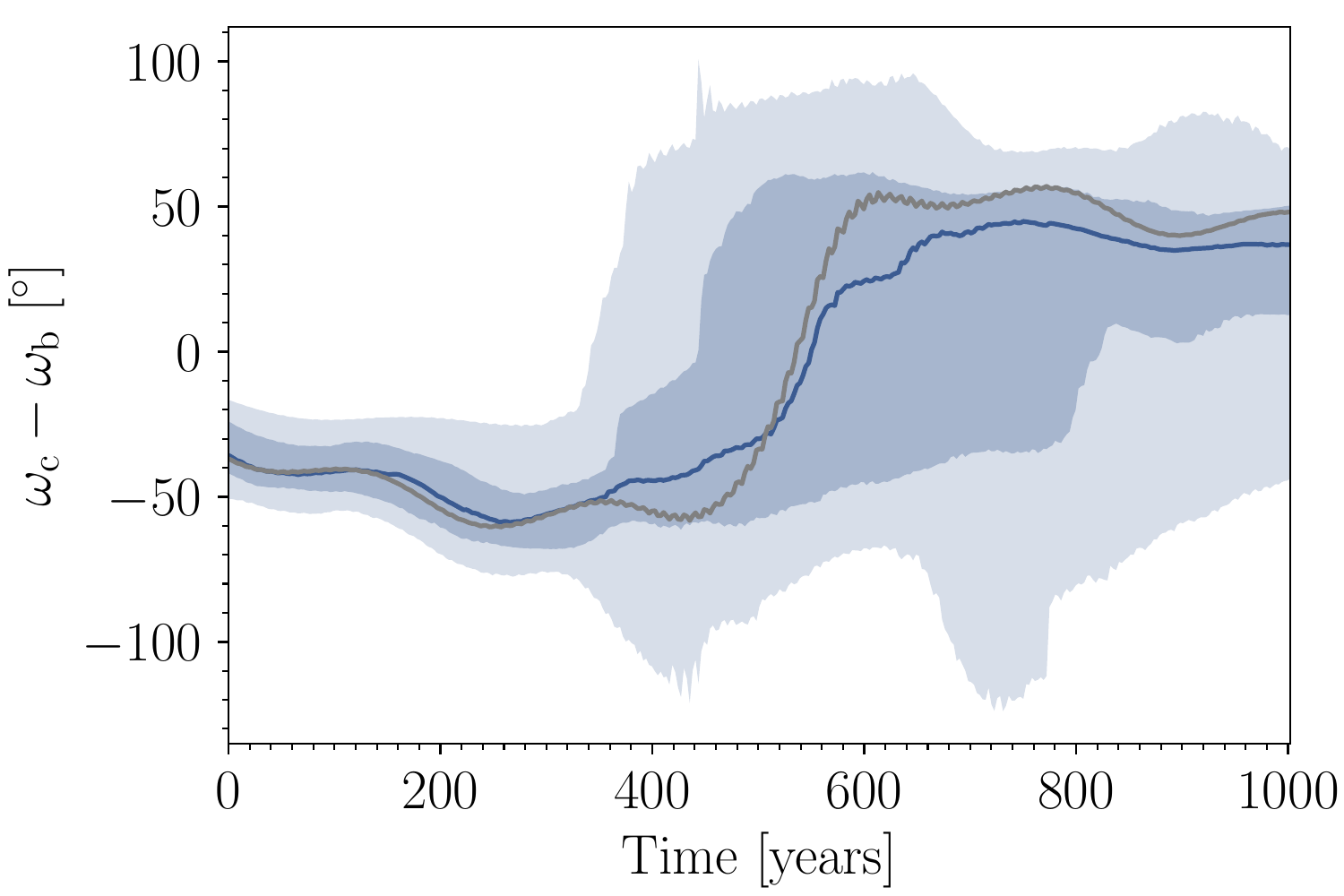}\includegraphics[height=3.2cm]{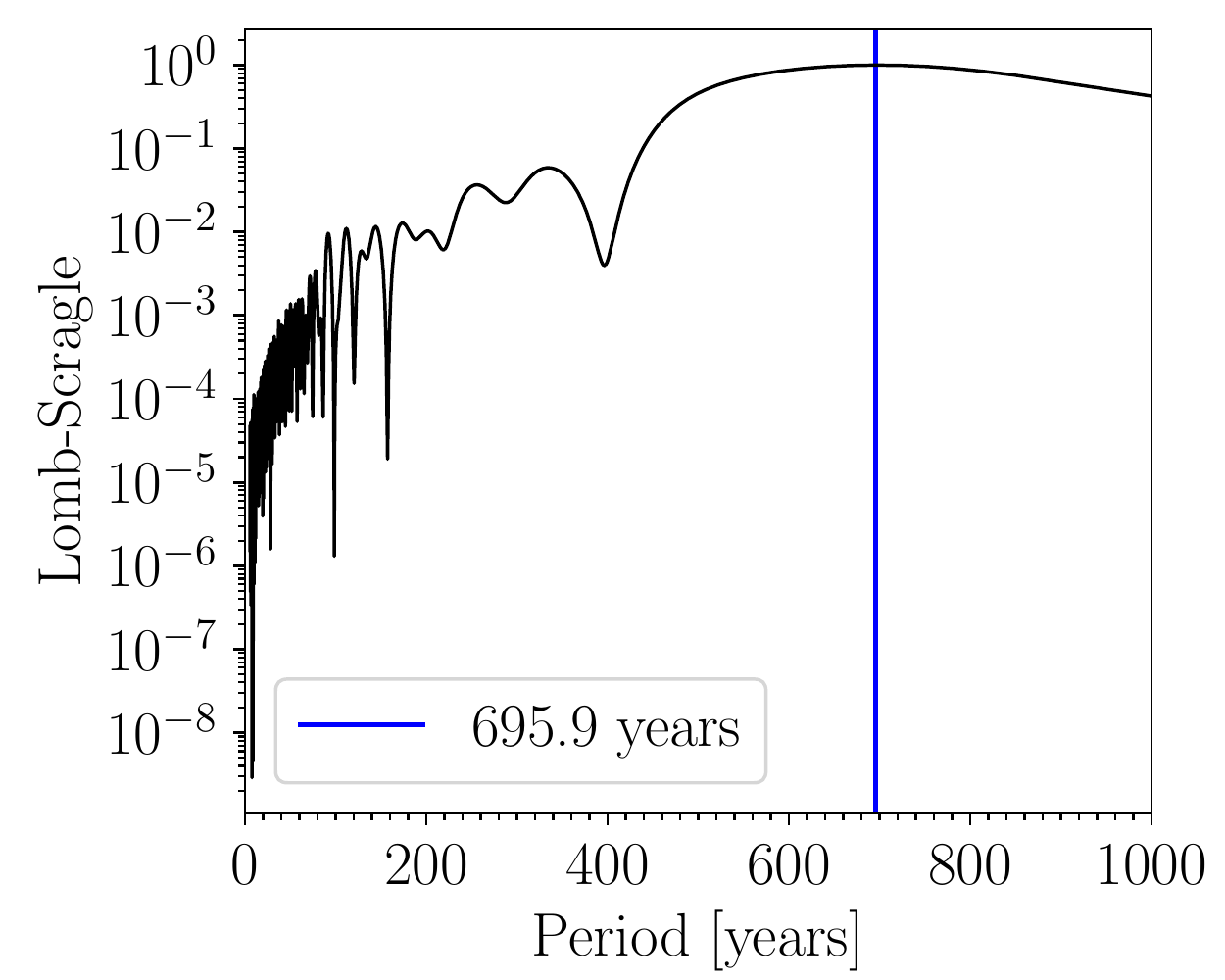}
\includegraphics[height=3.2cm]{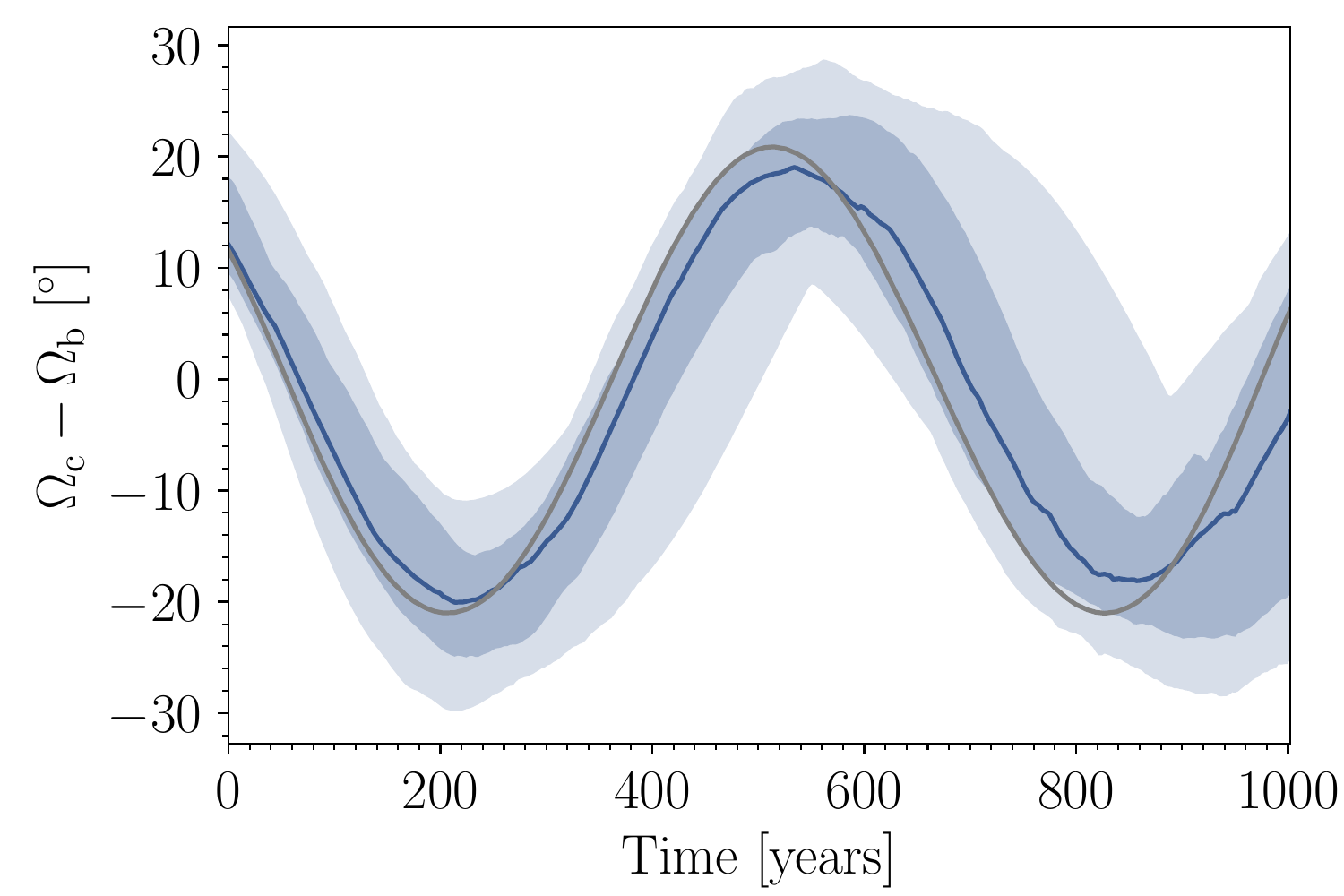}\includegraphics[height=3.2cm]{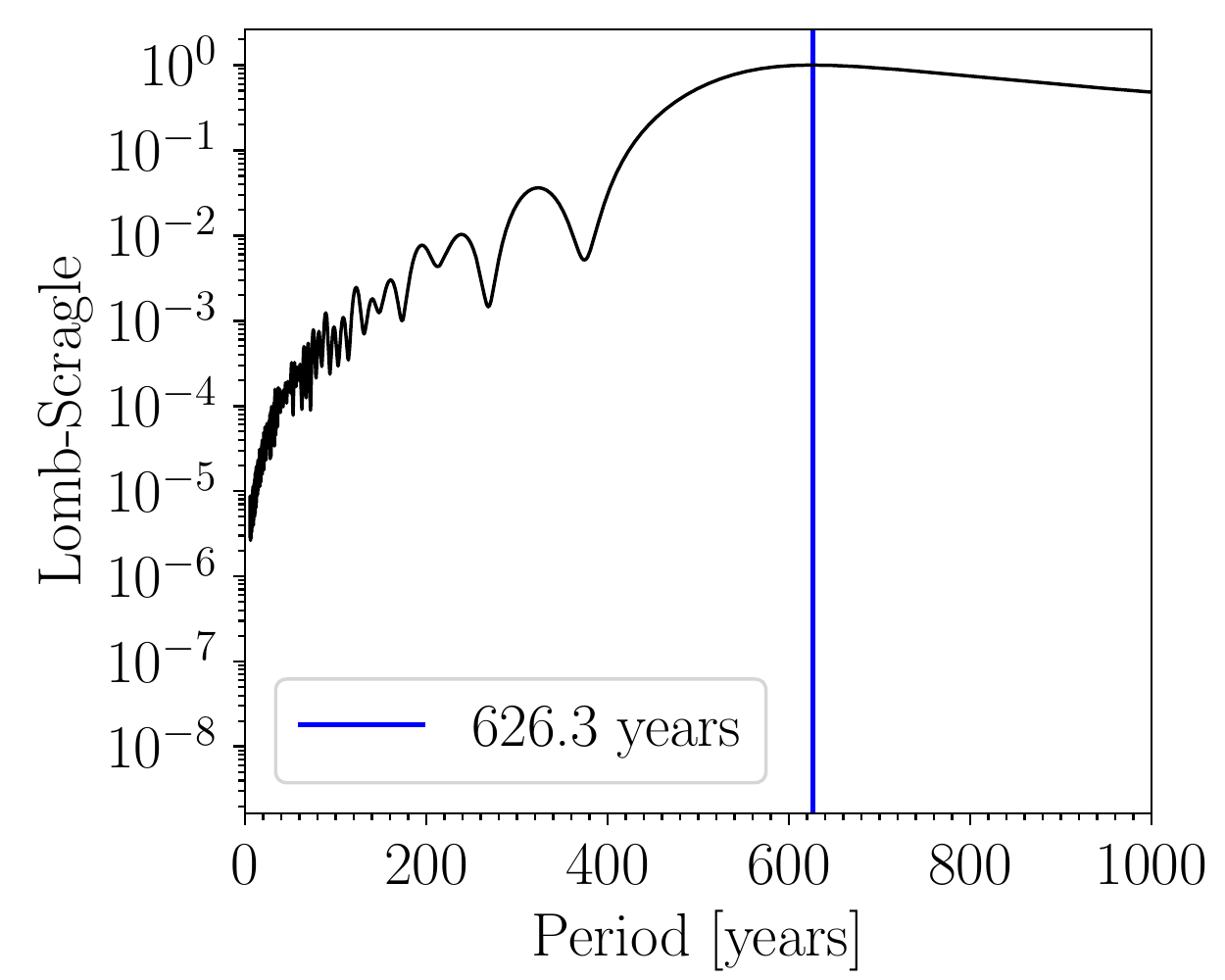}

\includegraphics[height=3.2cm]{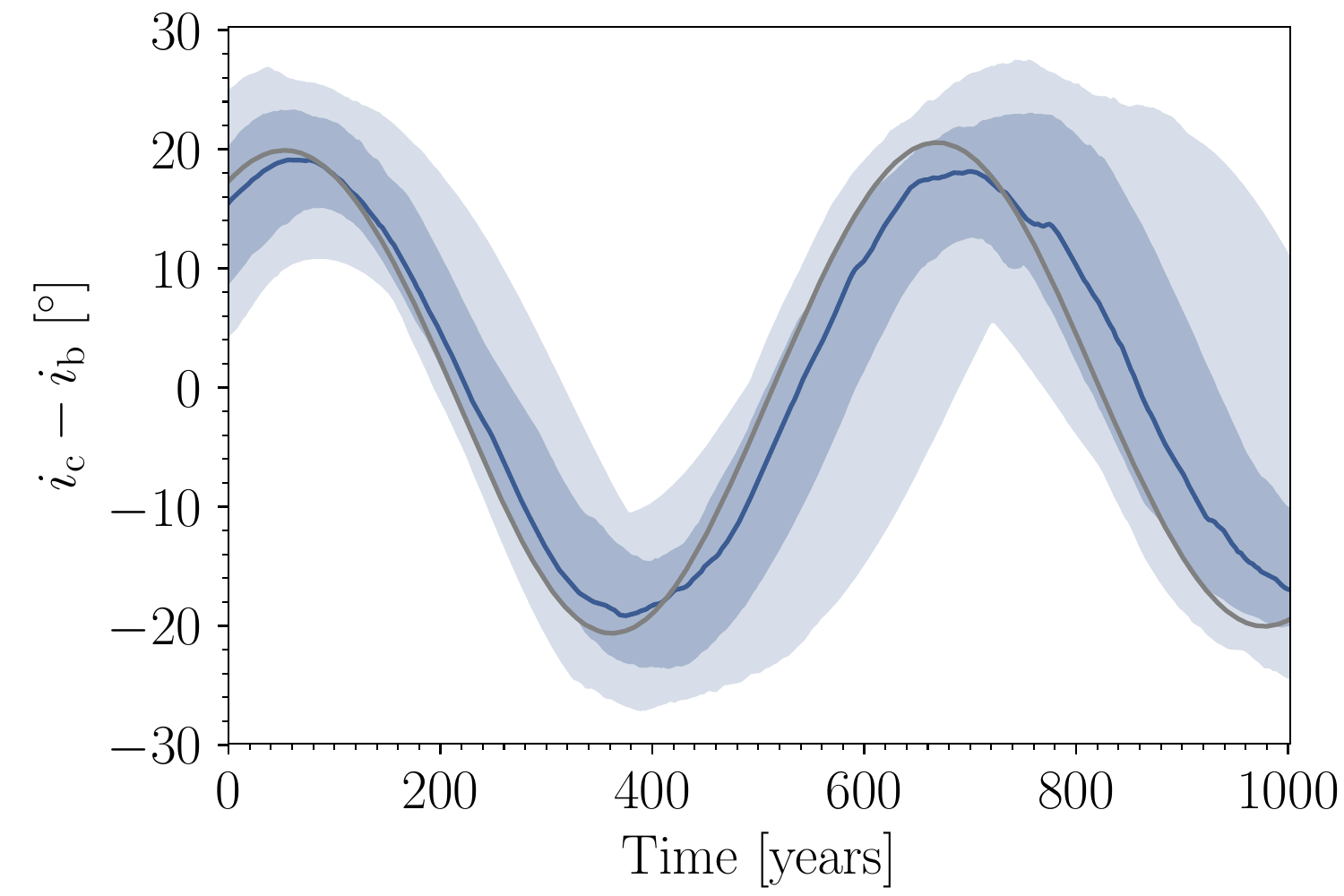}\includegraphics[height=3.2cm]{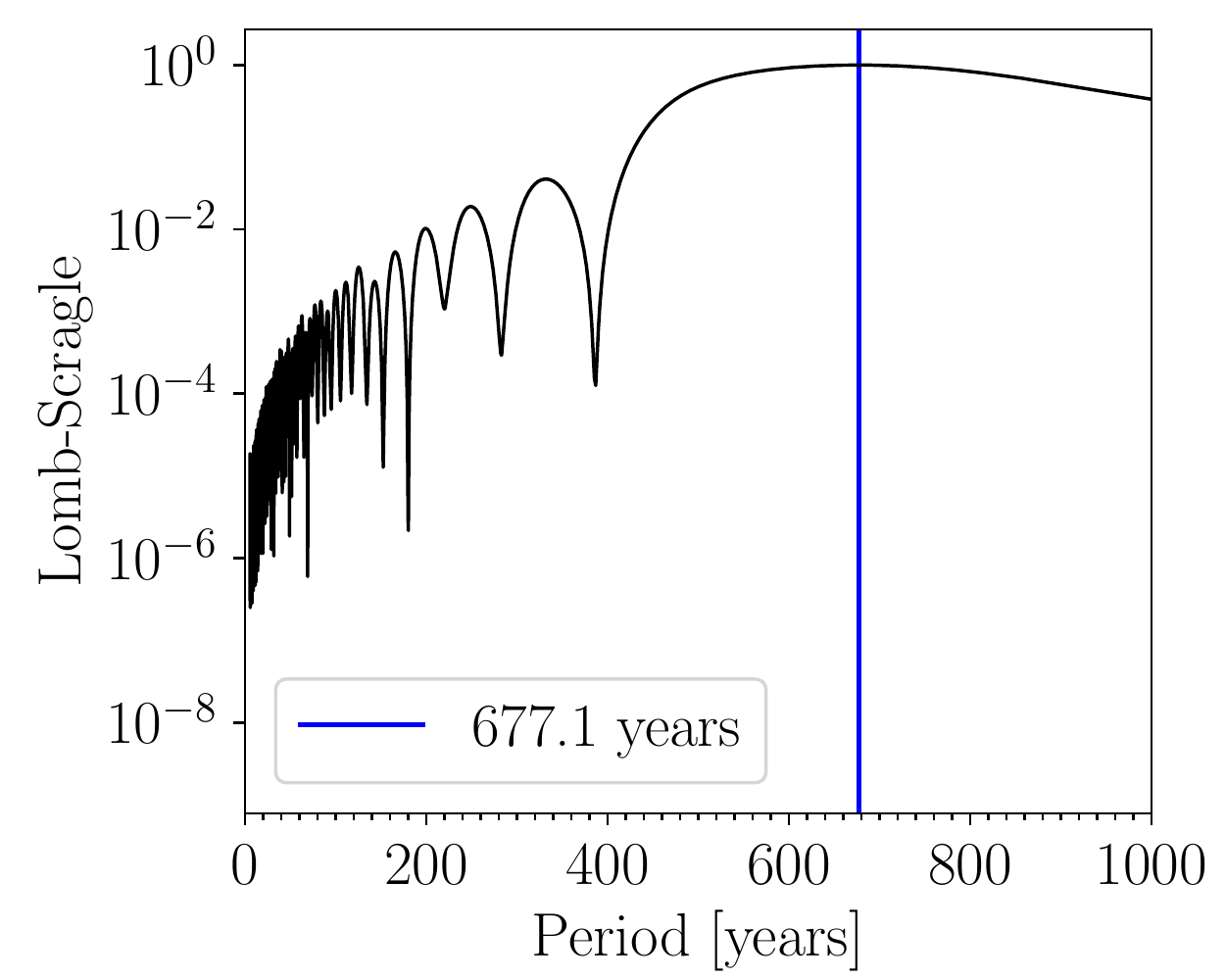}
\includegraphics[height=3.2cm]{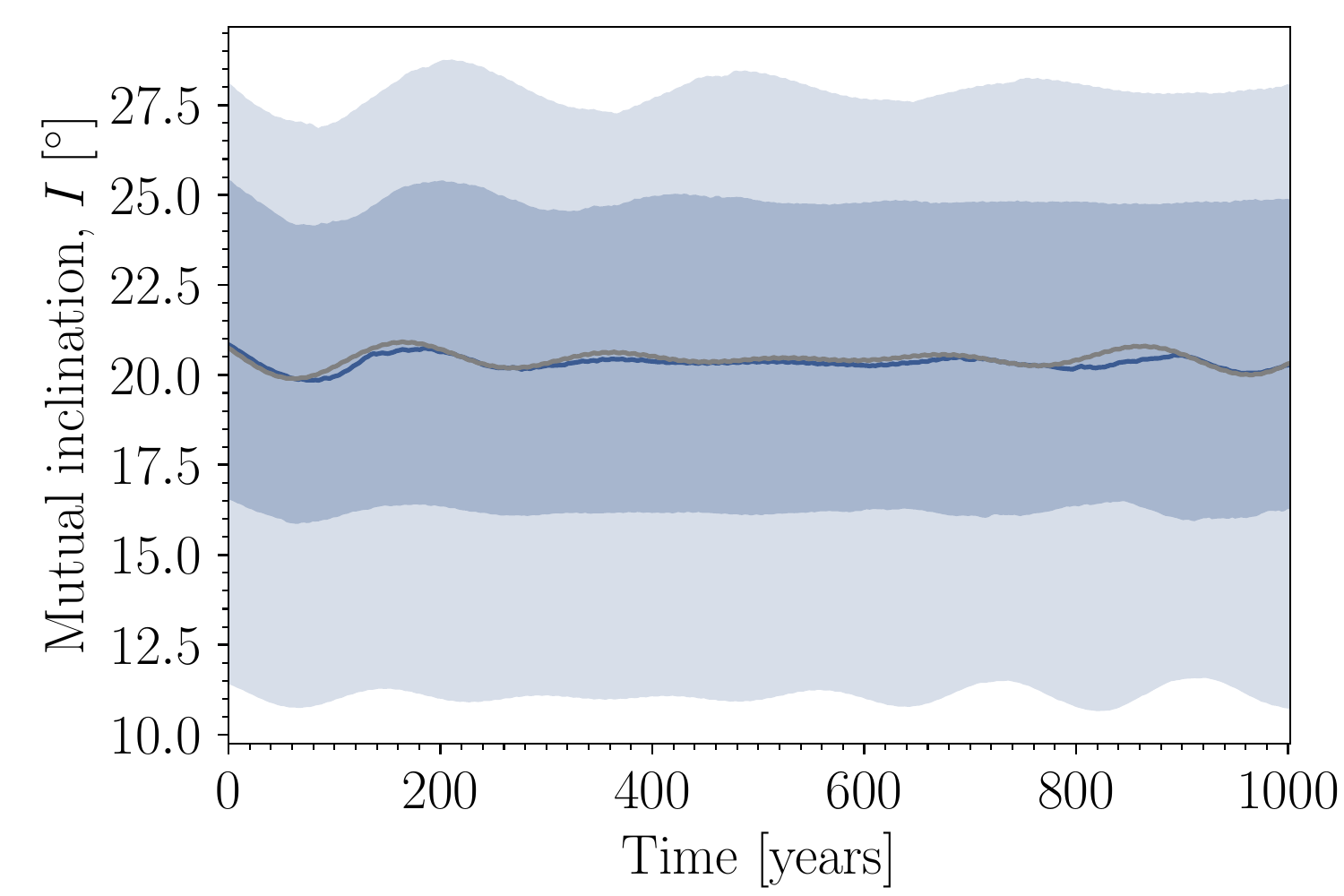}\includegraphics[height=3.2cm]{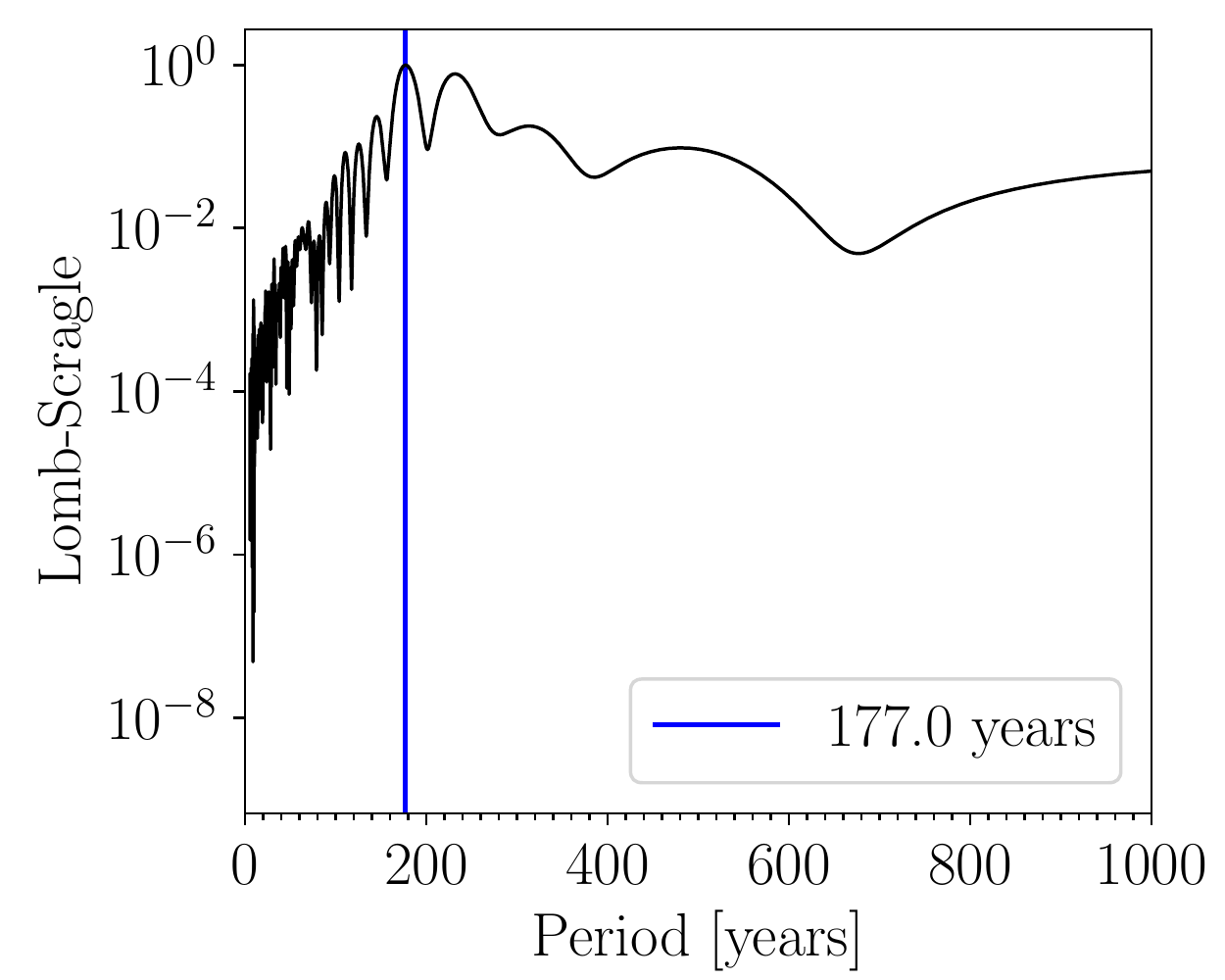}
\caption{Integrations over 1~kyr since the RISE transit from 1000~stable samples of the photodynamical modelling posterior. The 68.3\% and 95.4\% Bayesian CIs are plotted in different intensities. The solid colour curve marks the median of the posterior distribution. The solid grey curves correspond to the simulation based on the stable MAP values. The Lomb-Scargle periodogram of the median curve is shown, and the position of the most prominent peak is annotated.}
\label{fig.LongTermEvolution}
\end{figure*}

\newpage
\clearpage

\begin{figure*}
\includegraphics[height=7cm]{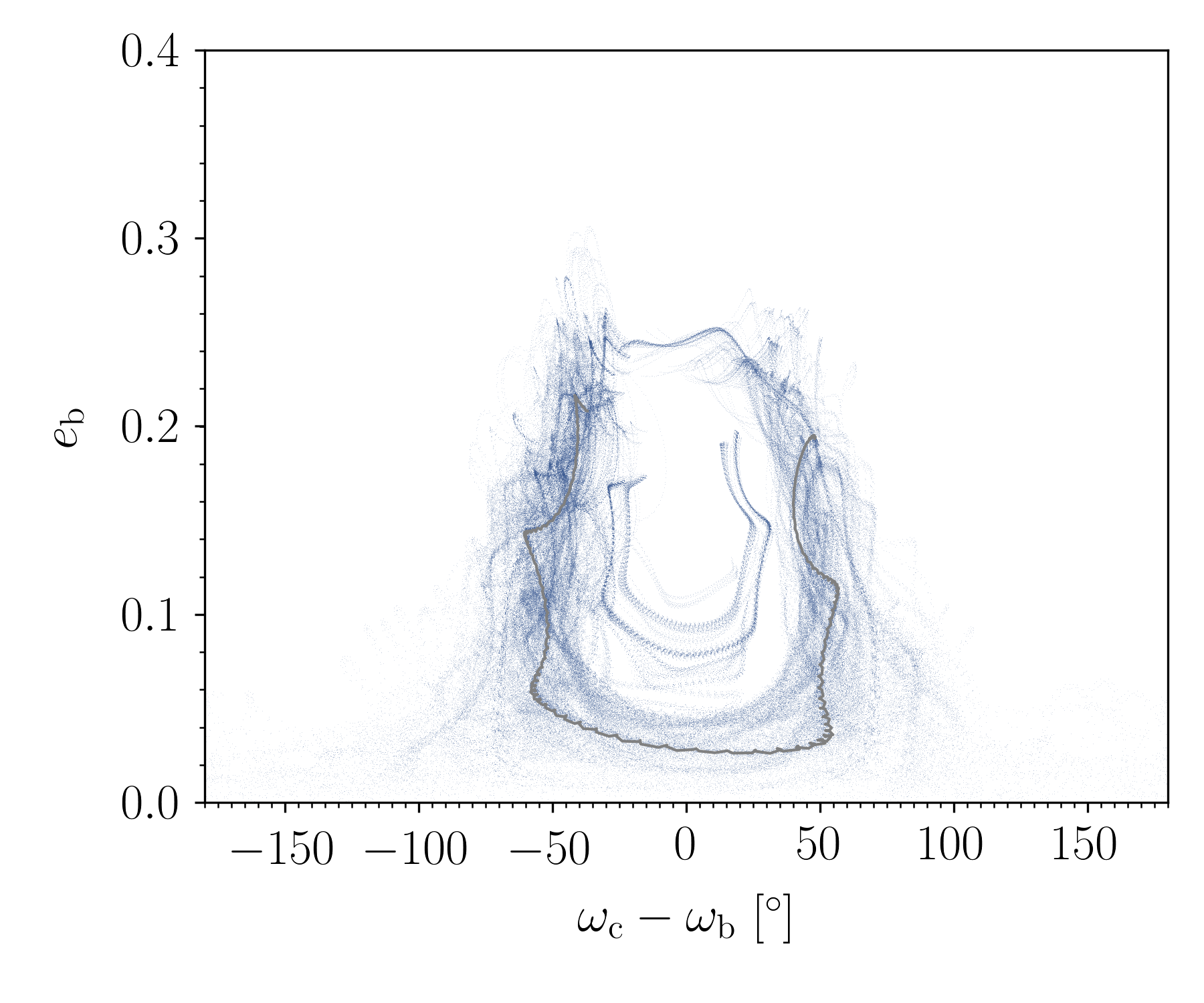}\includegraphics[height=7cm]{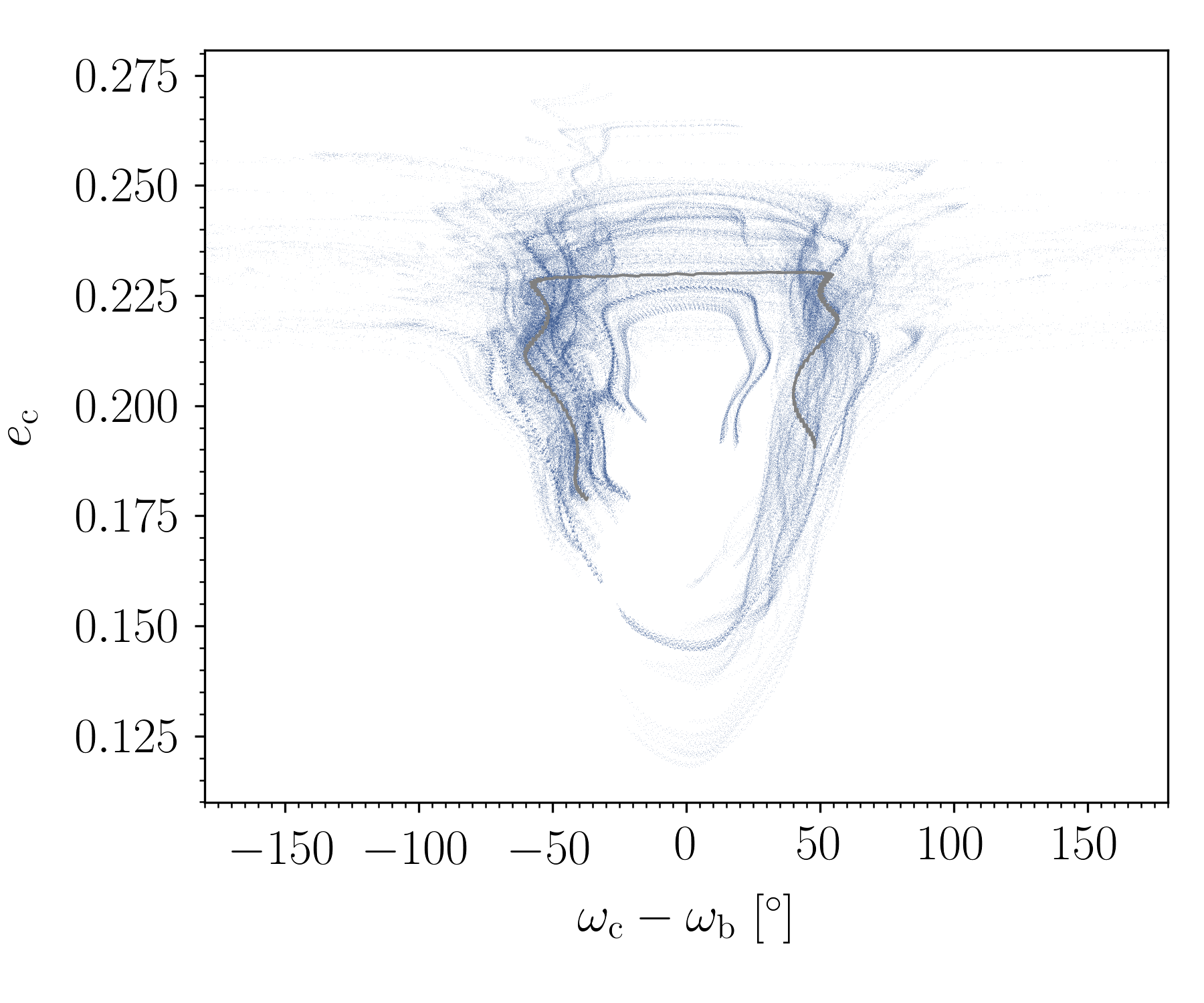}

\includegraphics[height=7cm]{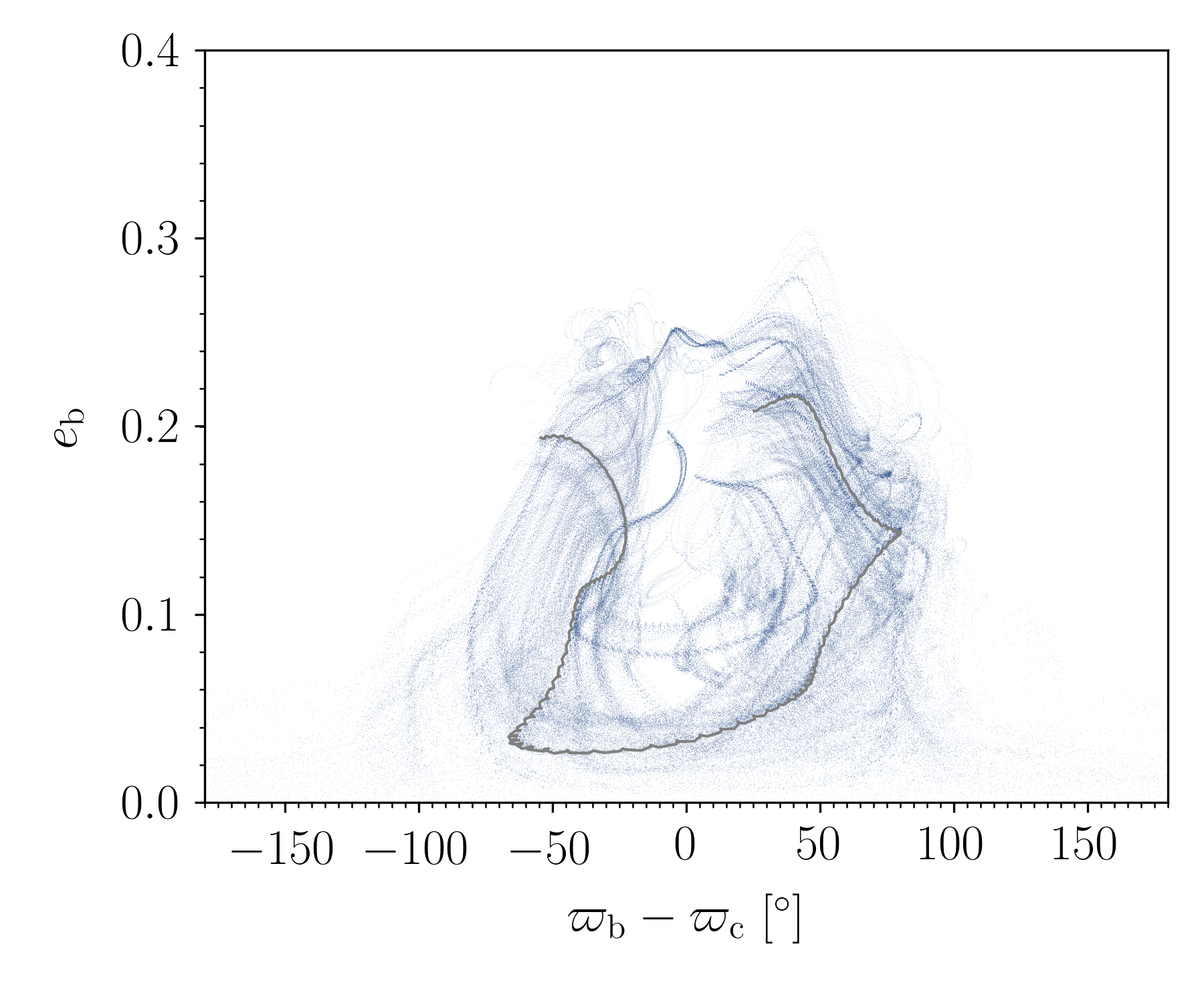}\includegraphics[height=7cm]{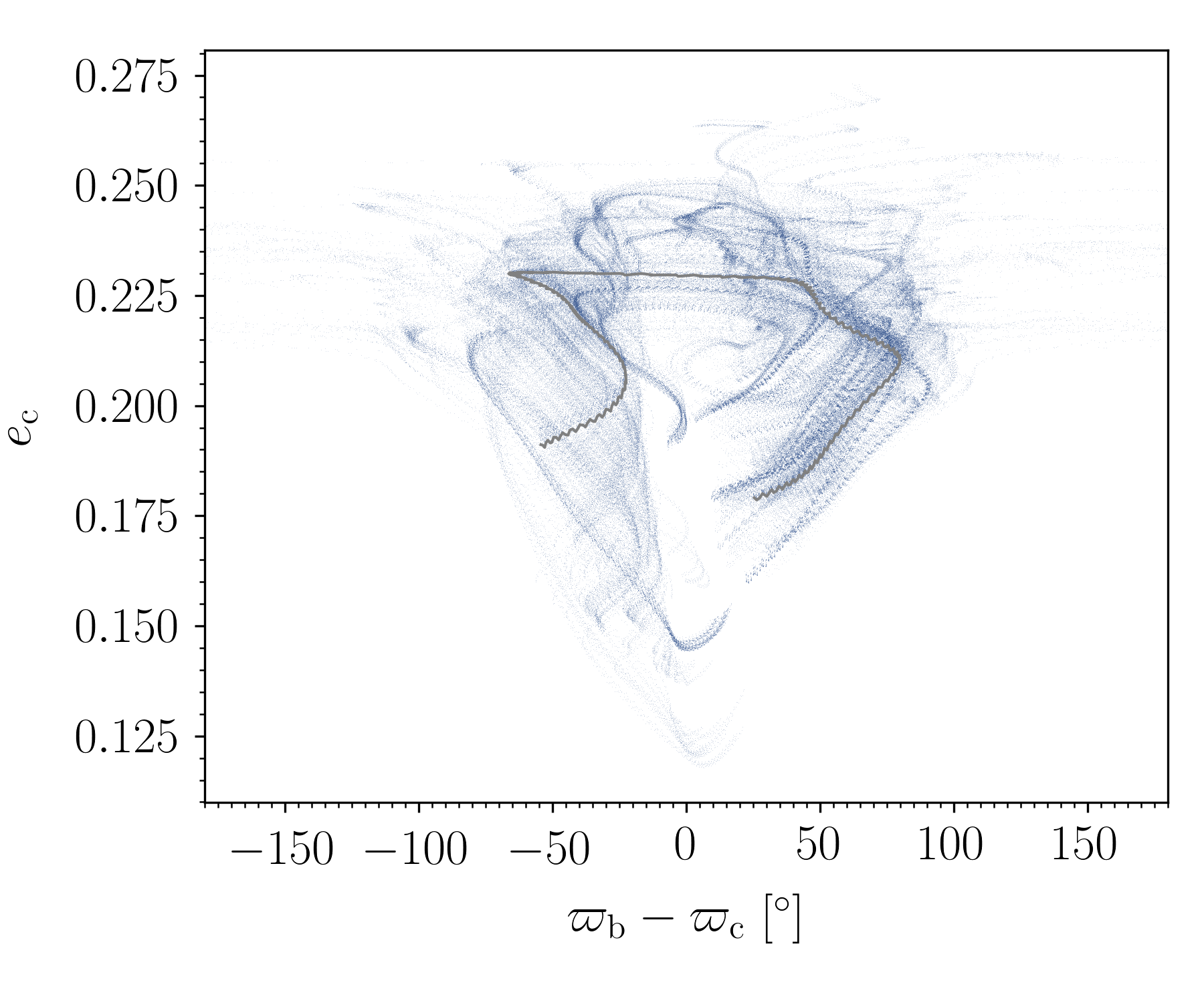}

\includegraphics[height=7cm]{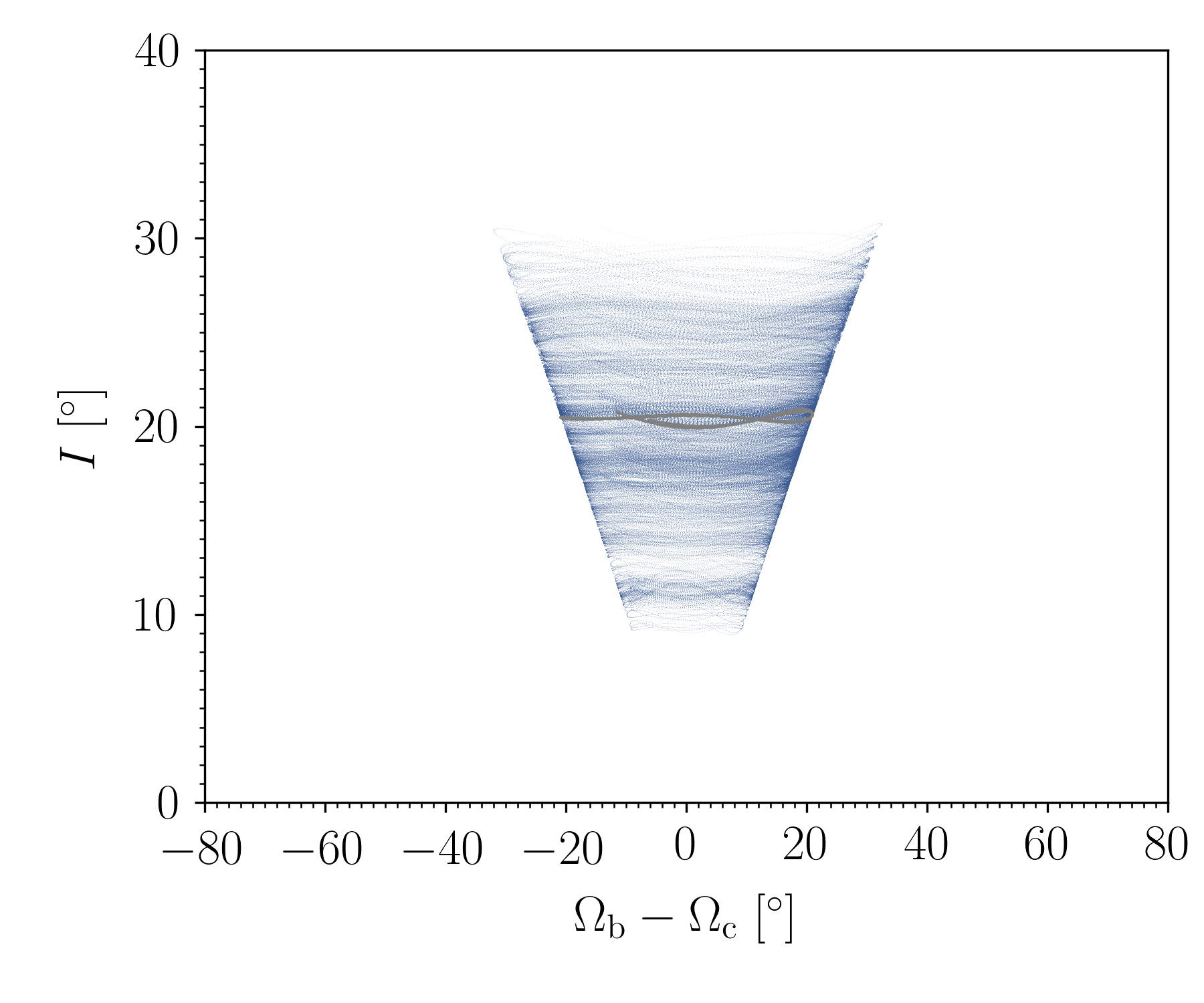}\includegraphics[height=7cm]{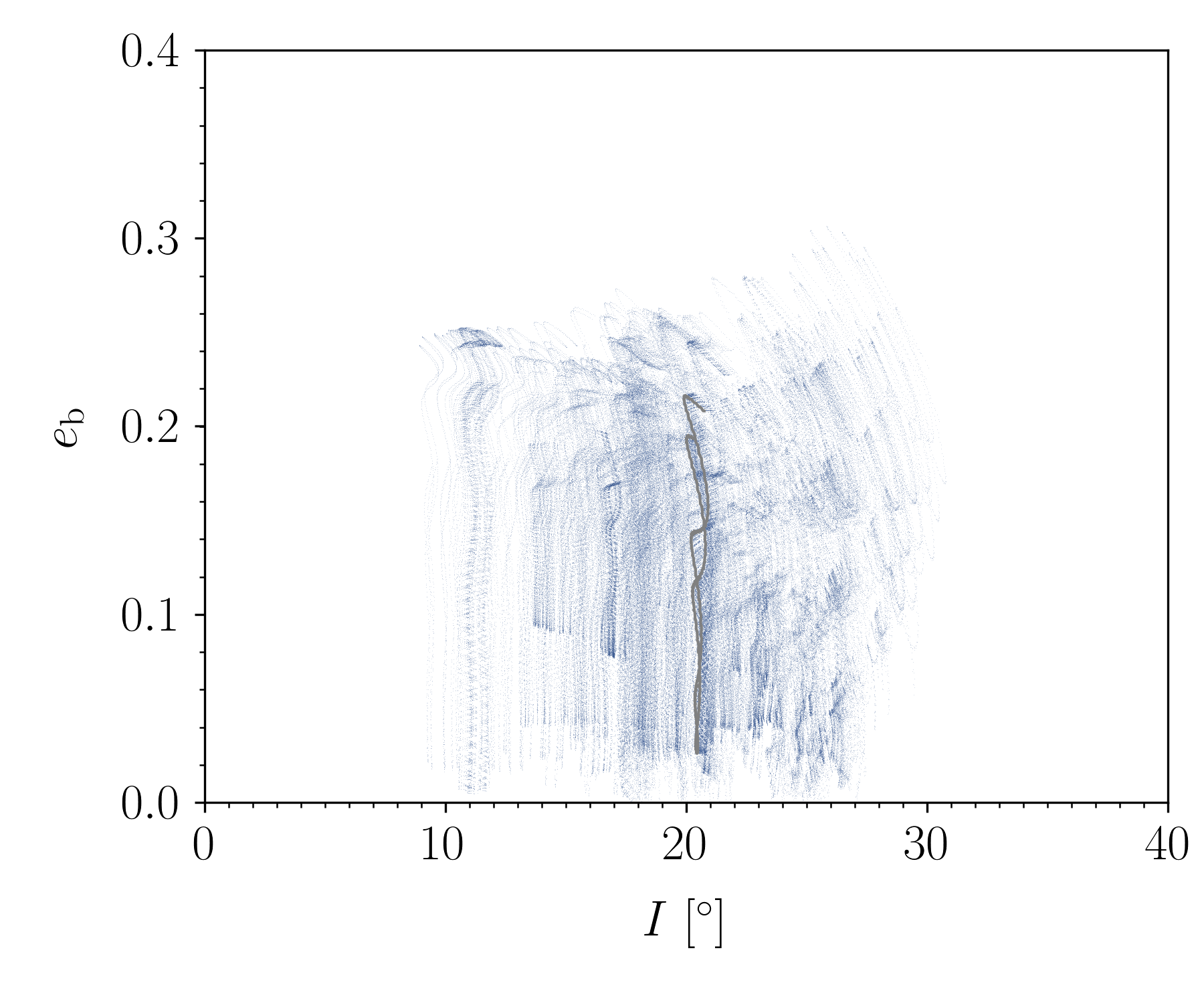}
\caption{Idem.\ as Figure~\ref{fig.LongTermEvolution}, but for correlations of selected parameters ($\varpi=\Omega+\omega$ is the longitude of the periapsis).}
\label{fig.LongTermEvolutionCorrelations}
\end{figure*}

\end{appendix}
\end{document}